\documentclass{svjour2}      
\smartqed 
\usepackage{xcolor}
\usepackage{graphicx}
\usepackage{color}
\usepackage[T1]{fontenc}
\usepackage[latin9]{inputenc}
\usepackage{mathptmx}      
\usepackage{geometry}
\usepackage{verbatim}
\usepackage{float}
\usepackage{amsmath}
\usepackage{amssymb}
\usepackage{esint}
\usepackage{caption}
\usepackage{ulem}
\journalname{Journal of Statistical Physics}

\begin{document}

\title{Equilibrium statistical mechanics and energy partition for the shallow water model}

\author{A. Renaud, A. Venaille, F. Bouchet}

\institute{Laboratoire de Physique de l'\'Ecole Normale Sup\'erieure de Lyon, Universit\'e de Lyon, CNRS, 46 All\'ee d'Italie, F-69364 Lyon CEDEX 07, France. antoine.renaud@ens-lyon.fr, antoine.venaille@ens-lyon.fr, freddy.bouchet@ens-lyon.fr }
\date{Received:  / Accepted: }

\maketitle

\begin{abstract}
The aim of this paper is to use large deviation theory in order to compute the entropy of macrostates for the microcanonical measure of the shallow water system. The main prediction of this full statistical mechanics computation is the  energy partition between a large scale vortical flow and small scale fluctuations related to inertia-gravity waves. We introduce for that purpose a {  semi-Lagrangian} discrete model of the continuous shallow water system, and compute the corresponding statistical equilibria. We argue that microcanonical equilibrium states of the discrete model in the continuous limit are equilibrium states of the actual shallow water system.

We show that the presence of small scale fluctuations selects a subclass of equilibria among the states that were previously computed by phenomenological approaches that were neglecting such fluctuations. In the limit of weak height fluctuations, the equilibrium state can be interpreted as two subsystems in thermal contact: one subsystem corresponds to the large scale  vortical flow, the other subsystem corresponds to small scale height and velocity fluctuations. It is shown that either a non-zero circulation or rotation and bottom topography are required to sustain a non-zero large scale flow at equilibrium.  

Explicit computation of the equilibria and their energy partition is presented in the quasi-geostrophic limit for the energy-enstrophy ensemble. The possible role of small scale dissipation and shocks is discussed. A geophysical application to the Zapiola anticyclone is presented.

\end{abstract}


\newpage

\tableofcontents

\section{Introduction\label{sec:Introduction}}

Geophysical turbulent flows have the propensity to self-organize into
large scale coherent structures such as cyclones, anticyclones and
jets. These coherent structures are long lived, but can also loose
energy, for instance through the radiation of waves that eventually
break into small scale structures. The aim of this paper is to understand
the energy partition into large scale structures and small scale fluctuations
in the framework of freely evolving shallow water dynamics, using
statistical mechanics arguments. Indeed, geophysical turbulent flows
involve a huge number of degrees of freedom coupled through non-linear
interactions, which strongly motivates a statistical mechanics approach.
This approach allows to reduce the study of self-organization and
energy partition down to a few parameters, such as the total energy
of the flow and its total enstrophy. 

In the case of the three dimensional Euler equations, equilibrium
statistical mechanics predicts that all the energy is lost into small
scales, consistently with the classical picture of a small scale energy
transfer. By contrast, two-dimensional flows are characterized by a 
large scale energy transfer, and equilibrium tools are appropriate to describe
the large scale structure resulting from self-organization at the
domain scale, in the absence of forcing and dissipation. The idea
goes back to Onsager \cite{Onsager:1949_Meca_Stat_Points_Vortex},
and has been mostly developed during the nineties after the work of
Miller-Robert-Sommeria \cite{Miller_Weichman_Cross_1992PhRvA,SommeriaRobert:1991_JFM_meca_Stat},
see also Refs. \cite{Majda_Wang_Book_Geophysique_Stat,Eyink_Sreenivasan_2006_Rev_Modern_Physics,BouchetVenaillePhysRep}
and references therein. Importantly, the theory predicts that the
contribution of small scale fluctuations to the total energy are negligible
in the two-dimensional case. Equilibrium statistical mechanics of
two-dimensional and quasi-geostrophic flows is now fairly well understood.
It has been applied to various problems in geophysical context such
as the description of Jovian vortices \cite{TurkingtonMHD:2001_PNAS_GRS,BouchetSommeria},
oceanic rings and jets \cite{VenailleBouchetJPO,Weichman_2006PhRvE},
equilibria on a sphere \cite{herbert2013additional}, and to describe
the vertical energy partition in continuously stratified quasi-geostrophic
flows \cite{Merryfield98JFM,venaille2012catalytic,venaille2012bottom}.

Due to the combined effect of stable stratification, thin aspect ratio
and rotation, geophysical flows are very different from classical
three-dimensional turbulence. However, such flows are not
purely two-dimensional. Here we consider the shallow water equations,
which is an intermediate model between three-dimensional and two-dimensional
turbulence. This model describes the dynamics of a thin layer a fluid
with homogeneous density. On the one hand, shallow water equations
admit conservation laws similar to two-dimensional Euler equations,
that lead to self-organization of the energy at large scale in the
Euler case. On the other hand, shallow water dynamics support the
presence of inertia-gravity waves that are absent from purely two-dimensional
turbulence. A small scale energy transfer may exist due to the existence
of these inertia-gravity waves in the shallow water system. The quasi-geostrophic
model is recovered as a limit case of the shallow water model, when
the Rossby parameter (comparing inertial terms to Coriolis forces)
is small. A small Rossby number corresponds to a strong rotation limit.
It is then natural to ask whether previously computed statistical
equilibrium states of the quasi-geostrophic models remain equilibrium
states of the shallow water model. More generally, given a certain
amount of energy in an unforced, undissipated geophysical flow, will
the flow self-organize into a large scale coherent structure, just
as in two dimensional turbulence? Or will the energy be transferred towards the small scales, just as in three dimensional
homogeneous turbulence? The aim of this paper is to answer these
questions by computing statistical equilibrium states of the inviscid
shallow water model.%

The first step before computing equilibrium states is to identify
a suitable phase space to describe microscopic configurations of the
system. The phase space variables must satisfy a Liouville theorem,
which ensures that the flow in phase space is divergence-less. Consequently,
a uniform measure on a constant energy-Casimirs shell of phase space
is invariant (microcanonical measure). The second step is to describe
the system at a macroscopic level. The macrostates will be the sets
of microstates sharing the same macroscopic behavior. The third step
is to find the most probable macrostate, and to show that almost all
the microstates correspond to this macrostate for given values of
the constraints. While these three steps may be straightforward for differential equations with a finite number of degrees of freedom, for continuous systems described by partial differential equations, these three steps require the introduction
of discrete approximations of the continuous field and of the invariant measure, and to study the continuous field limit of these discrete approximations. This point will be further discussed in the following.

This program has been achieved in the 90s for the two-dimensional Euler
equations. Indeed, a Liouville theorem is satisfied by the vorticity
field, which describes therefore a microscopic configuration of the
system. A macrostate can be defined as a probability field describing
the distribution of vorticity levels at a given point, either through
a coarse graining procedure \cite{Miller:1990_PRL_Meca_Stat,Miller_Weichman_Cross_1992PhRvA}
or directly by the introduction of Young measures \cite{Robert:1990_CRAS,Robert:1991_JSP_Meca_Stat,SommeriaRobert:1991_JFM_meca_Stat}.

% the use of a  discrete model in order to guess what is the invariant measure of the continuous dynamical system \\

For the shallow water dynamics, they are further issues that need
to be overcome. The existence of a Liouville theorem for the shallow
water flow was found by Warn \cite{Warn86} by describing the flow
configurations on a basis given by the eigenmodes of the linearized
dynamics. However, the constraints of the problem given by dynamical
invariants are not easily expressed in terms of the variables satisfying
this Liouville theorem (except in the weak flow limit discussed by
Warn \cite{Warn86}). This difficulty has been overcome by Weichman
and Petrich \cite{Weichman01} who considered first a Lagrangian representation
of the flow, and then used a formal change of variable to describe
the flow configurations with Eulerian variables convenient to express
the constraints of the problem. Using a different method that does
not require a Lagrangian representation of the fluid, we will show
the existence of a formal Liouville theorem. 

{ 
A second difficulty concerns the choice of a relevant discrete approximation that allows to keep as much as possible geometric conservations laws of the continuous dynamics. Those geometric conservation laws include the Liouville property, the Lagrangian conservation  laws (i.e. the conservation of the volume carried by each fluid particle), and global dynamical invariants.  Unfortunately, we are not aware of a discrete model that does not break at least one of those geometric conservation laws. However, we will argue that there is no logical need for the discrete model to satisfy all the conservation laws of the continuous dynamics in order to guess the correct microcanonical measure of the continuous system by considering the limit of a large number of degrees of freedom.% This is easily achieved if one can identify a set of modes that verify a detailed Liouville theorem.   Unfortunately, none of the known formal Liouville theorems for the shallow water equations are detailed Liouville theorems, 

A third difficulty is that local small scale fluctuations of the fields may have a substantial contribution to the total energy. This contrasts with 2d Euler dynamics, for which small scale fluctuations of the vorticity field  have a vanishingly small contribution to the total energy in the continuous limit. In the shallow water case, it is not a priori obvious that the contribution of these small scale fluctuations to the total energy can be expressed in terms of the macroscopic probability density field.

In order to overcome the second and the third difficulty, we introduce in this paper a semi-Lagrangian discretization : one the one hand, we consider fluid particles of equal volume, which allows to  keep track of the Lagrangian conservation laws of the dynamics. On the other hand, the particle positions are restricted to a uniform horizontal grid, and each grid node may contain many particles, which allows for an Eulerian representation of the macrostates and to keep track of global conservation laws, while  taking into account the presence of small scale fluctuations contributing to the total energy. 
%The particles positions are restricted to a uniform Eulerian grid in agreement with our formal Liouville theorem. 
We then derive the statistical mechanics theory for this discrete representation of the shallow water model, using large deviation theory.

We will argue that in the limit of a large number of degrees of freedom, the equilibrium states of the semi-Lagrangian discrete model are the equilibrium states of the actual shallow water system, and we will also show that we recover with this model results already obtained in several limiting cases.   While the statistical mechanics treatment of our discrete model is rigorous, there is some arbitrariness in our definition of the discrete model, which is not fully satisfactory.

 We stress that our approach to guess the invariant measure of the continuous shallow-water system is heuristic: there is to our knowledge no simple way to define mathematically the microcanonical measure or the Gibbs measure of an Hamiltonian infinite dimensional system. As far as we know the only example of a rigorous work for defining invariant measures in the class of deterministic partial differential equations of interest, is a work by \cite{bourgain1994periodic} on the periodic nonlinear Schrodinger equation. For the 2d Euler equations, the microcanonical measure seems clear from a physical point of view because different discretizations, for instance either Eulerian ones \cite{Miller:1990_PRL_Meca_Stat,Robert:CMP_StatMec2DEUl} or Lagrangian ones (see \cite{Dubinkina_Frank_2010JCoPh}, lead to a consistent picture in the thermodynamics limit, but even in that case no clear mathematical construction of the invariant measure exists.}

There exist only a very limited number of results on statistical equilibrium
states of the shallow water system. Warn \cite{Warn86} studied the
equilibrium states in a weak flow limit in the energy-enstrophy ensemble.
He showed that in the absence of lateral boundaries and in the absence
of bottom topography, all the initial energy of a shallow water flow
is transferred towards small scales. Here we relax the hypothesis
of a weak flow, generalizing the conclusions of Warn \cite{Warn86}
for any flow, and we discuss the effect of bottom topography. We show
that when there is a non zero bottom topography and when the flow
is rotating, there is a large scale flow associated with the equilibrium
states.

Refs. \cite{Merryfield01,Chavanis02} did compute statistical equilibrium
states of ``balanced'' shallow water flows, by assuming that all
the energy remain in the large scale flow. The equilibria described
by Merryfield et al \cite{Merryfield01} were obtained in the framework
of the energy-enstrophy theory, neglecting any other potential vorticity
moments than the potential enstrophy. Similar states were described
as minima of the potential enstrophy for the macroscopic potential
vorticity field by Sanson \cite{sanson2010}. 

Chavanis and Sommeria \cite{Chavanis02} proposed a generalization
of the 2D-Euler variational problem given by the Miller-Robert-Sommeria
theory to the shallow water case. Their main result is a relationship
between the large scale streamfunction and height field. This result
was very interesting and inspiring to us. However, Chavanis and Sommeria
did not derive their variational problem from statistical mechanics
arguments but proceeded through analogy. They were moreover neglecting
height fluctuations. This does not allow for energy partition between
vortical flow and fluctuations, and moreover leads to inconsistencies
for some range of parameters, as the Chavanis-Sommeria constrained
entropy can be shown to have no maxima for negative temperatures (the
negative temperature critical points of this variational problem are
saddles rather than maxima). As a consequence a range of possible
energy values can not be achieved in this phenomenological framework.
Nevertheless, with an approach based
on statistical mechanics, we will confirm, in this study, the form of the variational
problem proposed by Chavanis and Sommeria \cite{Chavanis02} for describing
part of the field, for a restricted range of parameters. We stress
however, that generically part of the energy will be carried by the
fluctuations, and thus that the Chavanis--Sommeria variational problem
should be considered with an energy value which is not the total one.
Determining which part of the energy should be taken into account
requires a full statistical mechanics treatment, taking into account
height fluctuations. We also note that for some other ranges of parameters,
all the energy will be carried by the fluctuations, and thus the Chavanis-Sommeria
variational problem then does not make sense.

In the preparation of this work, we have also been inspired by the
work of Weichman and Petrich \cite{Weichman01}. In this work, the
authors computed a class of statistical equilibrium states of the
shallow water system, starting from a grand canonical distribution.
The main result of their work is the same equation describing the
relation between the large scale stream function and height field,
as the one previously obtained by Chavanis and Sommeria \cite{Chavanis02}.
The Weichman--Petrich approach seemed more promising than the Chavanis--Sommeria
one as it was based on a statistical mechanics treatment. However
it also fails to predict which part of the energy goes into fluctuations,
and to recognize the range of parameters for which the mean field
equation for the largest scale is relevant. The reason is that, while
natural in condensed matter physics, the hypothesis of a Gibbs or
a grand canonical distribution, is questionable for an inertial flow
equation which is in contact neither with an energy bath, nor with
a potential vorticity bath. Related to this issue, Weichman and Petrich
had to assume an ad hoc scaling for the thermodynamical parameters,
in order to obtain statistical ensembles where entropy actually balances
conservation laws. It has been recognized for a long time that this
kind of theoretical difficulties are related to the Rayleigh-Jeans
paradox. The proper way to address these issues, in some fluid models
like for instance the two-dimensional Euler equation, is to start
from the microcanonical measure rather than from a canonical or grand
canonical one \cite{BouchetVenaillePhysRep}. We will overcome all these problems in this work by solving the microcanonical
problem, which may seem more difficult, but which can be handled using
large deviation theory. It also leads to a more precise description
of macrostate probabilities.\\

In this paper, starting from the microcanonical measure, we propose
a complete computation of the macrostate entropy for the shallow water
equations. In order to achieve this goal, we consider a large-deviation
approach that allows to perform the statistical mechanics computation
in an explicit and clear way, and that gives a precise description
of most-probable macrostates. Moreover, we compute explicitly both
the large scale flow and the small scales fluctuations of the equilibrium
states. Our results are thus a complete statistical mechanics treatment
of the shallow water equations. 

We also connect our general results to some of the partial treatment
discussed in the previous literature. For instance, we show that
only one subclass of the states described by \cite{Merryfield01,Weichman01,Chavanis02}
are actual statistical equilibria of the shallow water model. More
precisely, we will see that the large scale flow and the small scales
fluctuations can be interpreted as two subsystems in thermal contact,
and that the temperature of small scales fluctuations is necessarily
positive. The large scale flow is therefore also characterized by
a positive temperature.\\

The paper is organized as follows. The shallow water model and its
properties are introduced in section \ref{sec:Shallow-water-model}.
Equilibrium statistical mechanics of a discrete flow model arguably
relevant to describe the continuous shallow water system is derived
in section \ref{sec:Equilibrium-statistical-mechanic}. Computation of
the equilibrium states and a description of their main properties
are presented in section \ref{sec:Computation-of-the}. Energy partition
between a large scale flow and small scale fluctuations is computed
analytically in the quasi-geostrophic limit in section \ref{sec:Computation-of-energy-enstrophy},
which also includes a geophysical application to the Zapiola anticyclone.
The main results are summarized and discussed in the conclusion.

\section{Shallow water model\label{sec:Shallow-water-model}}

\subsection{Dynamics\label{sub:Dynamics}}

\begin{figure}
\centering\includegraphics[width=\textwidth]{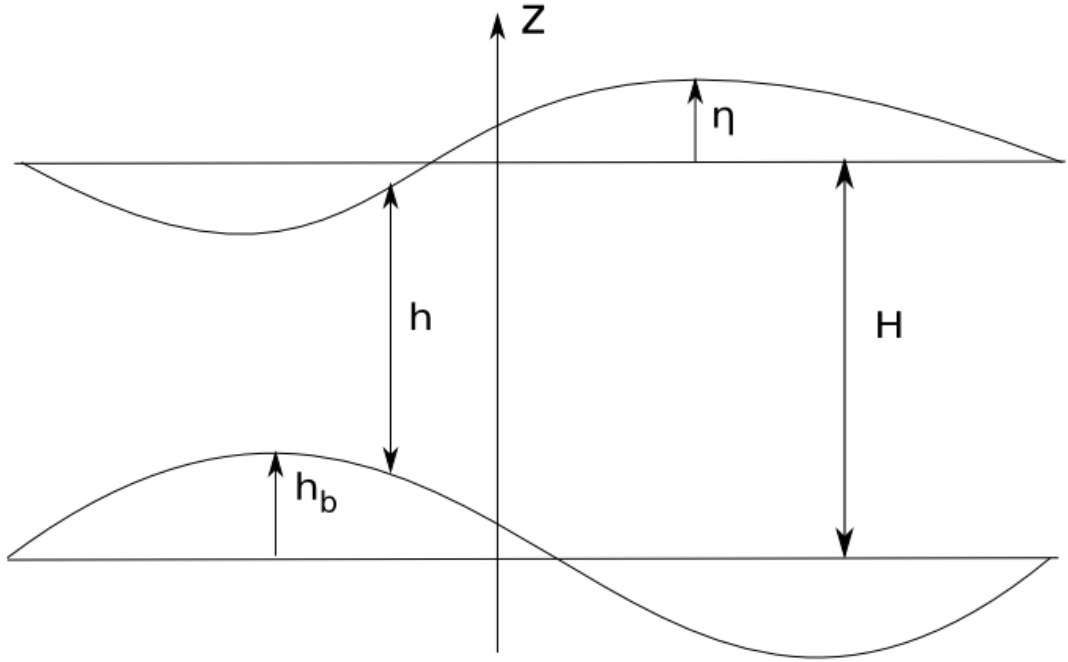}
\protect\caption{Scheme of a vertical slice of fluid, $\eta$ is the upper interface
displacement, $h_{b}$ is the bottom topography, $h$ is the height
field and $H$ is the mean height of the fluid ($\left\langle h\right\rangle $)\label{fig:Shallow-water-height}}
\end{figure}

The shallow water equations describe the dynamics of a fluid layer
with uniform mass density, in the limit where the layer depth is very
small compared to the horizontal length scales of the flow \cite{VallisBook,PedloskyBook}.
In this limit the vertical momentum equation yields hydrostatic
equilibrium, and the horizontal velocity field $(u,v)=\mathbf{u}(\mathbf{x},t)$
is depth independent, where $\mathbf{x}=(x,y)$ is any point of a
two\--dimensional simply connected domain $\mathcal{D}$. We consider the
Coriolis force in the f-plane approximation, i.e. with a constant
Coriolis parameter $f$ and a rotation axis along the vertical direction.
We denote $\eta\left(\mathbf{x},t\right)$ the vertical displacement
of the upper interface and $h_{b}\left(\mathbf{x}\right)$ the bottom
topography (see Fig. \ref{fig:Shallow-water-height}). The origin
of the $z$ axis is chosen such that 
\begin{equation}
\int_{\mathcal{D}}\mathrm{d}\mathbf{x}\ h_{b}\left(\mathbf{x}\right)=0,\label{eq:hb_moyen}
\end{equation}
and the vertical displacement is defined such that 
\begin{equation}
\int_{\mathcal{D}}\mathrm{d}\mathbf{x}\ \eta\left(\mathbf{x},t\right)=0\label{eq:eta}
\end{equation}
 as well. We introduce the total depth 
\begin{equation}
h=H-h_{b}+\eta\left(\mathbf{x},t\right),\label{eq:define_h_hb_eta}
\end{equation}
 where 
\begin{equation}
H=\frac{1}{\left|\mathcal{D}\right|}\int_{\mathcal{D}}\mathrm{d}\mathbf{r}\ h\left(\mathbf{r}\right)\label{eq:Hmean}
\end{equation}
 is the mean depth of the fluid, with $\left|\mathcal{D}\right|$
the area of the flow domain. The horizontal
and vertical length units can always be chosen such that the domain
area and the mean height $H$ are equal to one ($\left|\mathcal{D}\right|=1$,
$H=1$), and this choice will be made in the remainder of this paper. 

The dynamics is given by the horizontal momentum equations

\begin{equation}
\begin{cases}
\partial_{t}u+u\partial_{x}u+v\partial_{y}u-fv=-g\partial_{x}(h+h_{b})\\
\partial_{t}v+u\partial_{x}v+v\partial_{y}v+fu=-g\partial_{y}(h+h_{b})
\end{cases}\label{eq:SW_momentum_general}
\end{equation}
and the mass continuity equation 

\begin{equation}
\partial_{t}h+\nabla\cdot\left(h\mathbf{u}\right)=0,\label{eq:mass_continuity}
\end{equation}
where $\nabla=(\partial_{x},\partial_{y})$, with impermeability boundary
conditions
\begin{equation}
\mathbf{u}\cdot\mathbf{n}=0\quad\text{ on }\partial\mathcal{D},\label{eq:boundaries_cond}
\end{equation}
where $\mathbf{n}$ is the outward border-normal unit vector. 

By introducing the Bernoulli function 
\begin{equation}
B\equiv\frac{1}{2}\mathbf{u}^{2}+g\left(h+h_{b}-1\right),\label{eq:Bernoulli}
\end{equation}
the relative vorticity field 
\begin{equation}
\omega=\partial_{x}v-\partial_{y}u,\label{eq:omega_def}
\end{equation}
the divergence field
\begin{equation}
\zeta=\nabla\cdot\mathbf{u}=\partial_{x}u+\partial_{y}v,\label{eq:divergence_def}
\end{equation}
and the potential vorticity field
\begin{equation}
q=\frac{\omega+f}{h},\label{eq:PV_SWmodel_def}
\end{equation}
the momentum equations (\ref{eq:SW_momentum_general}) can be recast
into a conservative form 
\begin{equation}
\begin{cases}
\partial_{t}u-qhv=-\partial_{x}B\\
\partial_{t}v+qhu=-\partial_{y}B
\end{cases}.\label{eq:sw_equation_momentum}
\end{equation}
One can get from the momentum equations (\ref{eq:sw_equation_momentum})
a dynamical equations for the potential vorticity field (\ref{eq:PV_SWmodel_def})
and for the divergence field (\ref{eq:divergence_def}), respectively:
\begin{equation}
\partial_{t}q+\mathbf{u}\cdot\nabla q=0,\label{eq:sw_pv_dyn}
\end{equation}

\begin{equation}
\partial_{t}\zeta-\nabla^{\perp}\cdot\left(qh\mathbf{u}\right)=-\Delta B,\label{eq:sw_mu_dyn}
\end{equation}
where $\Delta=\nabla\cdot\nabla$ is the Laplacian operator and $\nabla^{\bot}=\left(-\partial_{y},\partial_{x}\right)$. In order
to show that the shallow water system described by $\left(h,\mathbf{u}\right)$
is also fully described by the triplet $\left(h,q,\zeta\right)$, we
introduce the velocity streamfunction $\psi(x,y,t)$ and the velocity
potential $\phi(x,y,t)$ through the Helmholtz decomposition of the
velocity%
\footnote{In the particular case of a bi-periodic domain, i.e. with periodic
boundary conditions for the velocity $\mathbf{u}$, one would need
to describe in addition the homogeneous part of the velocity field,
which is both divergence-less and irrotational.%
}:%
\begin{comment}
it may not be possible to fully described the shallow water system
with the field $\left(h,q,\zeta\right)$ only since it is possible to
have a homogeneous part of the velocity which is both irrotational
and divergence-less.
\end{comment}

\begin{equation}
\mathbf{u}=\mathbf{\nabla}^{\perp}\psi+\nabla\phi.\label{eq:v_helmholtz}
\end{equation}
The boundary condition for the velocity given in Eq. (\ref{eq:boundaries_cond})
yields 
\begin{equation}
\begin{cases}
\mathbf{n}\cdot\nabla^{\bot}\psi=0\\
\mathbf{n}\cdot\nabla\phi=0
\end{cases}\text{ on }\partial\mathcal{D},\label{eq:Bound_cond_psi_phi}
\end{equation}
where $\mathbf{n}$ is the boundary normal vector. The relative vorticity (\ref{eq:omega_def})
and the divergence (\ref{eq:divergence_def}) can be expressed in
terms of the velocity streamfunction and velocity potential as 

\begin{equation}
\omega=\Delta\psi,\quad\zeta=\Delta\phi\ .\label{eq:omega_vs_psiv_delta_vs_phiv}
\end{equation}
To the fields $\omega$ and $\zeta$ correspond a unique field $\psi$
and a field $\phi$ defined up to a constant with the boundary condition
(\ref{eq:Bound_cond_psi_phi}). Thus, to the fields $\omega=qh-f$
and $\zeta$ corresponds a unique velocity vector written formally as
\begin{equation}
\mathbf{u}=\mathbf{\nabla}^{\perp}\Delta^{-1}\left(qh-f\right)+\nabla\Delta^{-1}\zeta.\label{eq:velocity_hqnu_rel}
\end{equation}

It will also be useful to consider the field $\mu$ defined as 
\begin{equation}
\mu=\Delta^{1/2}\phi,\label{eq:mu_def}
\end{equation}
where $\Delta^{1/2}$ is the linear operator whose eigenmodes are
Laplacian eigenmodes in the domain $\mathcal{D}$ with the boundary
condition $\mathbf{n}\cdot\boldsymbol{\nabla}=0$ on $\partial\mathcal{D}$, and whose
eigenvalues are the negative square root of the modulus of Laplacian
eigenvalues. The field $\mu$ can be interpreted as a measure of the
amplitude of the potential contribution to the velocity field, given
that $\int_{\mathcal{D}}\mu^{2}=\int_{\mathcal{D}}\left(\nabla\phi\right)^{2}$.
This field will also be referred to as the divergence field in the
following. To the field $\mu$ correspond a unique field $\phi$ (up
to a constant). Thus, the shallow water system is fully described
by the triplet $\left(h,q,\mu\right)$, and the velocity vector can
be written formally in terms of these fields as 
\begin{equation}
\mathbf{u}=\nabla^{\bot}\Delta^{-1}\left[hq-f\right]+\nabla\Delta^{-1/2}\left[\mu\right].\label{eq:REL_u_hqmu}
\end{equation}

\subsection{Stationary states\label{sub:Stationary-states}}

We investigate in this subsection the conditions
for stationarity of the flow in the shallow water model following \cite{Chavanis02}. Using a Helmholtz
decomposition, one can define the mass transport streamfunction $\Psi(x,y,t)$
and the mass transport potential $\Phi(x,y,t)$ , not to be confused
with $\psi$ and $\phi$ defined in Eq. (\ref{eq:v_helmholtz}) and
(\ref{eq:Bound_cond_psi_phi}):\textbf{ }

\begin{equation}
h\mathbf{u}=\mathbf{\nabla}^{\perp}\Psi+\nabla\Phi.\label{eq:hv_helmholtz}
\end{equation}
The boundary condition for the velocity given in Eq. (\ref{eq:boundaries_cond})
yields
\begin{equation}
\begin{cases}
\mathbf{n}\cdot\nabla^{\bot}\Psi=0\\
\mathbf{n}\cdot\nabla\Phi=0
\end{cases}\text{ on }\partial\mathcal{D}.\label{eq:Bound_cond_psi_phi-1}
\end{equation}
It will be useful in the remainder
of the paper to express the vorticity field defined in Eq. (\ref{eq:omega_def})
and the divergent field defined in Eq. (\ref{eq:divergence_def})
in terms of these transport streamfunction $\text{\ensuremath{\Psi}}$
and transport potential $\Phi$ : 

\begin{equation}
\omega=\nabla\cdot\left(\frac{1}{h}\nabla\Psi\right)+J\left(\frac{1}{h},\Phi\right),\label{eq:omega_vs_psi_phi}
\end{equation}

\begin{equation}
\zeta=\nabla\cdot\left(\frac{\nabla\Phi}{h}\right)-J\left(\frac{1}{h},\Psi\right),\label{eq:delta_vs_psi_phi}
\end{equation}
where 
\begin{equation}
J\left(f,g\right)=\partial_{x}f\partial_{y}g-\partial_{y}f\partial_{x}g\label{eq:jacobian}
\end{equation}
 is the Jacobian operator. The dynamics in Eqs. (\ref{eq:mass_continuity}),
(\ref{eq:sw_pv_dyn}) and (\ref{eq:sw_mu_dyn}) can also be written
in terms of $\Psi$ and $\Phi$:

\begin{equation}
\partial_{t}h=-\Delta\Phi,\label{eq:h_dyn_phi}
\end{equation}

\begin{equation}
\partial_{t}q+\frac{1}{h}J(\Psi,q)+\frac{1}{h}\nabla\Phi\cdot\nabla q=0,\label{eq:q_dyn_phi_psi}
\end{equation}

\begin{equation}
\partial_{t}\zeta+J(\Phi,q)-q\Delta\Psi-\nabla q\cdot\nabla\Psi=-\Delta B.\label{eq:mu_dyn_phi_psi}
\end{equation}
We see that $\partial_{t}h=0$ implies $\Phi=Cst$, and $\partial_{t}q=0$
implies $J(\Psi,q)=0$, which means that isolines of potential vorticity
are the mass transport streamlines. This is the case if for instance
$q=F\left(\Psi\right)$. Reciprocally, if $\Phi=Cst$ and if $q$
and $\Psi$ have the same isolines, then $\partial_{t}h=0$ and $\partial_{t}q=0$,
which also implies $\partial_{t}\Psi=0$. Thus, through the decomposition
(\ref{eq:hv_helmholtz}), the velocity field is also stationary, with
$\mathbf{u}=\left(1/h\right)\nabla^{\bot}\Psi$. We conclude that
a necessary and sufficient condition for a shallow water flow $\left(h,\mathbf{u}\right)$
to be stationary is 
\begin{equation}
\Phi=Cst,\quad q\mbox{ and }\Psi\mbox{ have the same isolines.}\label{eq:condition_stationary flow}
\end{equation}

There is an additional relation verified by the stationary flow. This
relation may be obtained by considering stationarity of the kinetic
energy. The dynamics of the kinetic energy is obtained from Eq. (\ref{eq:sw_equation_momentum}):
\begin{equation}
\partial_{t}\left(\frac{\mathbf{u}^{2}}{2}\right)=-\frac{1}{h}J(\Psi,B)-\frac{1}{h}\nabla\Phi\cdot\nabla B.\label{eq:u2_B}
\end{equation}
Stationarity of the kinetic energy field and Eq. (\ref{eq:condition_stationary flow})
gives $J(\Psi,B)=0$: the Bernoulli function defined in Eq. (\ref{eq:Bernoulli})
and the mass transport streamfunction $\Psi$ have the same isolines. In addition,
in any subdomain where $q=F(\Psi)$,  the stationarity of the velocity
field and the momentum equations (\ref{eq:sw_equation_momentum})
give the relation 
\begin{equation}
q=\frac{\mathrm{d}B}{\mathrm{d}\Psi}.\label{eq:q_B_isoline}
\end{equation}

\subsection{Conserved quantities\label{sub:Conserved-quantities}}

Provided that the velocity and height fields remain differentiable,
the shallow water dynamics in Eqs. (\ref{eq:mass_continuity}),
(\ref{eq:sw_pv_dyn}) and (\ref{eq:sw_mu_dyn}) conserves the total
energy %
\begin{comment}
\begin{eqnarray*}
\frac{d\mathcal{E}}{dt} & = & \int\,\mathrm{d}{\bf x}\;\left[B\partial_{t}h+h{\bf u}.\partial_{t}{\bf u}\right]\\
 & = & -\int\,\mathrm{d}{\bf x}\;\left[B\nabla\left(h{\bf u}\right)+h{\bf u}.\nabla B+h{\bf u}.hq{\bf u}^{\bot}\right]\\
 & = & -\int\,\mathrm{d}{\bf x}\;\nabla\left(Bh{\bf u}\right)\\
 & = & -\oint\:\mathrm{d}{\bf S}.\left(Bh{\bf u}\right)\\
 & = & 0
\end{eqnarray*}
\end{comment}

\begin{equation}
\mathcal{E}\left[\mathbf{u},h\right]=\frac{1}{2}\int\mathrm{d}\mathbf{x}\ \left[h\mathbf{u^{2}}+g\left(h+h_{b}-1\right)^{2}\right],\label{eq:def_energy}
\end{equation}
which includes a kinetic energy contribution and a potential energy
contribution. It is known that the shallow water dynamics sometimes
leads to shocks that prevent energy conservation. We postpone a discussion
of this important point to the last section of this paper. \\

As a consequence of particle relabeling symmetry there is an infinite
number of other conserved quantities called Casimir functionals, see
e.g. \cite{Salmon_1998_Book}. These functionals are written %
\begin{comment}
\begin{eqnarray*}
\frac{d\mathcal{C}_{s}}{dt} & = & \int\,\mathrm{d}{\bf x}\;\left[s\left(q\right)\partial_{t}h+hs'\left(q\right)\partial_{t}q\right]\\
 & = & -\int\,\mathrm{d}{\bf x}\;\left[s\left(q\right)\nabla\left(h{\bf u}\right)+h{\bf u}.\nabla s\left(q\right)\right]\\
 & = & -\int\,\mathrm{d}{\bf x}\;\nabla\left(s\left(q\right)h{\bf u}\right)\\
 & = & -\oint\:\mathrm{d}{\bf S}.\left(s\left(q\right)h{\bf u}\right)\\
 & = & 0
\end{eqnarray*}
\end{comment}

\begin{equation}
\mathcal{C}_{s}=\int\mathrm{d}\mathbf{x}\ hs(q)\ ,\label{eq:casimirs_definition}
\end{equation}
where $s$ is any distribution. This conservation can be easily checked from Eqs. (\ref{eq:sw_pv_dyn}) and (\ref{eq:mass_continuity}).
The conservation of all the Casimirs implies the conservation of
all the potential vorticity moments 
\begin{equation}
\forall k\in\mathbb{N},\ \mathcal{Z}_{k}=\int\,\mbox{d}\mathbf{x}\ hq^{k}\ .\label{eq:casimirs_definition_moments}
\end{equation}
These Casimir functionals include the total mass conservation ($k=0$),
the conservation of the circulation%
\footnote{The actual circulation is usually defined as $\Gamma\equiv\int_{\partial D}\mathrm{d}\mathbf{l}\cdot\mathbf{u}$,
where $\mathrm{d}\mathbf{l}$ is a vector tangent to the domain boundary.
Stokes theorem yields to $Z_{1}=\Gamma+f$.%
} ($k=1$) and the conservation of the enstrophy ($k=2$). 

The conservation of all the Casimirs is equivalent to the conservation
of mass plus the conservation of the potential vorticity distribution
$\mathcal{D}\left[q\right]$ defined through
\begin{equation}
\forall\sigma,\quad\mathcal{D}\left[q\right]\left(\sigma\right)\mathrm{d}\sigma=\int\mathrm{d}{\bf x}\; h\mathcal{I}_{\left\{ \sigma\leq q\leq\sigma+\mathrm{d}\sigma\right\} }\label{eq:pv_distrib_definition}
\end{equation}
where $\mathcal{I}_{\left\{ \sigma\leq q\leq\sigma+\mathrm{d}\sigma\right\} }$
is the characteristic function, i.e. it returns one if $\sigma\leq q\left({\bf x}\right)\leq\sigma+\mathrm{d}\sigma$
and zero otherwise. It means that the global volume of each potential
vorticity level $\sigma_{q}$ is conserved through the dynamics. 

We will restrict ourselves to initial states where the potential vorticity
field has a distribution characterized by its moments only, then the
knowledge of its global distribution given in Eq. (\ref{eq:pv_distrib_definition})
and of the total mass is equivalent to the knowledge of the moments
given in Eq. (\ref{eq:casimirs_definition_moments}).%
\begin{comment}
We will thus consider the conservation of the potential vorticity
moments given Eq. (\ref{eq:casimirs_definition_moments}) in the following.
\end{comment}
\\

Depending on the domain geometry, there could be additional invariants.
For the sake of simplicity, we do not discuss the role of these additional
invariant in this paper, but it would not be difficult to generalize
our results by taking them into account. \\

In the quasi-geostrophic or two-dimensional Euler models, dynamical
invariants have important consequences such as the large scale energy
transfer, the self-organization at the domain scale, and the existence
of an infinite number of stable states for the dynamics, see for instance
Ref. \cite{BouchetVenaillePhysRep} and references therein. We show
in the present paper that these dynamical invariants play a similar
role in the shallow water case, allowing for a large-scale circulation
associated with the potential vorticity field, even if this process
may be associated with a concomitant loss of energy toward small scales.

\section{Equilibrium statistical mechanics of a discrete shallow water model\label{sec:Equilibrium-statistical-mechanic}}

The aim of this section is to compute statistical equilibrium states
of a discrete model of the shallow water system and to consider
the thermodynamic limit for this model. In the first subsection, a
formal Liouville theorem is given for the triplet of variables $\left(h,hu,hv\right)$
and then the triplet $\left(h,q,\mu\right)$ after a change of variables. 
The derivation given with more details in Appendix \ref{sec:Invariant-measure-and} is made in the Eulerian representation.
This allows to write formally the microcanonical measure of the shallow
water model for these triplets of fields. 

{  In the second subsection, a finite dimensional semi-Lagrangian discretization of the model is proposed to give
a physical meaning to the formal measure. Other discretizations of the shallow water system may have been considered but any other choices we have tried were not taking into account the fluid particle mass conservation and led to inconsistent results (for instance the equilibrium states were not stationary, not stable by coarse-graining, and unbalanced, see Appendix \ref{app:eulerian}).}

Then the macroscopic states
are defined through a coarse-graining procedure in a third subsection.
The variational problem leading to the equilibrium states, i.e. the
most probable macroscopic state, is introduced in the fourth subsection.
This variational problem generalizes the Miller-Robert-Sommeria equilibrium
theory to the shallow water system.

\subsection{Liouville theorem\label{sub:Liouville-theorem}}

The first step before computing equilibrium states is to define what
is a microscopic configuration of the system, which requires to identify
the relevant phase space. The simplest set of variables to consider
are those that satisfy a Liouville theorem. A Liouville theorem means
that the flow in phase space is non-divergent, which implies the time
invariance of the microcanonical ensemble. The existence of a 
Liouville theorem for the shallow water system was initially shown
by Warn \cite{Warn86} who considered a decomposition of the flow
fields on a basis given by the eigenmodes of the linearized dynamics.
A Liouville theorem was shown by Weichman and Petrich\cite{Weichman01}
for a Lagrangian representation of the dynamics.
Following a general method proposed by one of us and described by
Thalabard \cite{thalabard2013}, it is shown in Appendix \ref{sec:Invariant-measure-and}
that the triplet $\left(h,hu,hv\right)$ does satisfy a formal Liouville
theorem. At a formal level, the microcanonical measure can then be
written%
\footnote{The letter ``$\mu$'' appearing in the measure denoted $\mathrm{d}\mu$
is not related to the divergent field denoted $\mu$. %
} 
\begin{equation}
\mathrm{d}\mu_{h,hu,hv}\left(E,\left\{ Z_{k}\right\} _{k\ge0}\right)=\frac{1}{\Omega\left(E,\left\{ Z_{k}\right\} _{k\geq0}\right)}\mathcal{D}\left[h\right]\mathcal{D}\left[hu\right]\mathcal{D}\left[hv\right]\delta\left(\mathcal{E}-E\right)\prod_{k=0}^{+\infty}\delta\left(\mathcal{Z}_{k}-Z_{k}\right),\label{eq:microcanonical_measure_hhuhv}
\end{equation}
 with the phase space volume 
\begin{equation}
\Omega\left(E,\left\{ Z_{k}\right\} _{k\ge0}\right)=\int\mathcal{D}\left[h\right]\mathcal{D}\left[hu\right]\mathcal{D}\left[hv\right]\delta\left(\mathcal{E}-E\right)\prod_{k=0}^{+\infty}\delta\left(\mathcal{Z}_{k}-Z_{k}\right).\label{eq:phase_space_volume_hhuhv}
\end{equation}
Here $\mathcal{E}$ is the energy of a microscopic configuration defined
in Eq. (\ref{eq:def_energy}), and the $\left\{ \mathcal{Z}_{k}\right\} _{k\ge0}$
are the potential vorticity moments of a microscopic configuration
defined in Eq. (\ref{eq:casimirs_definition_moments}). Those constraints
are the dynamical invariants of the shallow water model. The notation
$\int\mathcal{D}\left[h\right]\mathcal{D}\left[hu\right]\mathcal{D}\left[hv\right]$
means that the integral is formally performed over each possible triplet
of fields $\left(h,\ hu,\ hv\right)$. The microcanonical measure
allows to compute the expectation of an observable $A\left[h,hu,hv\right]$
in the microcanonical ensemble as 

\begin{equation}
\left\langle \mathcal{A}\right\rangle _{d\mu_{h,hu,hv}}=\int\mathrm{d}\mu_{h,hu,hv}\ \mathcal{A}\left[h,hu,hv\right].\label{eq:define_average_measurehhuhv}
\end{equation}
Assuming ergodicity, the ensemble average $\left\langle \mathcal{A}\right\rangle _{d\mu_{h,hu,hv}}$
can finally be interpreted as the time average of the observable $\mathcal{A}$.
\\

The triplet $(h,hu,hv)$ is not a convenient one to work with, since
the Casimir functionals $\left\{ \mathcal{Z}_{k}\right\} _{k\ge0}$
defined in Eq. (\ref{eq:casimirs_definition_moments}) are not easily
expressed in terms of these fields. Indeed, the expression of the
Casimir functionals $\left\{ \mathcal{Z}_{k}\right\} _{k\ge0}$ involve
not only the triplet $(h,hu,hv)$, but also the triplet of the horizontal
derivatives of these fields. We showed in subsection \ref{sub:Dynamics}
that the triplet of fields $(h,q,\mu)$ fully describes the shallow
water dynamics in a closed domain, and the functionals $\left\{ \mathcal{Z}_{k}\right\} _{k\ge0}$
are much more easily expressed in terms of the fields $h,q$. It is
therefore more convenient to use these fields as independent variables.
The price to pay is that the simple form of the energy defined in
Eq. (\ref{eq:def_energy}) in terms of the triplet $\left(h,hu,vh\right)$
becomes more complicated when expressed in terms of the triplet $(h,q,\mu)$.
However, we will propose a simplified version of this energy functional,
and argue in Appendix \ref{sec:Energy-partition} that this is the
relevant form of the energy to consider to compute the equilibrium
state.\\

Unfortunately no direct proof of a Liouville theorem can be obtained for the triplet $(h,q,\mu)$.
However, it is still possible to start from the microcanonical measure
built with $\left(h,hu,hv\right)$ in Eq. (\ref{eq:microcanonical_measure_hhuhv}),
and change variables at a formal level.%
\begin{comment}
This change of variable is similar to the strategy adopted by Weichman
and Petrich \cite{Weichman01}, who first considered a Lagrangian
description of the shallow water system, found canonical conjugate
variables satisfying a Liouville theorem (positions and velocity of
any labeled fluid particle), and then computed the Jacobian of the
transformation from Lagrangian to Eulerian variables\textbf{. }
\end{comment}
 It is shown in Appendix \ref{sec:Change-of-variables} that the Jacobian
of the transformation is%
\footnote{The $h^{3}$ term that appears after the change of variables
in the functional integral must be understood as a ``functional
product'' $\prod_{\mathbf{x}\in\mathcal{D}}h^{3}\left(\mathbf{x}\right)$,
see also the finite-dimensional representation of this measure given
in the next subsection.%
} 
\begin{equation}
\left|J\left[\frac{(h,hu,hv)}{(q,h,\mu)}\right]\right|=Ch^{3}.\label{eq:jacobian_hhuhv_hqnu}
\end{equation}
 where $C$ is a constant. We conclude that the microcanonical measure
can therefore be formally written 
\begin{equation}
\mathrm{d}\mu_{h,q,\mu}\left(E,\left\{ Z_{k}\right\} _{k\geq0}\right)=\frac{1}{\Omega\left(E,\left\{ Z_{k}\right\} _{k\geq0}\right)}h^{3}\mathcal{D}\left[h\right]\mathcal{D}\left[q\right]\mathcal{D}\left[\mu\right]\delta\left(\mathcal{E}-E\right)\prod_{k=0}^{+\infty}\delta\left(\mathcal{Z}_{k}-Z_{k}\right),\label{eq:microcanonical_measure_hqnu}
\end{equation}
with 
\begin{equation}
\Omega\left(E,\left\{ Z_{k}\right\} _{k\geq0}\right)=\int h^{3}\mathcal{D}\left[h\right]\mathcal{D}\left[q\right]\mathcal{D}\left[\mu\right]\delta\left(\mathcal{E}-E\right)\prod_{k=0}^{+\infty}\delta\left(\mathcal{Z}_{k}-Z_{k}\right).\label{eq:phase_space_volume_hqnu}
\end{equation}
In that case, the expectation of an observable $A\left[q,h,\mu\right]$
in the microcanonical ensemble is given by 
\begin{equation}
\left\langle \mathcal{A}\right\rangle _{d\mu_{q,h,\mu}}=\int\mathrm{d}\mu_{q,h,\mu}\mathcal{A}\left[q,h,\mu\right].\label{eq:define_average_measurehhuhv-1}
\end{equation}
A finite-dimensional projection of the fields will be given in the
next subsection to give a meaning to the formal notations of this
subsection.

\subsection{A discrete model\label{sub:A-discretized-model}}

In this subsection, we devise a discrete model of the shallow water
system based on a semi-Lagrangian representation. This allows to give a finite dimensional representation of
the formal measure given Eq. (\ref{eq:microcanonical_measure_hqnu}). 

In order to keep track of the conservation properties of the continuous
dynamics, we propose a convenient discretization of the shallow water
system in terms of fluid particles.  Since the fluid is considered
incompressible, it is discretized into equal volume particles. The
fluid particles do not have necessarily the same height, and the horizontal
velocity field may be divergent. It is not possible to build a uniform
grid with a single fluid particle per grid point since two particles
with equal volume and different height can not occupy the same area.
In order to bypass this difficulty, we define a uniform grid where
several fluid particles can occupy a given site. To keep track that the actual fluid is continuous and incompressible, we add the condition
that the area occupied by the particles inside a grid site fit the
available area of the grid site.

In a first step the ensemble of microscopic configurations (the ensemble
of the microstates of the discrete model) is defined. In a second
step, the constraints of the discrete shallow water system are
introduced, which allows to define the microcanonical measure of the
discrete model in a third step.

\subsubsection{Definition of the ensemble of microscopic configurations\label{sub:Definition-of-the}}

%\begin{figure}
%\centering\includegraphics[width=\textwidth]{Grid_def.eps}

%\protect\caption{Scheme of the $N_{x}\times N_{y}$ grid with a example of a one dimensional
%arrangement of fluid particles of volume $\delta V$ inside a given
%site $\left(i,j\right)$. The number of particle per site is $M_{ij}$
%and its average is $M$. The total number of fluid particle is $N_{x}N_{y}M=\sum_{ij}M_{ij}$.
%A particle's height is proportional to the inverse of its horizontal
%area, due to volume conservation. \label{fig:Grid-definition}}
%\end{figure}

For the sake of simplicity, the domain $\mathcal{D}$ where the flow
takes place is considered rectangular but generalizing the results
to any shape would be straightforward. We recall that the horizontal
and vertical length units have been chosen such that the domain area
and the mean height are equal to one ($\left|\mathcal{D}\right|=L_{x}L_{y}=1$,
$H=1$). The domain $\mathcal{D}$ is discretized into a uniform grid
with $N=N_{x}\times N_{y}$ sites. The area of a grid site is $\left|\mathcal{D}\right|/N=1/N$.
Each site can contain many fluid particles.
It is assumed that the fluid contains $N\times M$ fluid particles
of equal mass and volume (the fluid is incompressible), where $M$
is the average number of particles per site. The volume of a fluid
particle is therefore 
\begin{equation}
\delta V=\frac{\left|\mathcal{D}\right|H}{NM}=\frac{1}{NM}.\label{eq:deltaV}
\end{equation}
The grid sites are indexed by $(i,j)$ with $1\le i,j\leq N_{x},N_{y}$
and the fluid particles by $n$ with $1\leq n\leq NM$. \\

Each fluid particle is characterized by its position $(I_{n},J_{n})$
on the grid, by its potential vorticity $q_{n}\in\left[q_{min},q_{max}\right]$,
its divergence $\mu_{n}\in\left[-\mu_{max},\mu_{max}\right]$ and
its height $h_{n}\in\left[h_{min},h_{max}\right]$. The cutoffs on
the potential vorticity can be physically related to actual minimum
and maximum in the global distribution of potential vorticity level
defined in Eq. (\ref{eq:pv_distrib_definition}), since this distribution
is conserved by the dynamics. Such a justification does not exist
for the other cutoffs. We will consider first the limit of an infinite
number of fluid particles per grid site ($M\rightarrow+\infty$),
then the limit of an infinite number of grid site ($N\rightarrow\infty$)
and finally the limit of infinite height and divergence cut-off $\mu_{max}\rightarrow+\infty$,
$h_{max}\rightarrow+\infty$,  $h_{min}\rightarrow0$. We will see
that the result does not depend on those cut-off. \\

Let us introduce $M_{ij}$ the number of particles per grid site $\left(i,j\right)$,
defined as
\begin{equation}
M_{ij}=\sum_{n=1}^{NM}\delta_{I_{n},i}\delta_{J_{n},j}.
\end{equation}
 The set of the particles that belong to the site $\left(i,j\right)$
is denoted 
\begin{equation}
\mathcal{M}_{ij}=\left\{ 1\leq n\leq NM\left|\:\left(I_{n},J_{n}\right)=\left(i,j\right)\right.\right\} ,\label{eq:set_Mij}
\end{equation}
whose cardinal is $M_{ij}$. Mass conservation states that the total
number of particles filling the grid is a constant, which gives the
constraint 
\begin{equation}
\sum_{ij}M_{ij}=NM,\label{eq:mass_conservation_Mij}
\end{equation}
where $\sum_{ij}$ means that we sum over the all the sites of the
grid.  A fluid particle labeled
by $n$ and carrying the height $h_{n}$ occupies an area $\delta V/h_{n}$. 
The constraint that the area of each grid site is covered by fluid
particles leads to the constraint 
\begin{equation}
\forall i,j\qquad\frac{1}{M}\sum_{n\in\mathcal{M}_{ij}}\frac{1}{h_{n}}=1,\label{eq:area_constraint}
\end{equation}
where $\mathcal{M}_{ij}$ is the set defined in Eq. (\ref{eq:set_Mij}).

The ensemble of microstates of the discrete model is given by the
set of all reachable values of grid positions, potential vorticities,
divergences and heights of each fluid particle in accordance with
the constraint of particle filling the area of each grid site:%
\begin{comment}
An element of $X$, i.e. a microstate, is denoted $x$.
\end{comment}

\begin{equation}
X_{micro}\equiv\left\{ \chi_{micro}=\left\{ \left(I_{n},J_{n}\right),q_{n},\mu_{n},h_{n}\right\} _{1\le n\le MN}\left|\;\forall i,j\ 1\le i,j\le N_{x},N_{y}\quad\frac{1}{M}\sum_{n\in\mathcal{M}_{ij}}\frac{1}{h_{n}}=1\right.\right\} .\label{eq:X_def}
\end{equation}

\subsubsection{Coarse-graining\label{sub:Coarse-graining}}

Here we consider a microstate $\chi_{micro}=\left\{ \left(I_{n},J_{n}\right),q_{n},\mu_{n},h_{n}\right\} _{1\le n\le MN}$
and an arbitrary function 
\begin{equation}
g:\, n\rightarrow g_{n}=g\left(h_{n},q_{n},\mu_{n}\right).\label{eq:test_function}
\end{equation}
 We introduce two different coarse-graining procedures: an areal coarse-graining,
and a volumetric coarse-graining.\\

The \textit{areal coarse-graining} of the function $g$ is defined at
each grid point $(i,j)$ as 
\begin{equation}
\overline{g}_{ij}\equiv\frac{1}{M}\sum_{n\in\mathcal{M}_{ij}}\frac{1}{h_{n}}\, g_{n}\label{eq:CG_def}
\end{equation}
where $\mathcal{M}_{ij}$ is a set defined in Eq. (\ref{eq:set_Mij}).
The terms $1/h_{n}$ appearing in Eq. (\ref{eq:CG_def}) means that
we consider local average of $g_{n}$ weighted by the area occupied
by each fluid particle. Note that we will only consider function $g$
such that $\overline{g}_{ij}$ converges to a finite value in the
limit of large $M_{ij}\sim M$. This means that the terms $g_{n}$
should not be allowed to scale with $M$. This is the reason why we
will consider first the large $M$ limit, and then the limit of large
cut-off $\mu_{max}$ and $h_{max}$ for the fields $\mu$ and $h$.

The area filling constraint in Eq. (\ref{eq:area_constraint}) can
then be written in terms of this areal coarse graining: 

\begin{equation}
\overline{1}_{ij}=1.\label{eq:area_filling_constraint}
\end{equation}
for any $i,j$. We also notice that the areal coarse-grained height
field is simply the ratio of the number of particles in the site $(i,j)$
over the averaged number of particles per site:
\begin{equation}
\overline{h}_{ij}=\frac{M_{ij}}{M}.\label{eq:hij_Mij}
\end{equation}

The {\itshape volumetric coarse-graining }of the function $g$ is defined
at each grid point $(i,j)$ as 
\begin{equation}
\left\langle g\right\rangle _{ij}\equiv\frac{1}{M_{ij}}\sum_{n\in\mathcal{M}_{ij}}\, g_{n}.\label{eq:CG_def_2}
\end{equation}
This field corresponds to the average of the function $g$ carried
by a fluid particle on site $\left(i,j\right)$. The volumetric coarse-graining
is related to the areal one through%
\begin{equation}
\left\langle g\right\rangle _{ij}=\frac{\overline{hg}_{ij}}{\overline{h}_{ij}}.\label{eq:CG_vol_and_areal}
\end{equation}

\subsubsection{Definition of a velocity field on the grid \label{sub:Definition-of-the-1}}

Let us now define  a large scale velocity field (or mean flow) on
the uniform grid of the discrete model. We will introduce later
a field accounting for small scale fluctuations of the velocity at
each grid point. In the case of the actual shallow water model, for
a given triplet of continuous fields $h,\ q,\ \mu$, the velocity
field is computed by using Eq. (\ref{eq:REL_u_hqmu}), which involves
two spatial differential operators, namely $\nabla^{\bot}\Delta^{-1}$
and $\nabla\Delta^{-1/2}$. Discrete approximations of these
spatial operators are well defined on the uniform grid of the discrete
model. Discrete approximations of Eq. (\ref{eq:REL_u_hqmu}) can therefore
be used to define a velocity field on the uniform grid. Let us consider
a coarse-grained vorticity field $\widetilde{\omega}=\overline{h}\left\langle q\right\rangle -f$
and a coarse-grained divergent field $\widetilde{\mu}=\left\langle \mu\right\rangle $
defined on the same uniform grid. Discrete approximations of the operators
appearing in Eq. (\ref{eq:REL_u_hqmu}) can be written respectively
as $\left\{ \nabla^{\bot}\Delta^{-1}\left[\widetilde{\omega}\right]\right\} _{ij}=\sum_{kl}\mathbf{G}_{ij,kl}^{\omega}.\widetilde{\omega}_{kl}$
and $\left\{ \nabla^{\bot}\Delta^{-1}\left[\widetilde{\mu}\right]\right\} _{ij}=\sum_{kl}\mathbf{G}_{ij,kl}^{\mu}.\widetilde{\mu}_{kl}$,
where the sum $\sum_{kl}$ is performed over each grid site $(k,l)$.
In the remainder of this paper, we do not need the explicit expression
of the kernels $\left\{ \mathbf{G}^{\omega}\right\} $ and $\left\{ \mathbf{G}^{\mu}\right\} $
, which depend only on the domain geometry. Using these notations,
we define the large scale velocity field as 

\begin{equation}
\mathbf{u}_{mf,ij}=\left\{ \nabla^{\bot}\Delta^{-1}\left[\overline{h}\left\langle q\right\rangle -f\right]\right\} _{ij}+\left\{ \nabla\Delta^{-1/2}\left[\left\langle \mu\right\rangle \right]\right\} _{ij}\ ,\label{eq:discrete_velocity_0}
\end{equation}
where the index ``$mf$'' stands for ``mean flow''.%

At this point one may wonder why the relevant coarse-grained fields
used to define the large scale flow $\mathbf{u}_{mf}$ should be $\overline{h},\left\langle \mu\right\rangle ,\left\langle q\right\rangle $
(which, using Eq. (\ref{eq:CG_vol_and_areal}) , is equivalent to
either the triplet $\overline{h},\overline{h\mu},\overline{hq}$ or
to the triplet $\left\langle h^{-1}\right\rangle ,\left\langle \mu\right\rangle ,\left\langle q\right\rangle $).
Our motivation for such a choice is twofold. First, we will see\textit{
a posteriori} that this allows to recover previous results derived
in several limit cases (weak flow limit, quasi-geostrophic limit),
and to obtain consistent results in the general case. Second, in physical
space, the number of fluid particles at each grid site is given by
the areal-coarse-grained height field $\overline{h}$, according to
the previous section. The relevant macroscopic potential vorticity
field or divergent field is then given by the volumetric coarse-graining
$\left\langle q\right\rangle =\overline{hq}/\overline{h}$ and $\left\langle \mu\right\rangle =\overline{h\mu}/\overline{h}$.
In that respect, the phase space variable $h$ does not play the same
role as $q$ and $\mu$ when considering macroscopic quantities in
physical space, and this is why we do not consider the triplet $\left\langle h\right\rangle ,\left\langle \mu\right\rangle ,\left\langle q\right\rangle $
to describe the system at a macroscopic level. %
\begin{comment}
Choosing $\overline{h}$ $\left\langle q\right\rangle $ and $\left\langle \mu\right\rangle $
as the relevant coarse-grained fields to describe the large scale
flow therefore ensures that the equilibrium state is stable to 
\end{comment}

\subsubsection{Definition of the microcanonical ensemble for the discrete model. }

Here we introduce a set of constraints associated with the discrete
model in order to define the microcanonical ensemble. These constraints
are a discrete version of the Casimirs functional and the energy
of the continuous shallow water model, defined in Eqs. (\ref{eq:casimirs_definition_moments})
and (\ref{eq:def_energy}), respectively. Additional assumptions on
the form of the energy will be required, and we will discuss the relevance
of such assumptions. 

Note that we introduce here constraints for the discrete shallow
water model, but we do not define what would be the dynamics of the
discrete model. Indeed, it is not necessary to know the dynamics
in order to compute the equilibrium state of the system. Only the
knowledge of constraints provided by dynamical invariants is required.\\

Let us consider a given microstates $\chi_{micro}=\left\{ \left(I_{n},J_{n}\right),q_{n},\mu_{n},h_{n}\right\} _{1\leq n\leq MN}$
, which belongs to the ensemble $X_{micro}$ of possible configurations
of the discrete model defined in Eq. (\ref{eq:X_def}). By construction
of the ensemble $X_{micro}$, each element $\chi_{micro}$ satisfies
the areal filling constraint given in Eq. (\ref{eq:area_constraint}).
\\

The potential vorticity moments of the discrete model are defined
as 
\begin{equation}
\forall k\ge0,\quad\mathcal{Z}_{k}\equiv\frac{1}{NM}\sum_{n=1}^{NM}q_{n}^{k}=\frac{1}{N}\sum_{ij}\overline{hq^{k}}_{ij}.\label{eq:PV_Dist_Discr_def}
\end{equation}
The notation $\sum_{ij}$ means that the sum is performed over each
grid point, with $1\le i\le N_{x}$, $1\le j\le N_{y}$. It is shown
in Appendix \ref{sec:Energy-partition} that those discrete potential
vorticity moments tend to the potential vorticity moments of the continuous
dynamics defined in Eq. (\ref{eq:casimirs_definition_moments}) in
the limit of large number of fluid particles $NM$. \\

We also define the total energy of the discrete model as 

\begin{equation}
\mathcal{E}\equiv\frac{1}{NM}\sum_{n=1}^{NM}\left(\frac{1}{2}\mathbf{u}_{mf,I_{n}J_{n}}^{2}+\frac{1}{2}\left(\mu_{n}-\left\langle \mu\right\rangle _{I_{n}J_{n}}\right)^{2}+g\left(\frac{h_{n}}{2}+h_{bI_{n}J_{n}}\right)\right)-\mathcal{E}_{cst}.\label{eq:energy_def_discret}
\end{equation}
where $h_{b,ij}$ is the discrete topography%
\footnote{Here, the bottom topography is assumed sufficiently smooth to be considered
constant over a grid site. A fluctuating topography would require
further discussion.%
}, where $\mathbf{u}_{mf,ij}$ is the mean flow defined in Eq. (\ref{eq:discrete_velocity_0}),
and where $\mathcal{E}_{cst}\left[h_{bij}\right]$ is an unimportant
functional of $h_{b}$ chosen such that $\mathcal{E}=0$ at rest (i.e.
when $\mu_{n}=0$, $\mathbf{u}_{mfI_{n}J_{n}}=0$, $h_{n}=1-h_{bI_{n}J_{n}}$
for any $n$ and $M_{ij}=M$ for any $(i,j)$):
\begin{equation}
\mathcal{E}_{Cst}=\frac{1}{N}\sum_{ij}\frac{g}{2}\left(1-h_{bij}^{2}\right).\label{eq:Ecst}
\end{equation}
 %
\begin{comment}
We assume that this energy is a conserved quantity of the discrete
shallow water model. 
\end{comment}
A simple interpretation for this form of the total energy is that
each fluid particle carries a kinetic energy associated with the mean
flow $\mathbf{u}_{mf}$, as well as a kinetic energy associated with
local fluctuations of the divergence field and finally a potential
energy (the height of the center of mass of the fluid particle is
$h_{bI_{n}J_{n}}+h_{n}/2$). 

It is argued in Appendix \ref{sec:Energy-partition} that with only
a few reasonable assumptions on the properties of the equilibrium
state, the energy of the discrete model defined in Eq. (\ref{eq:energy_def_discret})
would also be the energy of the equilibrium state of the actual shallow
water model defined in Eq. (\ref{eq:def_energy}) in the limit of
large number of fluid particles $NM$. Note also that according to
Eq. (\ref{eq:energy_def_discret}), the vortical part of the velocity
field does not contribute to local small scale kinetic energy, which
is analogous to previous statistical mechanics results for non-divergent
flow models such as two-dimensional Euler equations or quasi-geostrophic
equations \cite{Miller_Weichman_Cross_1992PhRvA,SommeriaRobert:1991_JFM_meca_Stat}.
Qualitatively, this is due to the fact that inverting the Laplacian
operator smooth out local fluctuations of the relative vorticity $\omega=hq-1$
so that the streamfunction associated with the microscopic vorticity
field is the same as the streamfunction associated with the coarse-grained
relative vorticity field, see Appendix B for more details.

The expression of the energy in Eq. (\ref{eq:energy_def_discret})
involves a sum over the $NM$ fluid particles. This sum can be recast
into a sum over the $N$ points of the grid, by using the definition
of the areal coarse-graining in Eq. (\ref{eq:CG_def}) and the definition
of volumetric coarse-graining in Eq. (\ref{eq:CG_def_2}): 
\begin{equation}
\mathcal{E}=\frac{1}{N}\sum_{ij}\left(\frac{1}{2}\overline{h}_{ij}\mathbf{u}_{mf,ij}^{2}+\frac{1}{2}\overline{h}_{ij}\left(\left\langle \mu^{2}\right\rangle _{ij}-\left\langle \mu\right\rangle _{ij}^{2}\right)+\frac{g}{2}\left(\left(\overline{h}_{ij}+h_{b,ij}-1\right)^{2}+\left(\overline{h^{2}}_{ij}-\overline{h}_{ij}^{2}\right)\right)\right).\label{eq:energy_def_discret-eulerian}
\end{equation}

This Eulerian representation of the energy allows to identify three
different contributions. 

One first contribution to the total energy is given by the sum over
each grid point $(i,j)$ of the kinetic energy of the mean flow $\mathbf{u}_{mf}$
(which is carried by $M_{ij}=M\overline{h}_{ij}$ fluid particles),
and of the potential energy of the areal coarse-grained height field
$\overline{h}_{ij}$ : 
\begin{equation}
\mathcal{E}_{mf}\equiv\frac{1}{2N}\sum_{ij}\left[\overline{h}_{ij}{\bf u}_{mf,ij}^{2}+g\left(\overline{h}_{ij}+h_{b,ij}-1\right)^{2}\right].\label{eq:meanenergy_discr_def}
\end{equation}
 This contribution will be referred to as the total mean flow energy,
or the energy of the large scale flow.

A second contribution to the total energy is given by the sum over
each grid points of the variance of the divergence levels $\mu$ carried
by fluid particles at site $(i,j)$ , times the number of fluid particles
($M_{ij}=M\overline{h}_{ij}$):

\begin{equation}
\mathcal{E}_{\delta\mu}\equiv\frac{1}{2N}\sum_{ij}\overline{h}_{ij}\left(\left\langle \mu^{2}\right\rangle _{ij}-\left\langle \mu\right\rangle _{ij}^{2}\right).\label{eq:kinenergy_discr_def}
\end{equation}
This term can be interpreted as a subgrid-scale (or small scale) kinetic
energy term due entirely to the divergent part of the velocity field
(see Appendix \ref{sec:Energy-partition}).

The last contribution to the total energy is the sum over each grid
point of the potential energy associated with local fluctuations of
the height field: 
\begin{equation}
\mathcal{E}_{\delta h}\equiv\frac{g}{2N}\sum_{ij}\left(\overline{h^{2}}_{ij}-\overline{h}_{ij}^{2}\right).\label{eq:potenergy_discr_def}
\end{equation}
This term can be interpreted as a subgrid-scale (or small scale) potential
energy term.%
\begin{comment}
Note that the small scale kinetic energy contribution involves the
variance of the divergence with respect of the volumetric coarse-graining,
while the potential energy term involves the variance of height fluctuations
with respect to the areal coarse-graining. The reason for this is
that the divergence
\end{comment}

One can finally check that total energy defined in Eq. (\ref{eq:energy_def_discret-eulerian})
is the sum of the three contributions given in Eqs. (\ref{eq:meanenergy_discr_def}),
(\ref{eq:kinenergy_discr_def}) and (\ref{eq:potenergy_discr_def})
: 
\begin{equation}
\mathcal{E}=\mathcal{E}_{mf}+\mathcal{E}_{\delta\mu}+\mathcal{E}_{\delta h}.\label{eq:energy_sum_3_contributions}
\end{equation}

It is well known that for the 2D Euler model or the quasi-geostrophic model, there is no contribution to the energy from the sub-grid fluctuations of the potential vorticity in the limit of vanishing grid size (see \cite{Miller:1990_PRL_Meca_Stat,SommeriaRobert:1991_JFM_meca_Stat}). For the shallow water model, this is also the case. However the sub-grid fluctuations of height and divergence do contribute to the energy (see Eqs. (\ref{eq:kinenergy_discr_def}), (\ref{eq:potenergy_discr_def}) and (\ref{eq:energy_sum_3_contributions})). A qualitative reason for this contribution of local height and divergence fluctuations to the total energy is that those two fields may be decomposed on the basis of inertia-gravity waves, which are known to develop a small scale energy transfer \cite{Warn86}, and consequently to a loss of energy at subgrid-scales in the discretized model.\\

Now that we have defined the configurations space in Eq. (\ref{eq:X_def}),
the potential vorticity moments $\left\{ \mathcal{Z}_{k}\left[\chi_{micro}\right]\right\} _{k\ge0}$
of the discrete model in Eq. (\ref{eq:PV_Dist_Discr_def}) and
the energy $\mathcal{E}\left[\chi_{micro}\right]$ of the discrete
model in Eq. (\ref{eq:energy_def_discret-eulerian}), we introduce
the microcanonical ensemble as the restriction of the configurations
space $X_{micro}$ to configurations with fixed values of energy $E$
and Casimirs $\left\{ Z_{k}\right\} _{k\ge0}$: 
\begin{equation}
\left\{ \chi_{micro}=\left\{ \left(I_{n},J_{n}\right),q_{n},\mu_{n},h_{n}\right\} _{1\leq n\leq MN}\in X_{micro}\left|\:\mathcal{E}\left[\chi_{micro}\right]=E,\:\forall k\in\mathbb{N}\quad\mathcal{Z}_{k}\left[\chi_{micro}\right]=Z_{k}\right.\right\} .\label{eq:mic_ens}
\end{equation}
In our discrete model, we assume that the microcanonical measure over the ensemble of configurations $X_{micro}$ is :
\begin{multline}
\mathrm{d}\mu^{N,M}_{E,\{Z_{k}\}_{k\geq 0}}(\chi_{\mathrm{micro}})=\frac{\delta(\mathcal{E}[\chi_{\mathrm{micro}}]-E)}{\Omega(E,\{Z_{k}\}_{k\geq 0})}\prod_{k=0}^{\infty}\delta(\mathcal{Z}_{k}[\chi\mathrm{micro}]-Z_{k})\prod_{ij}\delta(\overline{1}_{ij}[\chi\mathrm{micro}]-1)\\
\times\prod_{n=1}^{NM}h_{n}^{3}\mathrm{d}h_{n}\mathrm{d}q_{n}\mathrm{d}\mu_{n}\mathrm{d}I_{n}\mathrm{d}J_{n},\label{eq:mic_proba}
\end{multline}
 where $\Omega\left(E,\left\{ Z_{k}\right\} \right)$ is the
phase space volume defined as 
\begin{multline}
\Omega(E,\{Z_{k}\}_{k\geq 0})=\sum_{I_{1}=1}^{N_{x}}\cdots\sum_{I_{n}=1}^{N_{x}}\sum_{J_{1}=1}^{N_{y}}\cdots\sum_{J_{n}=1}^{N_{y}}\int\bigg[\prod_{n=1}^{NM} h_{n}^{3}dh_{n}dq_{n}d\mu_{n}\bigg]\delta(\mathcal{E}[\chi_{\mathrm{micro}}]-E)\prod_{k=0}^{\infty}\delta(\mathcal{Z}_{k}[\chi_{\mathrm{micro}}]-Z_{k})\\
\times\prod_{ij}\delta(\overline{1}_{ij}[\chi_{\mathrm{micro}}]-1).\label{eq:mic_ens_weight}
\end{multline}

The terms $\mathrm{d}I_{n}$ and $\mathrm{d}J_{n}$ are discrete measures with support on the grid coordinates: $\mathrm{d}I_{n}(f)=\sum_{i=1}^{N_{x}}f_{i}$, $\mathrm{d}J_{n}(f)=\sum_{j=1}^{N_{y}}f_{j}$. The product $\prod_{ij}$ is performed over the grid sites $(i,j)$
with $1\le i\le N_{x}$ and $1\le j\le N_{y}$. The constraint $\delta\left(\overline{1}_{ij}\left[\chi_{micro}\right]-1\right)$
corresponds to the area filling constraint defined in Eq. (\ref{eq:area_filling_constraint}),
which must be satisfied by each microstate $\chi_{micro}\in X_{micro}$. 

Note that Eq. (\ref{eq:mic_proba}) is a discrete version of the formal microcanonical measure given in Eq. (\ref{eq:microcanonical_measure_hqnu}) for the continuous case. The expectation of an observable $\mathcal{A}\left[\chi_{micro}\right]$
in the microcanonical ensemble is 
\begin{equation}
\langle\mathcal{A}\rangle^{N,M}_{E,\{Z_{k}\}_{k\geq 0}}=\mathrm{d}\mu^{N,M}_{E,\{Z_{k}\}_{k\geq 0}}(\mathcal{A})=\int\mathrm{d}\mu^{N,M}_{E,\{Z_{k}\}_{k\geq 0}}(\chi_{\mathrm{micro}})\mathcal{A}[\chi_{\mathrm{micro}}]\label{eq:avarage_mic}.
\end{equation}
The problem is now to compute the macrostate entropy, which accounts
for the logarithm of the number of microstates corresponding to the
same macrostate. Although we do not need to consider this problem in the present work, it would be very interesting to have continuous or discrete approximation of the shallow water equation that have the invariant measure (\ref{eq:mic_proba}), in an analogous way as what was achieved for the 2D Euler equations \cite{dubinkina2010statistical,frank2003hamiltonian}.

\subsection{Macrostates and their entropy\label{sub:Macrostates-and-the}}

The aim of this subsection is to compute the equilibrium state of
the discrete model introduced in the previous subsection. The first
step is to define the macrostates of the system, the next step is
to compute the most probable macrostates, using large deviation
theory, which yields also a concentration property asymptotically
for large $N$ and $M$ (almost all the microstates correspond to
the most probable macrostate).

\subsubsection{Definition of the empirical density field}

Let us consider a microstate $\chi_{micro}=\left\{ \left(I_{n},J_{n}\right),q_{n},\mu_{n},h_{n}\right\} _{1\leq n\leq MN}$
picked in the ensemble of configurations $X_{micro}$ defined in Eq.
(\ref{eq:X_def}). The empirical density field of this microstate
is defined for each grid point $(i,j)$ as 
\begin{equation}
p_{ij}\left(\sigma_{h},\sigma_{q},\sigma_{\mu}\right)\left[\chi_{\mathrm{micro}}\right]\equiv\overline{\delta\left(h-\sigma_{h}\right)\delta\left(q-\sigma_{q}\right)\delta\left(\mu-\sigma_{\mu}\right)}_{ij},\label{eq:empirical_density_def}
\end{equation}
where the overline operator is the areal coarse-graining defined in
(\ref{eq:CG_def}). This field contains all the statistical information
of the system at the grid level. The constraint that each grid site
$(i,j)$ is covered by particles is given by Eq. (\ref{eq:area_filling_constraint}),
which ensures the normalization:
\begin{equation}
\int\mathrm{d}\sigma_{h}\mathrm{d}\sigma_{q}\mathrm{d}\sigma_{\mu}\: p_{ij}=1.\label{eq:normalization_pij}
\end{equation}

Let us consider a function $g(h_{n},q_{n},\mu_{n})$ depending on
the height, potential vorticity and divergence carried by a fluid
particle. Let us then consider the discrete microscopic field $n\rightarrow g_{n}=g(h_{n},q_{n},\mu_{n})$
with $1\le n\le NM$. Following Eqs. (\ref{eq:CG_def}) and (\ref{eq:empirical_density_def}),
the corresponding coarse-grained field $\overline{g}_{ij}$ is expressed
solely in terms of the empirical density field $p_{ij}$: 
\begin{equation}
\overline{g}_{ij}=\int\mathrm{d}\sigma_{h}\mathrm{d}\sigma_{q}\mathrm{d}\sigma_{\mu}\: g(\sigma_{h},\sigma_{q},\sigma_{\mu})p_{ij}\left(\sigma_{h},\sigma_{q},\sigma_{\mu}\right).\label{eq:pij_fCG}
\end{equation}

If we consider for instance the function $g(h,q,\mu)=q$ , then $g_{n}=q_{n}$
the microscopic potential vorticity field, and coarse-grained potential
vorticity field $\overline{q}_{ij}$ is obtained by a direct application
of Eq. (\ref{eq:pij_fCG}): $\overline{q}_{ij}=\int\mathrm{d}\sigma_{h}\mathrm{d}\sigma_{q}\mathrm{d}\sigma_{\mu}\:\sigma_{q}p_{ij}$. 

Importantly, the constraints $\left\{ \mathcal{Z}_{k}\left[\chi_{micro}\right]=Z_{k}\right\} _{k\ge0}$
and $\mathcal{E}\left[\chi_{micro}\right]=E$ defined in Eq. (\ref{eq:PV_Dist_Discr_def})
and (\ref{eq:energy_def_discret-eulerian}) depend only on the empirical
density field $p_{ij}$ (since they depend only on local areal coarse-grained
moments of the different fields). The empirical density is therefore
a relevant variable to fully characterize the system at a macroscopic
level.

\subsubsection{Definition of the macrostates}

The macrostates are defined as the set of microscopic configurations
leading to a given value $p_{ij}=\rho_{ij}$ of the empirical density
field: 
\begin{equation}
\rho\equiv\left\{ \chi_{micro}\in X_{micro}\left|\:\forall i,j\quad p_{ij}\left[\chi_{micro}\right]=\rho_{ij}\right.\right\} .\label{eq:Macrostates_def}
\end{equation}
For the sake of simplicity, we make a small abuse of notation by denoting
$\rho$ both the macrostate defined in Eq. (\ref{eq:Macrostates_def})
and the field $\rho=\left\{ \rho_{ij}\right\} $. The values of the
constraints are the same for all microstates within a given macrostate
since they depend only on the local coarse-grained moments of the
different fields, which remain unchanged for a prescribed empirical
density field. The energy and the Casimirs, defined in Eqs. (\ref{eq:energy_def_discret}) and (\ref{eq:PV_Dist_Discr_def}) respectively, have the same values
for all the microstates within a single macrostate and will therefore
be denoted by $\mathcal{E}\left[\rho\right]$ and $\left\{ \mathcal{Z}_{k}\left[\rho\right]\right\} _{k\ge0}$
.

\subsubsection{Macroscopic observables and empirical density}

Let us now consider an observable $\mathcal{A}\left[\chi_{micro}\right]$
on the configuration space $X_{micro}$ defined in Eq. (\ref{eq:X_def})
such that their dependance on the microscopic configuration $\chi_{micro}$
occurs only through the empirical density field: 
\begin{equation}
\mathcal{A}\left[\chi_{micro}\right]=\mathcal{A}\left[\left\{ p_{ij}\left[\chi_{micro}\right]\right\} \right].
\end{equation}
This is actually the case for any observable written as a sum over
the fluid particles, i.e. for any observable appearing in Eq. (\ref{eq:avarage_mic}).
It is therefore possible to change variables from $\chi_{micro}$
to the empirical density field values $\left\{ \rho_{ij}\right\} _{1\le i,j\le N_{x,}N_{y}}$
in Eq. (\ref{eq:avarage_mic}): 
\begin{multline}
\langle\mathcal{A}\rangle^{N,M}_{E,\{Z_{k}\}_{k\geq 0}}=\int\bigg[\prod_{ij}\mathcal{D}\left[\rho_{ij}\right]\bigg]\;\mathcal{A}\left[\left\{ \rho_{ij}\right\} \right]\frac{\Omega\left(\rho\right)}{\Omega\left(E,\left\{ Z_{k}\right\} _{k\ge0}\right)}\delta\left(\mathcal{E}\left[\rho\right]-E\right)\prod_{k=0}^{\infty}\delta\left(\mathcal{Z}_{k}\left[\rho\right]-Z_{k}\right)\\
\times\prod_{ij}\delta\left(\overline{1}_{ij}\left[\rho_{ij}\right]-1\right),\label{eq:avarage_mac}
\end{multline}
where
\begin{equation}
\Omega(\rho)=\sum_{I_{1}=1}^{N_{x}}\cdots\sum_{I_{n}=1}^{N_{x}}\sum_{J_{1}=1}^{N_{y}}\cdots\sum_{J_{n}=1}^{N_{y}}\int\bigg[\prod_{n=1}^{NM} h_{n}^{3}dh_{n}dq_{n}d\mu_{n}\bigg]\prod_{ij}\big[\hat{\delta}(p_{ij}[\chi_{\mathrm{micro}}]-\rho_{ij})\big]\label{eq:macro_weight}
\end{equation}
is the volume of a macrostate $\rho$ defined in Eq. (\ref{eq:Macrostates_def})
in the configuration space $X_{micro}$ defined in Eq. (\ref{eq:X_def}),
and where $\Omega\left(E,\left\{ Z_{k}\right\} _{k\ge0}\right)$ is
the total volume in phase space defined in Eq. (\ref{eq:mic_ens_weight}).
The constraint $\delta\left(\overline{1}_{ij}\left[\rho_{ij}\right]-1\right)$
is a normalization constraint for the $\left\{ \rho_{ij}\right\} $,
since $\overline{1}_{ij}\left[\rho_{ij}\right]=\int\mathrm{d}\sigma_{h}\mathrm{d}\sigma_{q}\mathrm{d}\sigma_{\mu}\:\rho_{ij}$.
This normalization constraint also corresponds to the constraint that
the fluid particles within a site must fit the available area, see
Eq. (\ref{eq:area_constraint}). The term $\int\mathcal{D}\left[\rho_{ij}\right]$
means that the integral is performed over all the possible functions
$\rho_{ij}$. The term $\hat{\delta}\left(p_{ij}-\rho_{ij}\right)$
is a Dirac delta distribution on the functional space of the empirical
density field values $\rho_{ij}(\sigma_{h},\sigma_{q},\sigma_{\mu})$,
with $p_{ij}\left(\sigma_{h},\sigma_{q},\sigma_{\mu}\right)\left[\chi_{micro}\right]$
the empirical density field defined in Eq. (\ref{eq:empirical_density_def}).

\subsubsection{Asymptotic behavior of the macrostates volume and derivation of its
entropy}

Let us compute the asymptotic form of $\Omega\left(\rho\right)$ defined
in Eq. (\ref{eq:macro_weight}), by considering the limit $M\rightarrow\infty$,
where $M$ is the average number of particles per grid site $\left(i,j\right)$.
For a given set of macrostates $\text{\ensuremath{\rho}}$ defined
in Eq. (\ref{eq:Macrostates_def}), the number of fluid particles
per grid site is 
\begin{equation}
M_{ij}=M\int\mathrm{d}\sigma_{h}\mathrm{d}\sigma_{q}\mathrm{d}\sigma_{\mu}\:\sigma_{h}\rho_{ij}.\label{eq:M_ij_COND}
\end{equation}
This is the only constraint on the particle positions in the grid
for a microstate that belongs to the macrostates $\rho$. All the
realizations of the particle positions that satisfies (\ref{eq:M_ij_COND})
count equally in the ensemble of macrostates $\rho$. Through combinatorial
calculation, the number of realizations of $\{\left(I_{n},J_{n}\right)\}$ that satisfy (\ref{eq:M_ij_COND})
is $\left(NM\right)!/\prod_{ij}M_{ij}!$. Using Eq. (\ref{eq:hij_Mij})
to express $M_{ij}$ in terms of $\overline{h}_{ij}$, we get 
\begin{equation}
\Omega\left(\rho\right)=\frac{\left(NM\right)!}{\prod_{ij}\left(M\overline{h}_{ij}\right)!}\prod_{ij}\Omega_{ij},\label{eq:Mac_Weight_fact_first}
\end{equation}
where
\begin{equation}
\Omega_{ij}\equiv\int\bigg[\prod_{m=1}^{M\overline{h}_{ij}}h_{m}^{3}\mathrm{d}h_{m}\mathrm{d}q_{m}\mathrm{d}\mu_{m}\bigg]\: \hat{\delta}\color{black}\left(\frac{1}{M}\sum_{m=1}^{M\overline{h}_{ij}}\frac{1}{h_{m}}\delta\left(h_{m}-\sigma_{h}\right)\delta\left(q_{m}-\sigma_{q}\right)\delta\left(\mu_{m}-\sigma_{\mu}\right)-\rho_{ij}\right)\label{eq:Om_ij}
\end{equation}
is the number of possible configurations for a given set of $M_{ij}=M\overline{h}_{ij}$
fluid particles at site $(i,j)$.\\

The asymptotic behavior of the pre-factor in Eq. (\ref{eq:Mac_Weight_fact_first})
is computed through the Stirling formula:

\begin{equation}
\log\left(\frac{\left(NM\right)!}{\prod_{ij}\left(M\overline{h}_{ij}\right)!}\right)\underset{M\rightarrow\infty}{\sim}MN\left(-\frac{1}{N}\sum_{ij}\overline{h}_{ij}\log\left(\overline{h}_{ij}\right)+\log\left(N\right)\right).\label{eq:asympt_Mij}
\end{equation}

The asymptotic behavior with $M$ of $\Omega_{ij}$ defined in Eq.
(\ref{eq:Om_ij}) can be computed by using Sanov's theorem. 

Before applying this theorem to our problem, let us consider the simpler
case of $K$ independent and identically distributed variables $\left\{ \chi_{k}\right\} _{1\leq k\leq K}$
with common probability density function $F\left(\chi\right)$. Those variable take values in a bounded interval of $\mathbb{R}$. \footnote{The fact that the variable $\chi_{k}$ have to be bounded is the reason why we set the cutoffs on the values of $h_{n}$, $q_{n}$ and $\mu_{n}$.} Sanov's
theorem describes the large deviation of the empirical density distribution
\begin{equation}
f_{K}\equiv\frac{1}{K}\sum_{k=1}^{K}\delta\left(\mathrm{\chi}_{k}-\mathrm{\chi}\right),\label{eq:sanov_fk}
\end{equation}
which can be considered as an actual probability distribution for the variable $\chi$.
 The probability distribution functional of this empirical density
function is 
\begin{equation}
\mathcal{P}\left[f\right]\equiv\int\prod_{k=1}^{K}F\left(\mathrm{\chi}_{k}\right)\mathrm{d\chi}_{k}\;\hat{\delta}\left(f-f_{K}\right).\label{eq:sanov_Pf}
\end{equation}
Sanov's theorem is a statement about the asymptotic behavior of the
logarithm of $\mathcal{P}\left[f\right]$. For a given function $f\left(\chi\right)$,
Sanov's theorem states 

\begin{equation}
\log\left(\mathcal{P}\left[f\right]\right)\underset{K\rightarrow+\infty}{\sim}-K\int{\rm d}\chi\; f\left(\chi\right)\log\left(\frac{f\left(\chi\right)}{F\left(\chi\right)}\right)\label{eq:sanov_logP}
\end{equation}
if $\int\mathrm{d\chi}\: f\left(\chi\right)=1$ and $\log\left(\mathcal{P}\left[f\right]\right)\sim-\infty$
otherwise. An heuristic discussion of Sanov's theorem is given in
Ref \cite{touchette2009}. 

Combining Eqs. (\ref{eq:sanov_fk}), (\ref{eq:sanov_Pf}) and (\ref{eq:sanov_logP})
and generalizing this result to $K$ independent and identically distributed
L\--tuple of variables $\left\{ \left\{ \chi_{l,k}\right\} _{1\le l\le L}\right\} _{1\leq k\leq K}$
with common probability density function $F\left(\left\{ \chi_{l}\right\} _{1\le l\le L}\right)$,
Sanov's theorem is written in compact form as

\begin{equation}
\log\left(\int\prod_{k=1}^{K}F\left(\left\{ \mathrm{\chi}_{l,k}\right\} _{1\le l\le L}\right)\prod_{l=1}^{L}\mathrm{d\chi}_{l,k}\;\hat{\delta}\left(f-\frac{1}{K}\sum_{k=1}^{K}\prod_{l=1}^{L}\delta\left(\mathrm{\chi}_{l,k}-\mathrm{\chi}_{l}\right)\right)\right)\underset{K\rightarrow\infty}{\sim}-K\int\prod_{l=1}^{L}\mathrm{d\chi}_{l}\; f\log\left(\frac{f}{F}\right)\label{eq:Sanov_Statement}
\end{equation}
if $\int\prod_{l=1}^{L}\mathrm{d\chi_{l}}\: f\left(\left\{ \chi_{l}\right\} _{1\le l\le L}\right)=1$. 

Let us come back to the asymptotic behavior of $\Omega_{ij}$ defined
in Eq. (\ref{eq:Om_ij}), in the large $M$ limit. Before applying
Sanov's theorem, it is needed to recast this equation into a form
similar to the argument of the logarithm in the lhs of Eq. (\ref{eq:Sanov_Statement}).
Because of the $1/h_{m}$ term appearing in the delta Dirac function
of Eq. (\ref{eq:Om_ij}), one needs to perform first a change of variable
from $\rho_{ij}$ to
\begin{equation}
\pi_{ij}\equiv\frac{\sigma_{h}}{\overline{h}_{ij}}\rho_{ij}\label{eq:change_var}
\end{equation}
The formal Jacobian $\mathcal{J}$ arising from this change of variable in the functional  delta Dirac function depend only on $\overline{h}_{ij}$ which does not depend on 
$M$ and will therefore not matter for the asymptotic behavior of
$\log\left(\Omega_{ij}\right)$. In addition, the factor $h_{m}^{3}$
appearing in front of the Lebesgue measure in Eq. (\ref{eq:Om_ij})
must be divided by a normalization factor $Z=\int h_{m}^{3}\mathrm{d}h_{m}\mathrm{d}q_{m}\mathrm{d}\mu_{m}$
so that $P(h_{m},q_{m},\mu_{m})=h_{m}^{3}/Z$ can be interpreted as a probability
distribution function.\footnote{Here, $Z$ depends on the cutoffs introduced in subsection \ref{sub:A-discretized-model}. But we will see that it will vanish from the expression of the entropy in the end.} Then $\Omega_{ij}$ writes :
\begin{equation}
\Omega_{ij}=\mathcal{J}Z^{M\overline{h}_{ij}}\int\prod_{m=1}^{M\overline{h}_{ij}}\left(\frac{h_{m}^{3}}{Z}\right)\mathrm{d}h_{m}\mathrm{d}q_{m}\mathrm{d}\mu_{m}\:\hat{\delta}\left(\pi_{ij}-\frac{1}{M\overline{h}_{ij}}\sum_{m=1}^{M\overline{h}_{ij}}\delta\left(h_{m}-\sigma_{h}\right)\delta\left(q_{m}-\sigma_{q}\right)\delta\left(\mu_{m}-\sigma_{\mu}\right)\right).\label{eq:Om_ij_changevar}
\end{equation}
\begin{comment}
The change of variable performed in Eq. (\ref{eq:change_var}) corresponds
to change from the ``areal'' to the ``volumetric'' point of view.
Just as empirical density field $p_{ij}$ is defined through the areal
coarse-graining in Eq. (\ref{eq:CG_def}), an empirical distribution
field $P_{ij}$ can be defined through the volumetric coarse-graining
in Eq. (\ref{eq:CG_def_2}):
\begin{equation}
P_{ij}\left(\sigma_{h},\sigma_{q},\sigma_{\mu}\right)\equiv\left\langle \delta\left(h-\sigma_{h}\right)\delta\left(q-\sigma_{q}\right)\delta\left(\mu-\sigma_{\mu}\right)\right\rangle _{ij}\qquad\forall i,j.\label{eq:Pij}
\end{equation}
Injecting Eq. (\ref{eq:Pij}) in Eq. (\ref{eq:Om_ij_changevar}) yields
\begin{equation}
\Omega_{ij}=\mathcal{J}Z^{M\overline{h}_{ij}}\int\prod_{m=1}^{M\overline{h}_{ij}}\left(\frac{h_{m}^{3}}{Z}\right)\mathrm{d}h_{m}\mathrm{d}q_{m}\mathrm{d}\mu_{m}\:\delta\left(P_{ij}-\pi_{ij}\right).\label{eq:omega_ij_sanov}
\end{equation}
\end{comment}
A direct application of Sanov's theorem to Eq. (\ref{eq:Om_ij_changevar})
then yields
\begin{equation}
\log\left(\Omega_{ij}\right)\underset{M\rightarrow\infty}{\sim}M\overline{h}_{ij}\left[-\int\mathrm{d}\sigma_{h}\mathrm{d}\sigma_{q}\mathrm{d}\sigma_{\mu}\:\pi_{ij}\log\left(\frac{\pi_{ij}}{\sigma_{h}^{3}}\right)\right].\label{eq:Om_ij_sanov}
\end{equation}
The normalization constraint on the distributions $\pi_{ij}$ is already
fulfilled by the definition of the coarse-grained height fields $\overline{h}_{ij}=\overline{h}_{ij}\int\mathrm{d}\sigma_{h}\mathrm{d}\sigma_{q}\mathrm{d}\sigma_{\mu}\:\pi_{ij}$
(see Eq. (\ref{eq:pij_fCG})). %
\begin{comment}
The normalization condition required for using Sanov's theorem is
necessary fulfilled from the definition of $\overline{h}_{ij}$ in
Eq. (\ref{eq:M_ij_COND}). We recognize in Eq. (\ref{eq:Om_ij_sanov})
a classical form of the entropy such as the one derived for the 2D
Euler model or for the quasi-geostrophic model \cite{Miller:1990_PRL_Meca_Stat}\cite{SommeriaRobert:1991_JFM_meca_Stat}.
The $\sigma_{h}^{3}$ term in the logarithm corresponds to the $h^{3}$
pre-factor that occurs in the invariant measure defined in Eq. (\ref{eq:microcanonical_measure_hqnu}).
\end{comment}
The inverse change of variable $\rho_{ij}=\overline{h}_{ij}\pi_{ij}/\sigma_{h}$
in Eq. (\ref{eq:Om_ij_sanov}) yields 
\begin{equation}
\log\left(\Omega_{ij}\right)\underset{M\rightarrow\infty}{\sim}M\left[-\int\mathrm{d}\sigma_{h}\mathrm{d}\sigma_{q}\mathrm{d}\sigma_{\mu}\:\sigma_{h}\rho_{ij}\log\left(\frac{\rho_{ij}}{\sigma_{h}^{2}}\right)+\overline{h}_{ij}\log\left(\overline{h}_{ij}\right)\right].\label{eq:asympt_sanov}
\end{equation}
Combining Eq. (\ref{eq:Mac_Weight_fact_first}) with Eqs. (\ref{eq:asympt_Mij})
and (\ref{eq:asympt_sanov}), we obtain the asymptotic behavior of
$\Omega\left(\rho\right)$ with M:
\begin{equation}
\log\left(\Omega\left(\rho\right)\right)\underset{M\rightarrow\infty}{\sim}MN\,\mathcal{S}\left[\rho\right],\label{eq:LargeDev_OmRho}
\end{equation}
where 
\begin{equation}
\mathcal{S}\left[\rho\right]={\displaystyle -\frac{1}{N}\sum_{ij}\int\mathrm{d}\sigma_{h}\mathrm{d}\sigma_{q}\mathrm{d}\sigma_{\mu}\:\sigma_{h}\rho_{ij}\log\left(\frac{\rho_{ij}}{\sigma_{h}^{2}}\right)}\label{eq:Mixing_Entrop_LD}
\end{equation}
is the macrostate entropy\footnote{Here we dropped the term $\log N$ coming from Eq. (\ref{eq:asympt_Mij}) as it is constant that can be discarded by redefining $\Omega_{E,\{Z_{k}\}_{k\geq 0}}$.}.

\subsection{Continuous limit\label{sub:continuous-limit}}

\subsubsection{Expressions of the macrostate entropy, energy and potential vorticity
moments}

Considering now the limit of an infinite number of grid site ($N\rightarrow\infty$),
the site coordinates $\left(i,j\right)$ tend toward the continuous
space coordinates $\mathbf{x}$ and the macrostate entropy $\mathcal{S}$
derived in Eq. (\ref{eq:Mixing_Entrop_LD}) becomes 
\begin{equation}
\mathcal{S}\left[\rho\right]={\displaystyle -\int\mathrm{d}\mathbf{x}\mathrm{d}\sigma_{h}\mathrm{d}\sigma_{q}\mathrm{d}\sigma_{\mu}\:\sigma_{h}\rho\left(\mathbf{x},\sigma_{h},\sigma_{q},\sigma_{\mu}\right)\log\left(\frac{\rho\left(\mathbf{x},\sigma_{h},\sigma_{q},\sigma_{\mu}\right)}{\sigma_{h}^{2}}\right)}.\label{eq:Mixing_Entropy}
\end{equation}
The empirical density has become a probability density function (pdf).
For any function $g:\, n\rightarrow g_{n}=g\left(h_{n},q_{n},\mu_{n}\right)$,
its continuous coarse-grained field is now computed through 
\begin{equation}
\overline{g}\left(\mathbf{x}\right)=\int\mathrm{d}\sigma_{h}\mathrm{d}\sigma_{q}\mathrm{d}\sigma_{\mu}\:\rho\left(\mathbf{x},\sigma_{h},\sigma_{q},\sigma_{\mu}\right)g\left(\sigma_{h},\sigma_{q},\sigma_{\mu}\right),\label{eq:continuous_CG}
\end{equation}
The discrete mean flow defined in Eq. (\ref{eq:discrete_velocity_0})
becomes (by construction):
\begin{equation}
\mathbf{u}_{mf}\left[\overline{h},\overline{hq},\overline{h\mu}\right]=\nabla^{\bot}\Delta^{-1}\left(\overline{hq}-f\right)+\nabla\Delta^{-1/2}\left(\frac{\overline{h\mu}}{\overline{h}}\right),\label{eq:u_mf_continuous}
\end{equation}
where all the coarse-grained fields have been expressed in terms of
an areal coarse-graining, using Eq. (\ref{eq:CG_vol_and_areal}).
Similarly, the potential vorticity moments defined in Eq. (\ref{eq:PV_Dist_Discr_def})\textbf{
}for the discrete model become

\begin{equation}
\forall k\in\mathbb{N}\quad\mathcal{Z}_{k}\left[\rho\right]=\int\mathrm{d}{\bf x}\;\overline{hq^{k}}.\label{eq:Total_moments_funct}
\end{equation}
The total energy defined in Eq. (\ref{eq:energy_def_discret-eulerian})
for the discrete model becomes 

\begin{equation}
\mathcal{E}\left[\rho\right]=\frac{1}{2}\int\mathrm{d}{\bf x}\;\left[\overline{h}{\bf u}_{mf}^{2}+\left(\overline{h\mu^{2}}-\frac{\overline{h\mu}^{2}}{\overline{h}}\right)+g\overline{\left(h+h_{b}-1\right)^{2}}\right].\label{eq:Total_energy_funct}
\end{equation}
We note that the total energy is a functional of $\overline{h},\overline{h^{2}},\overline{hq},\overline{h\mu},\overline{h\mu^{2}}$
. Just as in the discrete case, this energy can be separated into
a mean flow contribution (a large scale contribution including kinetic
and potential energy), as well as a contribution from small scale
kinetic energy due to local fluctuations of the divergent velocity
field and a contribution from small scale potential energy due to
local height fluctuations. The large scale (or mean flow) energy defined
in Eq. (\ref{eq:meanenergy_discr_def}) for the discrete model
becomes

\begin{equation}
\mathcal{E}_{mf}\left[\overline{h},\overline{hq},\overline{h\mu}\right]=\frac{1}{2}\int\mathrm{d}{\bf x}\;\left[\overline{h}{\bf u}_{mf}^{2}+g\left(\overline{h}+h_{b}-1\right)^{2}\right].\label{eq:mean_field_energy_funct}
\end{equation}
The small scale (or subgrid-scale) kinetic energy due to local fluctuations
of the divergent part of the velocity field defined in Eq. (\ref{eq:kinenergy_discr_def})
becomes 
\begin{equation}
\mathcal{E}_{\delta\mu}\left[\overline{h},\overline{\mu},\overline{\mu^{2}}\right]=\frac{1}{2}\int\mathrm{d}{\bf x}\;\left(\overline{h\mu^{2}}-\frac{\overline{h\mu}^{2}}{\overline{h}}\right),\label{eq:kinetic_energy_fluct_funct}
\end{equation}
 and the small scale (or sub-grid scale) potential energy due to local
height fluctuations defined in Eq. (\ref{eq:potenergy_discr_def})
becomes 
\begin{equation}
\mathcal{E}_{\delta h}\left[\overline{h},\overline{h^{2}}\right]=\frac{g}{2}\int\mathrm{d}{\bf x}\;\left(\overline{h^{2}}-\overline{h}^{2}\right).\label{eq:Potential_energy_fluct_funct}
\end{equation}
 One can check that the total energy in Eq. (\ref{eq:Total_energy_funct})
is the sum of the three contributions given in Eq (\ref{eq:mean_field_energy_funct}),
(\ref{eq:kinetic_energy_fluct_funct}) and (\ref{eq:Potential_energy_fluct_funct}):

\begin{equation}
\mathcal{E}=\mathcal{E}_{mf}+\mathcal{E}_{\delta\mu}+\mathcal{E}_{\delta h}.\label{eq:total_energy_func_as_a_sum}
\end{equation}

\subsubsection{Microcanonical variational problem for the probability density field}

Let us come back to the average of a macroscopic observable $\mathcal{A}$
defined in Eq. (\ref{eq:avarage_mac}). Using the asymptotic estimate
for $\Omega\left[\rho\right]$ given in Eq. (\ref{eq:LargeDev_OmRho}),
the average of an observable $\mathcal{A}$ defined in Eq. (\ref{eq:avarage_mac})
becomes%
\footnote{Strictly speaking, the equal sign should be noted $\asymp$ which
means that the logarithm of the terms on both sides are equivalent,
see e.g. Ref. \cite{touchette2009}.%
} 
\begin{equation}
\left\langle \mathcal{A}\right\rangle _{d\mu_{q,h,\mu}^{M,N}}=\int\mathcal{D}\left[\rho\right]\;\mathcal{A}\left[\rho\right]\frac{\mathrm{e}^{NM\,\mathcal{S\left[\rho\right]}}}{\Omega\left(E,\left\{ Z_{k}\right\} \right)}\delta\left(\mathcal{E}\left[\rho\right]-E\right)\prod_{k=0}^{\infty}\delta\left(\mathcal{Z}_{k}\left[\rho\right]-Z_{k}\right)\prod_{\mathbf{x}}\delta\left(\overline{1}\left[\rho\right]-1\right).
\end{equation}
It comes to a Laplace-type integral where $NM\rightarrow\infty$.
Thus, the value of $\left\langle \mathcal{A}\right\rangle _{d\mu_{q,h,\mu}^{M,N}}$
will be completely dominated by the contribution of the pdf $\rho$
that maximizes the macrostate entropy defined in Eq. (\ref{eq:Mixing_Entropy})
while satisfying the normalization constraint 
\begin{equation}
\overline{1}\left[\rho\right]=\int\mathrm{d}\sigma_{h}\mathrm{d}\sigma_{q}\mathrm{d}\sigma_{\mu}\:\rho\left(\mathbf{x},\sigma_{h},\sigma_{q},\sigma_{\mu}\right)=1,\label{eq:normalization_constraint}
\end{equation}
and the microcanonical constraints $\mathcal{E}\left[\rho\right]=E$
, $\left\{ \mathcal{Z}_{k}\left[\rho\right]=Z_{k}\right\} _{k\ge0}$,
where the energy is defined in Eq. (\ref{eq:Total_energy_funct})
and the potential vorticity moments are defined in Eq. (\ref{eq:Total_moments_funct}).
This variational problem can be written in compact form as: 
\begin{equation}
\max_{\rho}\left\{ \mathcal{S}\left[\rho\right]\left|\;\mathcal{E}\left[\rho\right]=E,\;\forall k\in\mathbb{N}\quad\mathcal{Z}_{k}\left[\rho\right]=Z_{k},\;\forall\mathbf{x}\in\mathcal{D}\quad\overline{1}(\mathbf{x})\left[\rho\right]=1\right.\right\} .\label{eq:Max_prob_gen}
\end{equation}
The probability measure $\rho$ induced by the empirical density field has a concentration property. In other words, the average of an observable depending only on macrostates  is dominated by the most probable macrostates, which are solutions of the variational problem (\ref{eq:Max_prob_gen}).

An interesting limit case for the entropy in Eq. (\ref{eq:Mixing_Entropy})
is worth mentioning in order to relate the variational problem in Eq. (\ref{eq:Max_prob_gen})
with previous studies on the shallow water system. Let us assume
that there is neither height variation nor divergent fluctuations,
which would be the case if considering a quasi-geostrophic model or
2d incompressible Euler equations. Then the only height level and
the only divergence level are $\sigma_{h}=1$, $\sigma_{\mu}=0$,
respectively. Defining $\rho_{q}\left(\mathbf{x},\sigma_{q}\right)=\rho\left(\mathbf{x},1,\sigma_{q},0\right)$,
Eq. (\ref{eq:Mixing_Entropy}) becomes up to an unimportant constant:
\begin{equation}
\mathcal{S}\left[\rho_{q}(\mathbf{x},\sigma_{q})\right]=-\int\mathrm{d}\mathbf{x}\mathrm{d}\sigma_{q}\ \rho_{q}\log\rho_{q}.\label{eq:S_MRS}
\end{equation}
We recover in that case the macrostate entropy of the Miller-Robert-Sommeria
theory \cite{Miller:1990_PRL_Meca_Stat,RobertSommeria:1992_PRL_Relaxation_Meca_Stat}.

Let us now assume that the height varies with position but that there
is no local height fluctuations. Then at point $\mathbf{x}$ the only
height level is $\sigma_{h}=\overline{h}(\mathbf{x})=h(\mathbf{x})$.
Defining $\rho_{q\mu}\left(\mathbf{x},\sigma_{q},\sigma_{\mu}\right)=\rho\left(\mathbf{x},h(\mathbf{x}),\sigma_{q},\sigma_{\mu}\right)$,
the macrostate entropy in Eq. (\ref{eq:Mixing_Entropy}) becomes up
to an unimportant constant: 
\begin{equation}
\mathcal{S}\left[\rho_{q\mu}\left(\mathbf{x},\sigma_{q},\sigma_{\mu}\right),h\left(\mathbf{x}\right)\right]=-\int\mathrm{d}\mathbf{x}\mathrm{d}\sigma_{q}\mathrm{d}\sigma_{\mu}\: h\rho_{q\mu}\log\rho_{q\mu}.\label{eq:entropy_like_chavanis}
\end{equation}
This form of the entropy was proposed by \cite{Chavanis02},
without microscopic justification. Interestingly, \cite{Chavanis02}
obtained Eq. (\ref{eq:entropy_like_chavanis}) by assuming that the
macrostate entropy can be written as 
\begin{equation}
\mathcal{S}=\int\mathrm{d}\mathbf{x}\mathrm{d}\sigma_{q}\mathrm{d}\sigma_{\mu}h(\mathbf{x})s\left(\rho_{q\mu}\left(\mathbf{x},\sigma_{q},\sigma_{\mu}\right)\right),\label{eq:chavanissommeria_SZ}
\end{equation}
and by noting  that $s(\rho)=-\rho\log\rho$ is the
only function that leads to equilibrium states that are stationary.

\section{General properties of the equilibria and simplification of the theory
in limiting cases.\label{sec:Computation-of-the}}

General properties of equilibrium states, solutions of the variational
problem in Eq. (\ref{eq:Max_prob_gen}), are discussed in this section.
Critical points of the variational problem are given in the first
subsection ; they are computed in appendix \ref{sec:Critical-points-of}.
This allows to obtain an equation for the large scale flow, and to
show that equilibrium states of the shallow water model are positive
temperature states. A weak height fluctuation limit is considered
in a second subsection. It is found that the large scale flow and
the small scale fluctuations are decoupled in this limit, and that
there is equipartition between small scale potential energy and small
scale kinetic energy. The quasi-geostrophic limit is investigated
in a third subsection.

\subsection{Equilibrium states are stationary states with positive temperatures\label{sub:Critical-point-equations}}

\subsubsection{Properties of the critical points\label{sub:Properties-of-the}}

Critical points of the equilibrium variational problem defined in
Eq. (\ref{eq:Max_prob_gen}) are solutions of the equation
\begin{equation}
\forall\delta\rho,\quad\delta\mathcal{S}\left[\rho\right]-\beta\delta\mathcal{E}\left[\rho\right]-\sum_{k=0}^{+\text{\ensuremath{\infty}}}\alpha_{k}\delta\mathcal{Z}_{k}\left[\rho\right]-\int\mathrm{d}{\bf x}\:\xi\left({\bf x}\right)\int\mathrm{d}\sigma_{h}\mathrm{d}\sigma_{q}\mathrm{d}\sigma_{\mu}\:\delta\rho=0\label{eq:critpointsEq}
\end{equation}
where $\beta,\left\{ \alpha_{k}\right\} _{k\ge0}\text{ and }\xi\left({\bf x}\right)$
are the Lagrange multipliers associated with the conservation of the
energy, the potential vorticity moments, and with the normalization
constraint, respectively. The computation of the critical points
is performed in Appendix \ref{sec:Critical-points-of}. 

A first key result of Appendix \ref{sec:Critical-points-of} is that
solutions of Eq. (\ref{eq:critpointsEq}) factorize as 
\begin{equation}
\rho=\rho_{h}\left({\bf x},\sigma_{h}\right)\rho_{q}\left({\bf x},\sigma_{q}\right)\rho_{\mu}\left({\bf x},\sigma_{\mu}\right).\label{eq:rho_factorized}
\end{equation}

A second result of Appendix \ref{sec:Critical-points-of} is that
the pdf of the divergence field is a Gaussian

\begin{equation}
\rho_{\mu}\left({\bf x},\sigma_{\mu}\right)=\sqrt{\frac{\beta}{2\pi}}\exp\left(-\frac{\beta}{2}\left(\sigma_{\mu}-\overline{\mu}\right)^{2}\right).\label{eq:rho_mu}
\end{equation}
Recalling that $\beta=\partial S/\partial E$ is the inverse temperature,
we see that the variance of the pdf of the divergence field is given
by the temperature of the flow: 
\begin{equation}
\overline{\mu^{2}}-\overline{\mu}^{2}=\beta^{-1}.\label{eq:beta_fluct_rel}
\end{equation}
This equation has important physical consequences. First, fluctuations
of the divergent field do not vary with space. Second, the temperature
of the equilibrium state is necessary positive. This contrasts with
equilibrium states of two-dimensional Euler flow, which can be characterized
by negative temperature states \cite{SommeriaRobert:1991_JFM_meca_Stat,Miller_Weichman_Cross_1992PhRvA}. This was realized first by Onsager for the point vortex model
\cite{Onsager:1949_Meca_Stat_Points_Vortex}. In the context of 2D Euler equations in doubly periodic domains, the equilibrium states are always  characterized by negative temperature, just as in the case discussed by Onsager (this result is shown in \cite{Bouchet_These}). However, the existence of large scale flow structures characterized by positive temperatures are possible at low energy in the presence of lateral  boundaries and non-zero circulation, and/or in the presence of bottom topography in the framework of QG model. Such positive temperature states are  known and documented in the literature (see e.g. the original papers by \cite{Miller_Weichman_Cross_1992PhRvA}, \cite{RobertSommeria:1992_PRL_Relaxation_Meca_Stat} and  \cite{Majda_Wang_Bombardement_2006_Comm} or \cite{BouchetVenaillePhysRep} and references therein). 

In addition, it follows
from Eqs. (\ref{eq:kinetic_energy_fluct_funct}), (\ref{eq:rho_factorized})
and (\ref{eq:beta_fluct_rel}) that the temperature is directly related
to small scale kinetic energy due to the divergent velocity field:
\begin{equation}
\mathcal{E}_{\delta\mu}=\frac{1}{2\beta}.\label{eq:fluct_kin_en_crit_points}
\end{equation}

A third result of Appendix \ref{sec:Critical-points-of} is the expression
of the pdf of the height field: 
\begin{equation}
\rho_{h}\left({\bf x},\sigma_{h}\right)=\frac{1}{\mathbb{G}_{h}\left({\bf x}\right)}{\displaystyle \sigma_{h}^{2}\exp\left(-\beta\frac{g}{2}\sigma_{h}-\frac{\xi_{h}\left({\bf x}\right)}{\sigma_{h}}\right)},\qquad\mathbb{G}_{h}\left({\bf x}\right)=\int\mathrm{d}\sigma_{h}\;\sigma_{h}^{2}\exp\left(-\beta\frac{g}{2}\sigma_{h}-\frac{\xi_{h}\left({\bf x}\right)}{\sigma_{h}}\right),\label{eq:rho_h}
\end{equation}
where $\xi_{h}\left({\bf x}\right)$ a function related to $\bar{h}\left({\bf x}\right)$
through $\bar{h}=\int\mathrm{d}\sigma_{h}\:\sigma_{h}\rho_{h}$. 

A fourth result of Appendix \ref{sec:Critical-points-of} is the expression
of the pdf of the potential vorticity field: 
\begin{equation}
\rho_{q}\left({\bf x},\sigma_{q}\right)=\frac{1}{\mathcal{\mathbb{G}}_{q}\left(\beta\Psi_{mf}\right)}\exp\left(\text{\ensuremath{\beta\Psi_{mf}}}\sigma_{q}-\sum_{k=1}^{+\infty}\alpha_{k}\sigma_{q}^{k}\right),\quad\mathcal{\mathbb{G}}_{q}\left(\beta\Psi_{mf}\right)=\int\mathrm{d}\sigma_{q}\ \exp\left(\beta\Psi_{mf}\sigma_{q}-\sum_{k=1}^{+\infty}\alpha_{k}\sigma_{q}^{k}\right),\label{eq:rho_q}
\end{equation}
where $\Psi_{mf}\left[\overline{h},\overline{q},\overline{\mu}\right]$
is the mass transport streamfunction of the mean flow defined through
an Helmholtz decomposition of $\overline{h}\mathbf{u}_{mf}$: 
\begin{equation}
\overline{h}{\bf u}_{mf}=\nabla^{\bot}\Psi_{mf}+\nabla\Phi_{mf}.\label{eq:umf_psimf_phimf}
\end{equation}
According to Eq. (\ref{eq:rho_q}), the coarse-grained potential vorticity
field is a function of the mass transport streamfunction:
\begin{equation}
{\displaystyle \overline{q}=\frac{\mathcal{\mathbb{G}}_{q}^{\prime}(\beta\Psi_{mf})}{\mathcal{\mathbb{G}}_{q}(\beta\Psi_{mf})}}.\label{eq:overlineq_stat_state}
\end{equation}

A fifth result of Appendix \ref{sec:Critical-points-of} is that the
mass transport potential of the mean flow defined in Eq. (\ref{eq:umf_psimf_phimf})
vanishes: 
\begin{equation}
\Phi_{mf}=0,\label{eq:Pot_vanish}
\end{equation}
and the velocity field can now be written %
\begin{comment}
Furthermore, isoline of the mean-flow Bernoulli function $B_{mf}$
are isolines of the coarse-grained mass transport streamfunction $\Psi_{mf}$,
$B_{mf}=\mathbb{G}_{B}\left(\Psi_{mf}\right)$, where $\mathbb{G}_{B}$
is related to $\mathcal{\mathbb{G}}_{q}(\beta\Psi_{mf})$ by 
\begin{equation}
\frac{\mathrm{d}\mathbb{G}_{B}\left(\Psi_{mf}\right)}{\mathrm{d}\Psi_{mf}}=\frac{1}{\mathcal{\mathbb{G}}_{q}\left(\beta\Psi_{mf}\right)}\frac{\mathrm{d}\mathcal{\mathbb{G}}_{q}\left(\beta\Psi_{mf}\right)}{\mathrm{d}\left(\beta\Psi_{mf}\right)}.\label{eq:Q_Psi_rel_MF}
\end{equation}
\end{comment}

\begin{equation}
{\bf u}_{mf}=\frac{1}{\overline{h}}\nabla^{\bot}\Psi_{mf}.\label{eq:mean_u_Psi_rel}
\end{equation}

A sixth result of Appendix \ref{sec:Critical-points-of} concerns
the Bernoulli function of the mean flow defined as
\begin{equation}
B_{mf}=\frac{1}{2}\mathbf{u}_{mf}^{2}+g\left(\overline{h}+h_{b}-1\right).\label{eq:Bernouilli_Bar}
\end{equation}
According to Eq. (\ref{eq:bmf_final_app}), combining Eq. (\ref{eq:Bmf_inter_rhoh_app})
and Eq. (\ref{eq:Final_rho_h}) of Appendix \ref{sec:Critical-points-of}
allows to express $B_{mf}$ in terms of $\beta$, $\mathbb{G}_{q}$,
$\mathbb{G}_{h}$ and a constant $A_{0}$ that can be computed in
principle using the conservation of the total mass $\mathcal{Z}_{0}=1$:
\begin{equation}
B_{mf}=\beta^{-1}\log\left(\mathbb{G}_{q}\mathbb{G}_{h}\right)+g\overline{h}+A_{0}.\label{eq:bernouilli_MF_def}
\end{equation}

\subsubsection{Equation for the large scale flow}

Let us now establish the equations allowing to compute the large scale
flow. A first equation is obtained by injecting the expression for
the mean flow $\mathbf{u}_{mf}$ given in Eq. (\ref{eq:mean_u_Psi_rel})
into the expression of the Bernoulli function defined in Eq. (\ref{eq:Bernouilli_Bar}),
and by combining Eq. (\ref{eq:Bernouilli_Bar}) with Eq. (\ref{eq:bernouilli_MF_def}):
\begin{equation}
\frac{1}{2}\frac{1}{\overline{h}^{2}}\left(\nabla\Psi_{mf}\right)^{2}+gh_{b}=\frac{1}{\beta}\log\left(\mathbb{G}_{q}\mathbb{G}_{h}\right)+A_{1},\label{eq:Psi_Eq_Rel2-1}
\end{equation}
 with $A_{1}=A_{0}+g$. A second equation is obtained by taking first
the curl of Eq. (\ref{eq:u_mf_continuous}), which yields $\overline{hq}-f=\nabla^{\perp}\mathbf{u}_{mf}$.
Then, replacing $\mathbf{u}_{mf}$ by its expression given in Eq.
(\ref{eq:mean_u_Psi_rel}), remembering that the potential vorticity
field and the height field of the critical points of the variational
problem are decorrelated ($\overline{qh}=\overline{h}\overline{q}$),
and replacing $\overline{q}$ by its expression given in Eq. (\ref{eq:overlineq_stat_state})
yields 

\begin{equation}
\overline{h}\frac{\mathbb{G}_{q}'\left(\beta\Psi_{mf}\right)}{\mathbb{G}_{q}\left(\beta\Psi_{mf}\right)}-f=\nabla\left(\frac{\nabla\Psi_{mf}}{\overline{h}}\right).\label{eq:Psi_Eq_rel1-1}
\end{equation}

The closed system of partial differential equations (\ref{eq:Psi_Eq_Rel2-1})
and (\ref{eq:Psi_Eq_rel1-1}) must be solved for $\overline{h}$ and
$\Psi_{mf}$ for a given value of $\beta$ , $\mathbb{G}_{q}$, $\mathbb{G}_{h}$,
and $A_{1}$. The set of parameters $\beta,\left\{ \alpha_{k}\right\} _{k\ge1},A_{1}$
must \textit{in fine} be expressed in terms of the constraints of the
problem given by the energy $E$, and the potential vorticity moments
$\left\{ Z_{k}\right\} _{k\ge0}$. This may require a numerical resolution
in the general case. \\

We have seen in subsection \ref{sub:Stationary-states} that a flow
described by $(h,\mathbf{u})$ (or equivalently by $\left(h,q,\mu\right)$,
or by $\left(h,\Phi,\Psi\right)$) is a stationary state of the shallow
water model if and only if $J(q,\Psi)=0$ and $\text{\ensuremath{\Phi}}=0$.
%In addition, a stationary state is characterized by $q=\mathrm{d}B/\mathrm{d}\Psi$,
%where $B$ is the Bernoulli function of the flow. 

According to Eqs. (\ref{eq:overlineq_stat_state})
and (\ref{eq:Pot_vanish}), the large scale flow $\left(\overline{h},\mathbf{u}_{mf}\right)$ is a stationary state of the shallow water model. 

\subsubsection{Comparison with previous results \label{sub:comparison-with-prev}}

The set of equations (\ref{eq:Psi_Eq_Rel2-1}) and (\ref{eq:Psi_Eq_rel1-1}) describing the large scale flow is similar to the one obtained by Weichman and Petrich \cite{Weichman01} (through a Kac-Hubbard-Stratonovich
transformation) and by Chavanis and Sommeria \cite{Chavanis02} (through a phenomenological generalization of the Miller-Robert-Sommeria theory), excepted that the rhs of Eq. (\ref{eq:Psi_Eq_Rel2-1}) contains an
additional term $\frac{1}{\beta}\log\left(\mathbb{G}_{h}\right)+A_{1}$ which does not appear in these previous works. The reason for the presence of this additional term is that we have taken into account the presence of small scale fluctuations of height and velocity, which were neglected in Refs. \cite{Weichman01,Chavanis02}. We will see in subsection \ref{sub:The-quasi-geostrophic-limit} that this additional term becomes negligible with respect to the others in the quasi-geostrophic limit, in which case we recover exactly the set of equations for the large scale flow obtained in Refs. \cite{Weichman01,Chavanis02}.%

In addition, we have shown in subsection \ref{sub:Properties-of-the} that only positive temperature states are allowed, which shows that only one subclass of the states described by Eqs. (\ref{eq:Psi_Eq_Rel2-1}) and (\ref{eq:Psi_Eq_rel1-1}) are actual equilibrium states.

\subsection{Equipartition and decoupling in the limit of weak local height fluctuations }

We consider in this subsection the limit of weak local height fluctuations.
This step makes possible computation of explicit solutions of the
variational problem (\ref{eq:Max_prob_gen}). Meanwhile, it allows
to explore the consequence of the presence of these small scale fluctuations
on the structure of the large scale flow. By ``limit of weak local
height fluctuations'', we mean 
\begin{equation}
\forall\mathbf{x}\in\mathcal{D},\quad\left(\overline{h^{2}}-\overline{h}^{2}\right)^{1/2}\ll\overline{h}.\label{eq:condition_small_height_fluctuations}
\end{equation}

As already argued in \cite{Weichman01}, this limit of weak local
fluctuations is physically relevant, since the presence of shocks
in the actual dynamics tends to dissipate small scale fluctuations
of height or kinetic energy. This will be further discussed in subsection
\ref{sub:The-effect-of}.

\subsubsection{The height distribution}

Assuming in addition that height levels $\sigma_{h}$ such that $\left|\sigma_{h}-\overline{h}\right|\gg\left(\overline{h^{2}}-\overline{h}^{2}\right)^{1/2}$
do not contribute significantly to the pdf $\rho_{h}$ defined in
Eq. (\ref{eq:rho_h}), we perform an asymptotic development with $\left(\sigma_{h}/\overline{h}-1\right)\ll1$,
and obtain at lowest order in this small parameter a Gaussian shape
for the pdf: 
\begin{equation}
\rho_{h}\left({\bf x},\sigma_{h}\right)=\sqrt{\frac{g\beta}{2\pi\overline{h}\left({\bf x}\right)}}\exp\left(-\frac{g\beta}{2\overline{h}\left({\bf x}\right)}\left(\sigma_{h}-\overline{h}\left({\bf x}\right)\right)^{2}\right).\label{eq:rho_h_weakfluctlim}
\end{equation}
Similarly, the term $\mathbb{G}_{h}$ defined in Eq. (\ref{eq:rho_h})
can be computed explicitly in this limit. One gets at lowest order
\begin{equation}
\mathbb{G}_{h}=\overline{h}^{2}\left({\bf x}\right)\sqrt{\frac{2\pi\overline{h}\left({\bf x}\right)}{g\beta}}\exp\left(-g\beta\overline{h}\left({\bf x}\right)\right).\label{eq:Gh_Weak_Fluc}
\end{equation}
Using Eq. (\ref{eq:rho_h_weakfluctlim}), the weak local height fluctuation
limit given in Eq. (\ref{eq:condition_small_height_fluctuations})
can be interpreted as a low temperature limit $g\overline{h}\gg1/\beta$.
Injecting Eq. (\ref{eq:Gh_Weak_Fluc}) in the previously established
relation in Eq. (\ref{eq:bernouilli_MF_def}) yields 
\begin{equation}
B_{mf}=\frac{1}{\beta}\log\mathbb{G}_{q}+\frac{5}{2\beta}\log\overline{h}+A_{2},\label{eq:Bournilli_MF_smallfluct}
\end{equation}
where $A_{2}$ is a free parameter determined by the conservation
of the total volume. In the rhs of Eq. (\ref{eq:Bournilli_MF_smallfluct}),
there remains an additional term $\frac{5}{2\beta}\log\overline{h}$
which is not present in the large scale flow equation obtained by
\cite{Weichman01,Chavanis02}, but considering the weak flow limit
allowed to obtain an explicit expression for this additional term.
We will see in subsection \ref{sub:The-quasi-geostrophic-limit} that
this term becomes negligible with respect to $\frac{1}{\beta}\log\mathbb{G}_{q}$
in the quasi-geostrophic limit.%
\begin{comment}
We will see that in the quasi-geostrophic limit, we get at lowest
order $\overline{h}=1$ (and that this limit case is necessarily attained
when considering a weak-flow limit). In that case, we recover the 
\end{comment}
\begin{comment}
In the infinite $\beta$ limit, i.e. the small fluctuation limit,
$\overline{h}$ is of order one while $\log\mathbb{G}_{q}$ is of
order $\beta$ ( this can be seen by using a saddle point approximation
on the expression of $\mathbb{G}_{q}$ in Eq. (\ref{eq:E_fluct_def})),
then we get $B_{mf}=\beta^{-1}\log\mathbb{G}_{q}+A_{2}$ which give
the relation $\mathrm{d}B_{mf}/\mathrm{d}\Psi_{mf}=\overline{q}$
\end{comment}

\subsubsection{Equipartition and the limit of weak small scale energy}

Injecting Eq. (\ref{eq:rho_h_weakfluctlim}) in the expression of
the small scale potential energy defined in Eq. (\ref{eq:Potential_energy_fluct_funct})
yields 
\begin{equation}
\mathcal{E}_{\delta h}=\frac{1}{2\beta}.\label{eq:fluct_en_equipart}
\end{equation}
Comparing this result with Eq. (\ref{eq:fluct_kin_en_crit_points})
shows equipartition of the small scale energy between the potential
energy and the kinetic energy. The total energy due to small scale
fluctuations is 
\begin{equation}
E_{fluct}\equiv\mathcal{E}_{\delta h}+\mathcal{E}_{\delta\mu}=\frac{1}{\beta}.\label{eq:E_fluct_def}
\end{equation}

This equipartition result of the small scale energy was already obtained
by Warn when computing equilibrium states of the Galerking truncated
dynamics in a weak flow limit \cite{Warn86}. More precisely, Warn
decomposed the dynamics into vortical modes and inertia-gravity modes
(which are defined as the eigenmodes of the linearized dynamics),
and concluded that the energy of the equilibrium state should be equipartitioned
among inertia-gravity modes in the limit of infinite wavenumber cut-off.
This equipartition of energy among inertia-gravity modes would lead
to equipartition between potential and kinetic energy at small scales,
just as in our case. Here we have recovered this result with a different
approach and  we have generalized it beyond the weak flow limit.

Finally, we remark that the equipartition result of Eq. (\ref{eq:E_fluct_def})
shows that the low temperature limit $g\overline{h}\gg1/\beta$ corresponds
to a limit of weak small scale energy due to local fluctuation of
the height field and of the divergent field:%
\begin{comment}
\begin{equation}
\log\left(\mathbb{G}_{h}\right)=-E_{fluct}^{-1}g\overline{h}+\frac{1}{2}\left(\log\frac{g\overline{h}^{5}}{2\pi}-\log E_{fluct}\right).\label{eq:Psi_Eq_Rel2-1-2}
\end{equation}
\end{comment}
\begin{comment}

\subsubsection{Stationary states}

The system of equation (\ref{eq:Psi_Eq_Rel2-1}) and (\ref{eq:Psi_Eq_rel1-1})
for the large scale flow can be simplified in the weak local height
fluctuation limit by using the expression of $\mathbb{G}_{h}$ obtained
in Eq. (\ref{eq:rho_h}). Indeed, assuming $\overline{h}=\mathcal{O}(1)$
(which can be checked a posteriori), Eq. (\ref{eq:Psi_Eq_Rel2-1})
simplifies at lowest order into 
\begin{equation}
\frac{1}{2}\frac{1}{\overline{h}^{2}}\left(\nabla\Psi_{mf}\right)^{2}+g\left(\overline{h}+h_{b}-1\right)=E_{fluct}\log\mathbb{G}_{q}-A_{1}.\label{eq:Bmf_with_fluct}
\end{equation}
This equation is the same as the one obtained by \cite{Weichman01,Chavanis02}
up to a constant which has to be expressed a posteriori in terms of
the constraints of the problem (just as $\mathbb{G}_{q}$ and $\mathbb{G}_{h}$).
We will see that this constant $A_{1}$ is unimportant in the quasi-geostrophic
limit. 
\end{comment}

\begin{equation}
g\overline{h}\gg E_{fluct}.\label{eq:condition_small_height_term_of_energy}
\end{equation}

\subsubsection{Decoupling between the large scale flow and the fluctuations}

Still by considering the weak local height fluctuation limit, let
us assume that $E_{fluct}$ defined in Eq. (\ref{eq:E_fluct_def})
is given, which means that the temperature is given. According to
Eq. (\ref{eq:Psi_Eq_rel1-1}) and Eq. (\ref{eq:Psi_Eq_Rel2-1}), knowing
the coarse-grained height field $\overline{h}$ and the pdf of potential
vorticity levels $\rho_{q}(\mathbf{x},\sigma_{q})$ (which allows
to compute $\mathbb{G}_{q}$) is sufficient to determine the mass
transport streamfunction $\Psi_{mf}$, and hence the velocity $\mathbf{u}_{mf}$
by using Eq. (\ref{eq:mean_u_Psi_rel}). Then the mean-flow energy
defined in Eq. (\ref{eq:mean_field_energy_funct}) does not depend
on $\rho_{\mu}$:
\begin{equation}
\mathcal{E}_{mf}\left[\overline{h},\rho_{q}\right]=\frac{1}{2}\int\mathrm{d}{\bf x}\;\bigg[\overline{h}\mathbf{u}_{mf}^{2}\left[\overline{h},\overline{q}\right]+g\left(\overline{h}+h_{b}-1\right)^{2}\bigg].\label{eq:Emf_funct_rhoq_h}
\end{equation}
 %
\begin{comment}
(\ref{eq:Psi_Eq_rel1-1}) and an integration in the mean-field energy
defined in Eq. (\ref{eq:mean_field_energy_funct}) yields:

\begin{equation}
\mathcal{E}_{mf}\left[\overline{h},\rho_{q}\right]=\frac{1}{2}\int\mathrm{d}{\bf x}\;\left(f-\overline{q}\overline{h}\right)\Psi_{mf}+g\left(\overline{h}+h_{b}\right)^{2}.\label{eq:Emf_funct_rhoq_h-1}
\end{equation}
\end{comment}
The total energy defined in Eq. (\ref{eq:total_energy_func_as_a_sum})
is therefore the sum of the mean-field energy associated with a large
scale flow that depends only on $\overline{h}$ and $\rho_{q}$ ,
and of the energy of small scale fluctuations associated with height
and divergence fluctuations: 
\begin{equation}
\mathcal{E}=\mathcal{E}_{mf}\left[\overline{h},\rho_{q}\right]+E_{fluct}.\label{eq:energy_funct_Efluc_rhoq_h}
\end{equation}
In addition, injecting Eq. (\ref{eq:rho_factorized}) in Eq. (\ref{eq:Total_moments_funct})
gives the expression of the potential vorticity moments as a functional
of $\overline{h}$ and $\rho_{q}$ only:

\begin{equation}
\forall k\in\mathbb{N}\quad\mathcal{Z}_{k}\left[\overline{h},\rho_{q}\right]=\int\mathrm{d}{\bf x}\mathrm{d}\sigma_{q}\;\overline{h}\rho_{q}\sigma_{q}^{k}.\label{eq:Total_moments_funct_decorr}
\end{equation}

Let us now consider the macrostate entropy functional defined in Eq.
(\ref{eq:Mixing_Entropy}). Injecting Eqs. (\ref{eq:rho_mu}) and
(\ref{eq:rho_h_weakfluctlim}) in Eq. (\ref{eq:rho_factorized}),
and performing the asymptotic expansion of the integrand of the mean-field
entropy with $\left(\sigma_{h}/\overline{h}-1\right)\ll1$ leads at
lowest order to (up to a irrelevant constant)
\begin{equation}
\mathcal{S}=\mathcal{S}_{mf}\left[\bar{h},\rho_{q}\right]+S_{fluct}\left(E_{fluc}\right),\label{eq:entropy_decor}
\end{equation}

\begin{equation}
\mathcal{S}_{mf}\left[\overline{h},\rho_{q}\right]\equiv-\int\mathrm{d}{\bf x}\mathrm{d}\sigma_{q}\;\overline{h}\rho_{q}\log\frac{\rho_{q}}{\overline{h}^{5/2}},\label{entropy_q}
\end{equation}
\begin{equation}
S_{fluct}\left(E_{fluct}\right)\equiv\log\left(E_{fluct}\right).\label{eq:Entropy_fluct_def}
\end{equation}
Since $\rho_{q}$ and $\overline{h}$ are two fields allowing to compute
the large scale flow of energy $\mathcal{E}_{mf}$, and since the
fluctuations of the potential vorticity field do not contribute to
the small scale energy $E_{fluct}$, the entropy $\mathcal{S}_{mf}\left[\overline{h},\rho_{q}\right]$
will be referred to as the macrostate entropy of the large scale flow. 

The second contribution to the total entropy in Eq. (\ref{eq:entropy_decor})
depends only on the energy of the small scale fluctuations $E_{fluct}$.
Since this energy is solely due to the local variance of the height
field and of the divergent field, it will be referred to as the entropy
of the small scale fluctuations. Note that in that case the height
field is involved both in the large scale flow through its local mean value,
and in the small scale fluctuations through its local variance. \\

The decoupling of the energy and the macrostate entropy functional
into a part that depends only on $\rho_{q},\overline{h}$ and another
part that depends only on small scale height and divergent fluctuations
with energy $E_{fluct}$ has both a useful practical consequence and
an interesting physical interpretation. The variational problem in
Eq. (\ref{eq:Max_prob_gen}) can now be recast into two simpler variational
problems: 
\begin{equation}
S\left(E,\left\{ Z_{k}\right\}_{k\geq0} \right)=\max_{E_{fluct}}\left\{ S_{mf}\left(E-E_{fluct},\left\{ Z_{k}\right\} _{k\ge0}\right)+S_{fluct}\left(E_{fluct}\right)\right\} ,\label{eq:max_prob_weakfluct_decor}
\end{equation}
\begin{equation}
S_{mf}\left(E_{mf},\left\{ Z_{k}\right\} _{k\ge0}\right)=\max_{\rho_{q},\int\rho_{q}=1,\overline{h}}\left\{ \mathcal{S}_{mf}\left[\overline{h},\rho_{q}\right]\left|\;\mathcal{E}_{mf}\left[\overline{h},\rho_{q},\right]=E_{mf},\;\forall k\in\mathbb{N}\quad\mathcal{Z}_{k}\left[\rho_{q},\overline{h}\right]=Z_{k}\right.\right\} ,\label{eq:max_prob_weakfluct_decor_2}
\end{equation}
where $\mathcal{S}_{mf}$, $\mathcal{E}_{mf}$ and $\mathcal{Z}_{k}$
are the functional defined in Eqs. (\ref{entropy_q}), (\ref{eq:Emf_funct_rhoq_h}),
and (\ref{eq:Total_moments_funct_decorr}), respectively. The variational
problem in Eq. (\ref{eq:max_prob_weakfluct_decor}) describes two
subsystems in thermal contact. In order to compute the equilibrium
state, one can then compute independently the equilibrium state of
each subsystem, and then equating their temperature in order to find
the global equilibrium state. This classical argument follows directly
from the maximization of (\ref{eq:max_prob_weakfluct_decor}). If
the two subsystems can not have the same temperature (for instance
when the temperature of both subsystem have a different sign), all
the energy is stored in the subsystem with positive temperature in
order to maximize the global entropy. 

In the present case, a first subsystem is given by the large scale
flow of energy $E_{mf},$ which involves the field $\overline{h}$
and the potential vorticity field described by the pdf of vorticity
levels $\rho_{q}$. The equilibrium state of this subsystem is obtained
by solving the variational problem in Eq. (\ref{eq:max_prob_weakfluct_decor_2}). This variational problem corresponds to the one introduced by \cite{Chavanis02} except that large scale flow energy is not the total energy but only the available energy when the energy of the fluctuations has been removed. The entropy of the large scale flow (\ref{entropy_q}) in the variational problem (\ref{eq:max_prob_weakfluct_decor_2}) is closely related to the entropy introduced in \cite{Chavanis02} (up to a functional of $\overline{h}$).
The potential vorticity moment constraints apply only to this subsystem.
These additional constraints are essential since they allow for the
possible existence of a large scale flow, see e.g. \cite{BouchetVenaillePhysRep}. 

The second subsystem is given by the local small scale height fluctuations
and the local small scale divergent fluctuations with total energy
$E_{fluct}$, and with the entropy given in Eq. (\ref{eq:Entropy_fluct_def}).
The inverse temperature of this subsystem $\beta=\mathrm{d}S_{fluct}/\mathrm{d}E_{fluct}$,
which, using Eq. (\ref{eq:Entropy_fluct_def}), yields a temperature
of $\beta^{-1}=E_{fluct}$. 

In practice, it is easier to compute directly equilibrium states of
the large scale flow subsystem in the canonical ensemble, where the
energy constraint is relaxed. In that case the equilibrium states
of this subsystem depend on the temperature $E_{fluct}$ and on the
dynamical invariants $\left\{ Z_{k}\right\} _{k\ge0}$. One just need
to check a posteriori that this ensemble is equivalent to the microcanonical
one by verifying that each admissible energy $E_{mf}$ is reached
when varying $E_{fluct}$ form $0$ to $+\infty$ , for a given set
of potential vorticity moments $\left\{ Z_{k}\right\} _{k\ge0}$ .
In order to find the actual equilibrium state associated with the
total energy $E$, one then needs to solve the equation 
\begin{equation}
E=E_{mf}\left(E_{fluct},\left\{ Z_{k}\right\} _{k\ge0}\right)+E_{fluct}.\label{eq:E_Emf_Efluc}
\end{equation}

To conclude, it is now possible to study independently the large scale
flow subsystem, to consider if necessary any approximation on this
flow, such as the quasi-geostrophic limit or the Euler 2D limit, and
finally to couple this subsystem with the small scale height and divergence
fluctuations subsystem in order to select the actual equilibrium state. If one linearize the large scale flow entropy entropy (\ref{entropy_q}) and the large scale flow energy (\ref{eq:Emf_funct_rhoq_h}), this picture of two subsystems in thermal contact gives a justification to the variational problem introduced by \cite{Vallis92} and extended by \cite{nageswaran_turkington2010}. \cite{Vallis92,nageswaran_turkington2010} suggest that for a given frozen in space potential vorticity, the dynamics should relax through geostrophic adjustment to a state minimizing the total energy. This result is recovered from the coupled variational problem (\ref{eq:max_prob_weakfluct_decor}).

\subsubsection{Either a non-zero circulation or a non-zero bottom topography is
required to sustain a large scale flow at equilibrium \label{sub:In-a-case-without-}}

It is shown in appendix \ref{sec:Global-maximizers-of} that when
circulation is zero (i.e. when $Z_{1}=f$) and when topography is
zero ($h_{b}=0$), a state with a large scale flow at rest ($\mathbf{u}_{mf}=0$)
is a maximizer of the large scale flow macrostate entropy among all
the possible energies:
\begin{equation}
S(0,\left\{ Z_{k}\right\} _{k\ge0})=\max_{E_{mf}}\left\{ S_{mf}\left(E_{mf},\left\{ Z_{k}\right\} _{k\ge0}\right)\right\} \quad\mbox{when \ensuremath{Z_{1}=f} and \ensuremath{h_{b}=0}.}
\end{equation}
 According to Eq. (\ref{eq:Entropy_fluct_def}), the fluctuation entropy
$S_{fluct}$ increases with the fluctuation energy $E_{fluct}$. The
total macroscopic entropy $S_{mf}\left(E-E_{fluct},\left\{ Z_{k}\right\} _{k\ge0}\right)+S_{fluct}\left(E_{fluct}\right)$
is therefore maximal when all the energy is transferred into fluctuations
$(E_{fluct}=E)$. This generalizes to a wider range of flow parameters
and to a wider set of flow geometries a result previously obtained
by Warn in a weak flow limit for a doubly periodic domain without
bottom topography \cite{Warn86}. In addition, we will see in the
next section that when there is a non-zero bottom topography and rotation,
a large scale flow can be sustained at equilibrium. \\

\subsection{The quasi-geostrophic limit\label{sub:The-quasi-geostrophic-limit}}

We show in this subsection that the variational problem of the Miller
Robert Sommeria theory is recovered from Eq. (\ref{eq:max_prob_weakfluct_decor_2})
for the large scale flow subsystem when considering the quasi-geostrophic
limit, which applies to strongly rotating and strongly stratified
flows.

\subsubsection{Geostrophic balance\label{GeoBalsub}}

Let $E_{mf}^{1/2}$ be the typical velocity of the large-scale flow
and let $L=\sqrt{\left|\mathcal{D}\right|}$ be the typical horizontal
scale of the domain where the flow takes place. We introduce the Rossby
number and the Rossby radius of deformation respectively defined as
\begin{equation}
Ro\equiv\frac{E_{mf}^{1/2}}{fL},\quad R\equiv\frac{\sqrt{gH}}{f}.\label{eq:define_rossby}
\end{equation}
Here $f$ is the Coriolis parameter, $H=1$ is the mean depth and
$g$ the gravity, see subsection \ref{sub:Dynamics}. If $f\ne0$
we can always rescale time unit so that $f=1$, and we make this choice
in the following. It is also assumed that the aspect ratio of the
domain where the flow takes place is of order one, so that $L=1$
since we chose length unit so that $|\mathcal{D}|=1$. The quasi-geostrophic
limit corresponds to small Rossby number, and to a Rossby radius of
deformation that is not significantly larger than the domain length
scale: 
\begin{equation}
Ro\ll1,\quad R^{-1}=\mathcal{O}\left(1\right).\label{eq:condition_QG}
\end{equation}
By construction, the mean flow is of the order of the Rossby number:
$|\mathbf{u}_{mf}|\sim Ro$. The coarse-grained interface height is
given by 

\begin{equation}
\overline{\eta}=\overline{h}-1+h_{b}.\label{eq:coarse_grained_interface_height}
\end{equation}

Let us assume that the spatial variations in fluid depth are small
compared to the total depth $H=1$, with the scaling $\overline{\eta}\sim R^{-2}Ro$. At lowest order in $Ro$, the mean flow Bernoulli potential
defined in Eq. (\ref{eq:Bernouilli_Bar}) becomes $B_{mf}=R^{2}\overline{\eta}$. Remembering that we consider in addition to the quasi-geostrophic
limit a weak fluctuation limit given by Eq. (\ref{eq:condition_small_height_term_of_energy}),
which can be expressed as $R^{2}\gg\beta^{-1}$, Eq. (\ref{eq:Bournilli_MF_smallfluct})
yields at lowest order 
\begin{equation}
B_{mf}=R^{2}\overline{\eta}=\beta^{-1}\log\mathbb{G}_{q}+cst.\label{eq:Bmf_geo}
\end{equation}
This equation implies $\mathrm{d}B_{mf}/\mathrm{d}\Psi_{mf}=\overline{q}$,
consistently with what we expected for a stationary flow in the absence
of small scale fluctuations, see subsection \ref{sub:Stationary-states}.
Taking the curl of Eq. (\ref{eq:Bmf_geo}) and collecting the lowest
order terms yields geostrophic balance\footnote{For the shallow water model, the fluid is at hydrostatic balance. Thus the pressure in the fluid is $P(x,y,z,t)=P_{o}+\rho g(H+\eta(x,y,t)-z)$. Then the pressure horizontal gradient is simply proportional to the interface height horizontal gradient. Hence, the geostrophic balance simply writes $R^{2}\nabla^{\bot}\overline{\eta}=\mathbf{u}_{mf}$.} 
\begin{equation}
R^{2}\nabla^{\perp}\overline{\eta}=\mathbf{u}_{mf}\label{eq:GeoBal}.
\end{equation}
Eq. (\ref{eq:GeoBal}) also shows that the scaling hypothesis for $\overline{\eta}$ is self-consistent.

 {It is remarkable that equilibrium statistical mechanics predicts the emergence of geostrophic balance. We stress that those results, which have been obtained through the introduction of a semi-Lagrangian discrete model, are valid whatever the amplitude of bottom topography variations (i.e. beyond the usual approximation $h_b\sim Ro$ required to derive the quasi-geostrophic dynamics). By contrast, in the framework of a Eulerian discrete model, one would find that the large scale flow is not at geostrophic equilibrium unless $h_b\sim Ro$ or $h_b\ll Ro$, see Appendix E.}

\subsubsection{Quasi-geostrophic dynamics}

The geostrophic balance is not a dynamical equation. When  $h_{b}\sim Ro$,  the dynamics if given by the quasi-geostrophic equations. At lowest order in $Ro$, we
get $\overline{h}\mathbf{u}_{mf}=\mathbf{u}_{mf}$, and 
\begin{equation}
\psi_{mf}=\Psi_{mf},\quad\phi_{mf}=\Phi_{mf}=0.\label{eq:Psimg_psimf_qg}
\end{equation}
where $\Psi_{mf}$ and $\psi_{mf}$ are the transport streamfunction
and the streamfunction obtained through the Helmholtz decomposition
of $\overline{h}\mathbf{u}_{mf}$ and $\mathbf{u}_{mf}$, respectively.
In that case, the relative vorticity is 
\begin{equation}
\overline{\omega}=\Delta\psi_{mf}.\label{eq:omega_qg}
\end{equation}
 
The geostrophic balance (\ref{eq:GeoBal}) is equivalent to $\overline{\eta}=R^{-2}\psi_{mf}+C$. The
value of $\psi_{mf}$ at the domain boundary can always be chosen
such that the constant term vanishes, which yields 
\begin{equation}
\overline{\eta}=\frac{\psi_{mf}}{R^{2}}.\label{eq:geos_eq_macro}
\end{equation}
Mass conservation given in Eq. (\ref{eq:eta}) leads then to the
following constraint on the streamfunction:
\begin{equation}
\int\mathrm{d}\mathbf{x}\ \psi_{mf}=0.\label{eq:mass_conservation_psi_mf}
\end{equation}

Given a potential vorticity level $\sigma_{q}$, a change of variable
can be performed by introducing quasi-geostrophic potential vorticity
levels\footnote{This change of variable  is a guess guided by the fact that the QG potential vorticity is usually obtained by expanding the SW potential vorticity in the limit of small height variations minus a constant and unimportant term. Here we start by removing the unimportant constant (1 in our unit system) from the potential vorticity levels, and then perform the small scale expansion in height. 
}

\begin{equation}
\sigma_{g}\equiv\left(\sigma_{q}-1\right).\label{eq:qg_levels}
\end{equation}
 The pdf of quasi-geostrophic levels is 

\begin{equation}
\rho_{g}\left(\mathbf{x},\sigma_{g}\right)=\rho_{q}\left(\mathbf{x},1+\sigma_{g}\right).\label{eq:rho_g}
\end{equation}
The local quasi-geostrophic potential vorticity moments are defined
as 
\begin{equation}
\forall k\in\mathbb{N},\ \overline{q_{g}^{k}}\equiv\int\mathrm{d}\sigma_{g}\ \sigma_{g}^{k}\rho_{g}.\label{eq:qg_moments}
\end{equation}
At lowest order in $Ro$, the coarse-grained quasi-geostrophic potential
vorticity obtained by considering $k=1$ in Eq. (\ref{eq:qg_moments})
becomes 

\begin{equation}
\overline{q}_{g}=\overline{\omega}-\overline{\eta}+h_{b}.\label{eq:qg_expression_inter}
\end{equation}
which, using Eqs. (\ref{eq:omega_qg}) and (\ref{eq:geos_eq_macro}),
yields

\begin{equation}
\overline{q}_{g}=\Delta\psi_{mf}-\frac{\psi_{mf}}{R^{2}}+h_{b}\ .\label{eq:qg_pv_def}
\end{equation}
In the quasi-geostrophic limit, the large scale flow is fully described
by the streamfunction $\psi_{mf}$, which can be obtained by inverting
the coarse-grained potential vorticity field defined in Eq. (\ref{eq:qg_pv_def}).

\subsubsection{Quasi-geostrophic constraints and variational problem }

At lowest order in $Ro$, and after an integration by part, the expression
of the mean-flow energy in Eq. (\ref{eq:Emf_funct_rhoq_h}) is equal
to the quasi-geostrophic energy:

\begin{equation}
\mathcal{E}_{mf,g}\left[\overline{q}_{g}\right]\equiv\frac{1}{2}\int\mathrm{d}\mathbf{x}\ \left[\left(\nabla\psi_{mf}\right)^{2}+\frac{\psi_{mf}^{2}}{R^{2}}\right]\ ,\label{eq:Eqg_def}
\end{equation}
 where $\psi_{mf}$ can be expressed in terms of $\overline{q}_{g}$
through Eq. (\ref{eq:qg_pv_def}). Similarly, the conservation of
the potential vorticity moments defined in Eq. (\ref{eq:Total_moments_funct_decorr})
implies the conservation of quasi-geostrophic potential vorticity
moments 

\begin{equation}
\forall k\in\mathbb{N},\quad\mathcal{Z}_{g,k}\equiv\int\mathrm{d}\mathbf{x}\ \overline{q_{g}^{k}}\ ,\label{eq:PV_moments_qg}
\end{equation}
and the macrostate entropy of the large scale flow defined in Eq.
(\ref{entropy_q}) writes now
\begin{equation}
\mathcal{S}_{mf,g}\left[\rho_{g}\right]=-\int\mathrm{d}{\bf x}\mathrm{d}\sigma_{q}\;\rho_{g}\log\rho_{g}.\label{eq:entropy_qg_mf}
\end{equation}
Using Eq. (\ref{eq:Eqg_def}), (\ref{eq:PV_moments_qg}) and (\ref{eq:entropy_qg_mf}),
the variational problem in Eq. (\ref{eq:max_prob_weakfluct_decor_2})
is now recast into a simpler variational problem on the pdf $\rho_{g}$:
\begin{equation}
S_{mf,g}\left(E_{mf},\left\{ Z_{k}\right\} _{k\ge1}\right)=\max_{\rho_{g},\int\rho_{g}=1}\left\{ \mathcal{S}_{mf,g}\left[\rho_{g}\right]\left|\;\mathcal{E}_{mf,g}\left[\rho_{g}\right]=E_{mf},\;\forall k\in\mathbb{N}\quad\mathcal{Z}_{g,k}\left[\rho_{g}\right]=Z_{k}\right.\right\} .\label{eq:max_prob_qg_2}
\end{equation}
The entropy $\mathcal{S}_{mf,g}$, the energy $\mathcal{E}_{mf,g}$
and the potential vorticity moments $\left\{ \mathcal{Z}_{g,k}\right\} _{k\ge1}$
are defined in Eqs. (\ref{eq:entropy_qg_mf}), (\ref{eq:Eqg_def})
and (\ref{eq:PV_moments_qg}). The variational problem defined in Eq.
(\ref{eq:max_prob_qg_2}) is the variational problem of the Miller-Robert-Sommeria
statistical mechanics \cite{Miller:1990_PRL_Meca_Stat,SommeriaRobert:1991_JFM_meca_Stat}.%

\subsubsection{Maximum energy states and consistency of the quasi-geostrophic approximation. }

We have shown that in the weak height fluctuation limit and the quasi-geostrophic
limit, computation of the large scale flow associated with the equilibrium
state amounts to the computation of the solution of the variational
problem in Eq. (\ref{eq:max_prob_qg_2}), with the restriction that
the temperature is positive (due to the coupling with small scale
fluctuations of height and divergence, as discussed in subsection
\ref{sub:Properties-of-the}). Here we discuss the solutions of this
variational problem, which are energy maxima for a given set of potential
vorticity moments $\left\{ \mathcal{Z}_{g,k}\right\} _{k\ge1}$. Although
the initial weak fluctuation or quasi-geostrophic limit may not be
fulfilled for such states, they can always be computed, and it provides
an upper bound for the energy of the large scale flow obtained in
those limits.

It is known that for positive temperatures ($\beta^{-1}=E_{fluct}>0$),
the equilibrium entropy defined in Eq. (\ref{eq:max_prob_qg_2}) is
concave \cite{BouchetVenaillePhysRep}, and the energy increases when
$\beta$ decreases. This means that the state with a maximum energy
is reached when $\beta\rightarrow0$, see e.g. \cite{SommeriaRobert:1991_JFM_meca_Stat}.
\\

Injecting $\beta=0$ in the expression of $\rho_{q}$ in Eq. (\ref{eq:rho_q})
leads to a uniform mean potential vorticity field. According to Eq.
(\ref{eq:rho_g}) and (\ref{eq:qg_moments}), this implies that the
quasi-geostrophic potential vorticity field is also uniform: $\overline{q}_{g}=Z_{g,1}$
where $Z_{g,1}=\int\mathrm{d}\mathbf{x}\ \overline{q}_{g}$ is the
circulation. Since $\overline{q}_{g}$ is a constant, this state is
referred to as the ``mixed'' state. We get 
\begin{equation}
\Delta\psi_{mix}-\frac{\psi_{mix}}{R^{2}}=Z_{g,1}-h_{b}.\label{eq:psi_mf_mix}
\end{equation}
Let us call $E_{mix}$ the energy of the mixed state: 
\begin{equation}
E_{mix}\equiv\max_{\rho_{g},\int\rho_{g}=1}\left\{ \mathcal{E}_{mf}|\forall k\in\mathbb{N}\quad\mathcal{Z}_{g,k}=Z_{g,k}\right\} .\label{eq:Eb_def}
\end{equation}
Using Eq. (\ref{eq:Eqg_def}), it can formally be written 

\begin{equation}
E_{mix}=-\frac{1}{2}\int\left(Z_{g,1}-h_{b}\right)\left(\Delta-\frac{1}{R^{2}}\right)^{-1}\left(Z_{g,1}-h_{b}\right).\label{eq:Emix}
\end{equation}

For a given domain geometry, a given circulation $Z_{g,1}$ and a
given bottom topography field $h_{b}$, a non trivial large scale
flow can be observed whenever $E_{mix}>0$. We see from Eq. (\ref{eq:Emix})
and (\ref{eq:Eb_def}) that the condition for a large scale flow to
exist is that either the circulation $Z_{g,1}$ is non zero or the bottom
topography $h_{b}$ is non-zero. If both $Z_{g,1}=0$ and $h_{b}=0$,
then the energy of the large scale flow vanishes ($E_{mf}=0$), and
all the energy is lost in small scale fluctuations ($E=E_{fluct}$),
consistently with the results of subsection \ref{sub:In-a-case-without-}.
In this case, coupling thermally a large scale flow with fluctuations
leads to a state with all the energy lost in fluctuations. Note that
in the case $Z_{g,1}=0$ $R\sim1$, $E_{mix}$ can also be interpreted
as a norm of the topography field $h_{b}$. \\

Since $E_{mf}\le E_{mix,}$ and since $E_{mix}$ depends only on the
problem parameters (namely the circulation, the Rossby radius and
the bottom topography), a sufficient condition to have $Ro\ll1$ is
\begin{equation}
E_{mix}^{1/2}\ll1.\label{eq:Qg_InCondRange}
\end{equation}
If this condition is fulfilled, the quasi-geostrophic assumption is
self-consistent (as well as for the scaling $h_{b}\sim Ro$ in the
case $Z_{g,1}=0$).

\section{Explicit computation of phase diagrams and discussion\label{sec:Computation-of-energy-enstrophy}}

The aim of this section is to apply the results of the previous section
to the actual computation of equilibria and their energy partition.
In order to solve analytically the variational problem of the statistical
mechanics theory, we focus on a subclass of equilibria referred to
as the energy-enstrophy equilibrium states. This allows to build phase
diagrams in a two parameter space, and to discuss the energy partition
between a large scale flow and small scale fluctuations when these parameters
are varied. We finally discuss the role of shocks that occur in the
actual shallow water dynamics, and present a geophysical application
to the Zapiola anticyclone.

\subsection{Energy-enstrophy equilibria for the quasi-geostrophic model. }

We consider the variational problem
\begin{equation}
S_{g,mf}\left(E,Z_{2}\right)=\max_{\rho_{g},\int\rho_{g}=1}\left\{ \mathcal{S}_{mf,g}\left[\rho_{g}\right]\ \left|\ \mathcal{E}_{mf,g}\left[\rho_{g}\right]=E_{mf},\:\mathcal{Z}_{g2}\left[\rho_{g}\right]=Z_{2},\ \mathcal{Z}_{g1}\left[\rho_{g}\right]=0\right.\right\} ,\label{eq:max_prob_qg_2-1}
\end{equation}
where the functionals $\mathcal{S}_{mf,g}$, $\mathcal{E}_{mf,g}$,
$\mathcal{Z}_{g1,2}$ are defined in Eqs. (\ref{eq:entropy_qg_mf}),
(\ref{eq:Eqg_def}), and (\ref{eq:PV_moments_qg}), respectively. 

The peculiarity of this variational problem is that only two potential
vorticity moments (the circulation and the enstrophy) have been retained
as a constraint, in addition to the energy. Such energy-enstrophy
equilibria are a subclass of statistical equilibria solutions of the
more general variational problem given in Eq. (\ref{eq:max_prob_qg_2}),
see e.g. \cite{bouchet2008,NasoChavanisDubrulle}. For a given global
distribution of potential vorticity, several limit cases on the energy
allow to simplify the computation of the solutions of %

the variational problem in Eq. (\ref{eq:max_prob_qg_2}) into the
computation of the simpler variational problem in Eq. (\ref{eq:max_prob_qg_2-1}).
For instance, assuming that bottom topography is non zero, and that
the global potential distribution is such that the mixed state $\overline{q}=cst$
exists, the solutions of (\ref{eq:max_prob_qg_2-1}) are the solutions
of the more general variational problem Eq. (\ref{eq:max_prob_qg_2})
when $E\rightarrow E_{mix}$. 

The set of all potential vorticity fields $\overline{q}$ corresponding
to solutions of the variational problem in Eq. (\ref{eq:max_prob_qg_2-1})
have been previously described by \cite{ChavanisSommeria:1996_JFM_Classification,venaille2009,venaille2011solvable,NasoChavanisDubrulle}
in the case of a bounded geometry, and phase diagrams were obtained
with energy $E$ and circulation $Z_{1}$ as external parameters.
The role of enstrophy $Z_{2}$ was not discussed. The main reason
is that for a given large scale flow characterized by $E_{1}$ and
$Z_{1}$, changing $Z_{2}$ would only imply changes in the small
scale fluctuations of potential vorticity levels, see e.g. \cite{NasoChavanisDubrulle},
and such small scale fluctuations do not contribute to the total energy.
In the present case, $Z_{2}$ will play an important role in determining
the energy partition between the large scale (vortical) flow and small
scale fluctuations due to the height and divergent velocity field.
For the sake of simplicity we consider vanishing circulations $Z_{1}=0$.
\\

\begin{figure}
\centering\includegraphics[width=0.9\columnwidth]{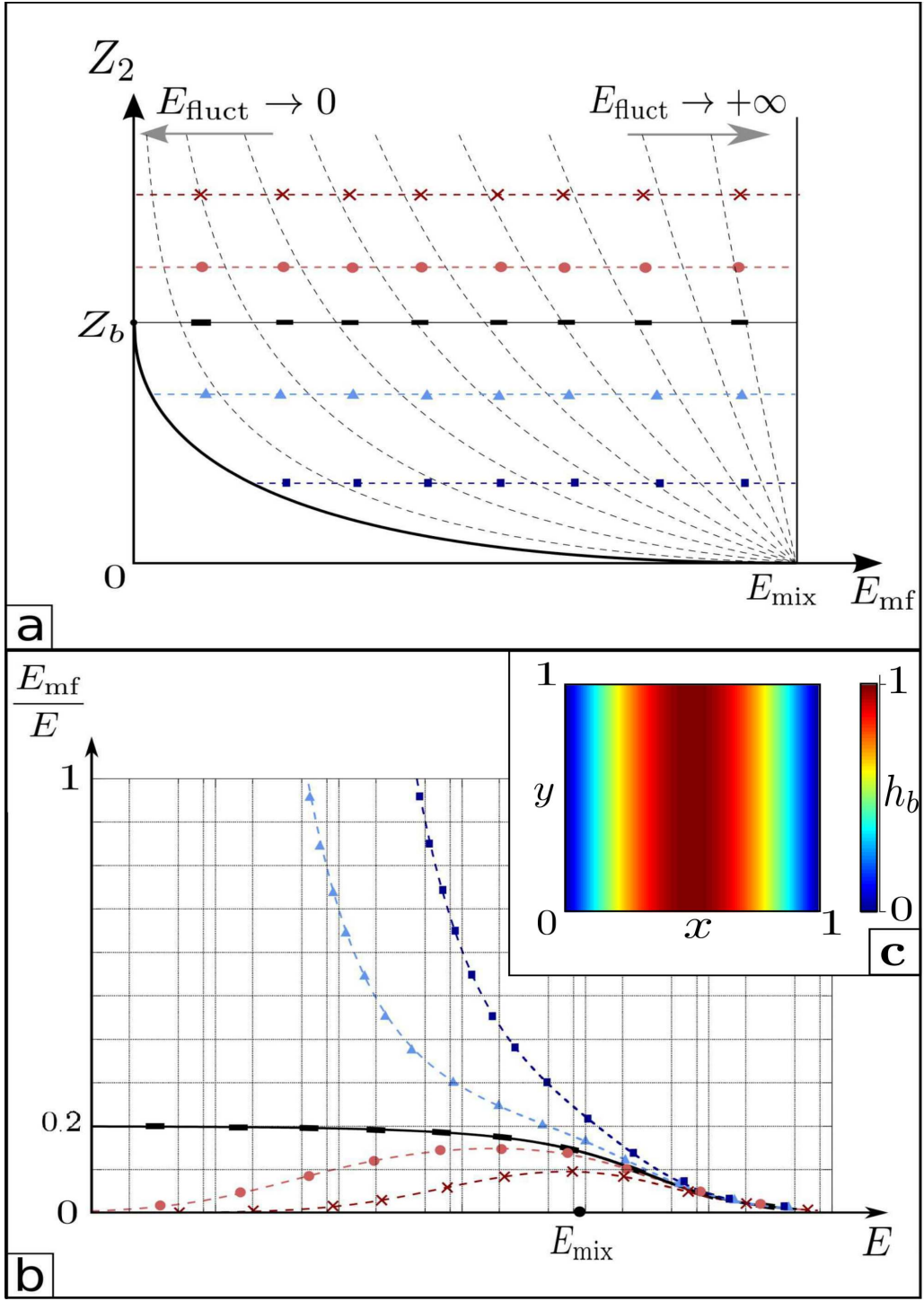}

\protect\caption{\label{fig:Phase_Diag_and_E_part} a) Phase diagram of the energy-enstrophy
ensemble in the plane $(Z_{2},E_{mf})$. $Z_{b}$ is the maximum reachable
value for the macroscopic enstrophy defined in Eq. (\ref{eq:Zb_def}).
$E_{mix}$ is the maximum reachable energy for the mean-flow defined
in Eq. (\ref{eq:Eb_def}). The dashed lines corresponds to isotherms,
on which the energy $E_{fluct}$ of the small scale fluctuations is
constant. The thick black line on the bottom left corner is a boundary
below which no equilibria exist. b) Ratio between the energy of the
large scale flow $E_{mf}$ over the total energy $E_{mf}/E$
as a function of the total energy $E=E_{mf}+E_{fluct}$. The
different curves correspond to different values of the initial microscopic
enstrophy $Z_{2}$ represented by horizontal marked lines on panel
a. The x-axis is on a logarithmic scale.   c) Colormap plot of the bottom topography $h_{b}$ used to compute the phase diagram in Fig. \ref{fig:Phase_Diag_and_E_part}-a-b and \ref{fig:Energy-and-enstrophy}. Here, $h_{b}=\sin (\pi x)$ is a single mode of the Laplacian operator. Thus the stream function $\psi$ and the potential vorticity field $q_{g}$ are simply proportional to $h_{b}$ for any values of the initial enstrophy $Z_{2}$ and the initial energy $E$. }
\end{figure}

The problem (\ref{eq:max_prob_qg_2-1}) is solved in Appendix \textbf{\ref{sec:Energy-Enstrophy-ensemble-detail}}
for positive temperature states, and we present here the main results.
A typical phase diagram is shown on Fig. \ref{fig:Phase_Diag_and_E_part}.
This figure is obtained by assuming that the bottom topography is
proportional to the first Laplacian eigenmode (see Fig. \ref{fig:Phase_Diag_and_E_part}-c) but it is explained
in Appendix \textbf{\ref{sec:Energy-Enstrophy-ensemble-detail}} that
this phase diagram is generic to any bottom topography.\footnote{For such a bottom topography, the topography, the stream function and the potential vorticity field are all proportional to each other for any initial condition for the enstrophy $Z_{2}$ and the energy $E$. That is why we do not show plots of the flows for different point of the phase diagram in Fig. \ref{fig:Phase_Diag_and_E_part}. We rather choose to consider the case of the Zapiola drift in subsection \ref{sub:Zapiola} to see the effect of different value for the initial energy.}

There are two important quantities related to the height field: the
maximum allowed energy $E_{mix}$ defined in Eq. (\ref{eq:Emix}),
and the available potential enstrophy
\begin{equation}
Z_{b}\equiv\int\mathrm{d}\mathbf{x}\ h_{b}^{2},\label{eq:Zb_def}
\end{equation}
which is the maximum reachable value for the macroscopic enstrophy
$\int\mathrm{d}\mathbf{x}\:\overline{q_{g}}^{2}$\textbf{ }, see Appendix
\textbf{\ref{sec:Energy-Enstrophy-ensemble-detail}}. Both $Z_{b}$
and $E_{mix}$ are a norm for the height field.\\

The phase diagram of the quasi-geostrophic energy-enstrophy ensemble
restricted to positive temperature states is presented in Fig. \ref{fig:Phase_Diag_and_E_part}-a.
As explained in subsection \ref{sub:The-quasi-geostrophic-limit},
the energy of the large scale flow $E_{mf}$ can not exceed the value
$E_{mix}$ defined in Eq. (\ref{eq:Eb_def}), due the restriction
of positive temperature states. Depending on the sign of $Z_{2}-Z_{b}$,
where the potential enstrophy $Z_{b}$ is defined in Eq. (\ref{eq:Zb_def}),
the system behaves differently: 
\begin{itemize}
\item When $Z_{2}>Z_{b}$ the minimum admissible large scale flow energy is zero. 
\item When $Z_{2}<Z_{b}$ , there exists a minimum reachable large scale flow energy $E_{min}(Z_{2})$
below which there is no equilibria. The curve $E_{min}$ increases
from $0$ to $E_{mix}$ when $Z_{2}$ decreases from $Z_{b}$ to $0$. 
\end{itemize}
The thick black line in Fig. \ref{fig:Phase_Diag_and_E_part}-a delimits
the domain of existence for the equilibria. The thin dashed black
curves represent the isotherm, i.e. the points of the diagram with
the same value of $E_{fluct}$. Note that there is no bifurcation
in this phase diagram; to each point $(E_{mf},Z_{2})$ corresponds
a single equilibrium state, whose expression is given explicitly in
Appendix \textbf{\ref{sec:Energy-Enstrophy-ensemble-detail}}. The
structure of an equilibrium large scale flow above a topographic bump
at low and high energy are presented in the last subsection. \\

The phase diagram in Fig. \ref{fig:Phase_Diag_and_E_part} allows
to discuss the energy partition between small scale fluctuations and 
large scale flow when the quasi-geostrophic flow is coupled to small
scale fluctuations of the height field and divergence field (through
the shallow water dynamics). The motivation behind the works of \cite{Warn86} and \cite{Weichman01} was the prediction of energy partition between large scale flow and small scale fluctuations. Here we provide for the first time an explicit expression for such energy partition.   

We have seen that in the weak height fluctuation limit, the temperature
of the large scale flow at equilibrium is given by the energy of the
small scale fluctuations of height and divergence fields. The parameters
are now the total energy $E$ and enstrophy $Z_{2}$. The mean flow
energy $E_{mf}$ and the fluctuation energy $E_{fluct}$ are found
by solving 
\begin{equation}
E=E_{mf}\left(E_{fluct},Z_{2}\right)+E_{fluct},\label{eq:E_E_fluct_Emf}
\end{equation}
which is easily done graphically using the diagram of Fig. \ref{fig:Phase_Diag_and_E_part}-a,
and performed numerically in practice, see Appendix \textbf{\ref{sec:Energy-Enstrophy-ensemble-detail}}
for more details. The ratio of the large scale energy $E_{mf}$ over
the total initial energy $E$ as a function of the total initial
energy $E$ is shown on Fig. \ref{fig:Phase_Diag_and_E_part}-b for
different values of initial enstrophy $Z_{2}$. According to computations
performed in Appendix \textbf{\ref{sec:Energy-Enstrophy-ensemble-detail},}
four cases for the energy partition in the low energy limit are distinguished
depending on the sign of $Z_{2}-Z_{b}$ and the scaling of $Z_{2}-Z_{b}$
with $E$: 
\begin{itemize}
\item When $Z_{2}<Z_{b}$, the ratio $E_{mf}/E$ tends to $1$ when $E$
tends to the minimal admissible energy $E_{min}(Z_{2})$. 
\item When $Z_{2}>Z_{b}$, the ratio $E_{mf}/E$ tends to $0$ when $E$
tends to zero. 
\item When $Z_{2}>Z_{b}$ with $Z_{2}-Z_{b}\sim E^{\alpha}$ with $\alpha>1/2$,
the ratio $E_{mf}/E$ tends to $1/5$ when $E$ tends to zero.
\item When $Z_{2}>Z_{b}$ with $Z_{2}-Z_{b}\sim E^{1/2}$, the ratio $E_{mf}/E$
tends to a finite value (depending on the proportionality coefficient
between $Z_{2}-Z_{b}$ and $E^{1/2}$ and the bottom topography).
\item When $Z_{2}>Z_{b}$ with $Z_{2}-Z_{b}\sim E^{\alpha}$ with $\alpha<1/2$,
the ratio $E_{mf}/E$ tends to $0$.
\end{itemize}
Whenever $Z_{2}\ge Z_{b}$, we found $E_{fluct}\sim E$. We see on
Fig. \ref{fig:Phase_Diag_and_E_part}-b that the ratio $E_{mf}/E$
converges to one when $E\rightarrow E_{min}(Z_{2}<Z_{b})$, that it
converges to zero for $Z_{2}>Z_{b}$, and that it converges to $1/5$
for $Z_{b}=Z_{2}$. Note also that $E_{mf}$ is bounded by $E_{mix}$
such that $E_{mf}/E$ tends to zero when $E$ tends to infinity. \\

\subsection{The effect of energy dissipation and enstrophy dissipation\label{sub:The-effect-of}}

The actual shallow water dynamics is known to be characterized by
shocks that prevent energy conservation. In addition, the presence
of viscosity, no matter how small it is, may lead to enstrophy dissipation,
and more generally would break the conservation of potential vorticity
moments. The aim of this subsection is to discuss qualitatively the
effect of these dissipative processes on the large scale flow, assuming
that the system evolves through a sequence a equilibrium states, which
is a natural hypothesis if there exists a separation of time scales.\\

\begin{figure}
\centering\includegraphics[width=0.9\columnwidth]{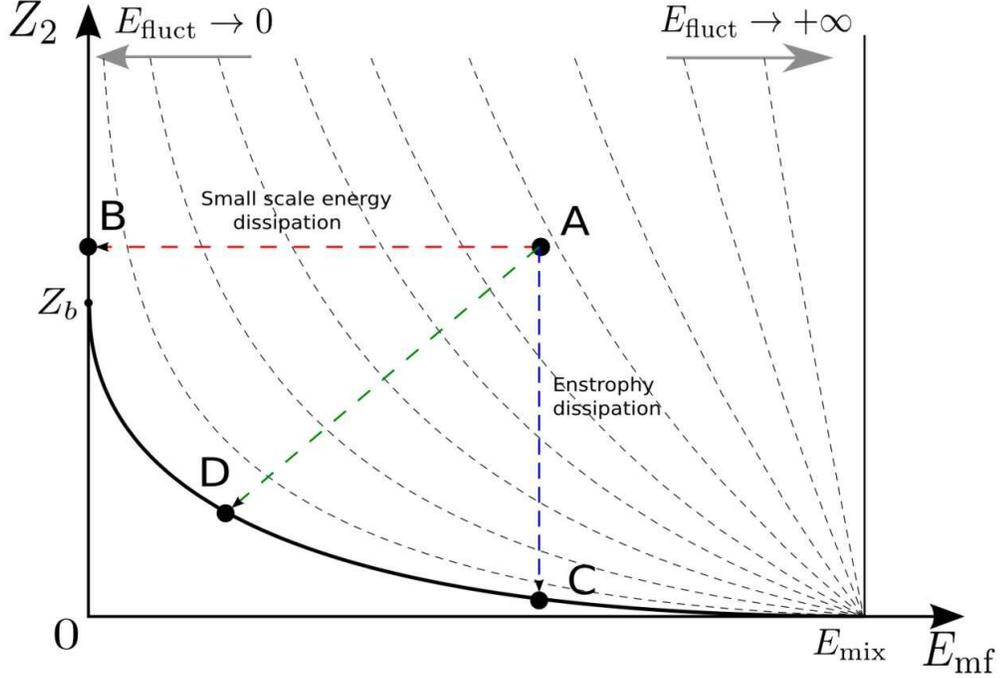}

\protect\caption{Phase diagram of Fig. \ref{fig:Phase_Diag_and_E_part}-a with hypothetical
trajectories of the shallow water system in the presence of dissipation.
Trajectory A-B: the dynamics dissipates small scale fluctuations of
height and divergence field only. Trajectory A-C: the dynamics dissipates
small scale fluctuations of potential vorticity. Trajectory A-D: the
dynamics dissipates small scale fluctuations of all the fields. \label{fig:Energy-and-enstrophy}}
\end{figure}

For the sake of simplicity, let us focus on the phase diagram obtained
in the energy-enstrophy ensemble and described in the previous subsection.
Let us consider that the system has reached at some time an arbitrary
equilibrium state denoted by $A$ in Fig. \ref{fig:Energy-and-enstrophy}.
\\

Let us first consider a case where energy is conserved, but enstrophy
is dissipated. Then the system will evolve form point $A$ to point
$C$ of Fig. \ref{fig:Energy-and-enstrophy}. In other words, the
small scale enstrophy $Z_{2}$ will be dissipated so that the enstrophy
of the system tends to the minimum admissible value of enstrophy $Z_{2\: min}(E_{mf})$.
Note that the curve $Z_{2\: min}(E_{mf})$ is nothing but the curve
of energy minima for a given $Z_{2}<Z_{b}$. We saw previously that the
mean-flow energy $E_{mf}$ dominates the fluctuation energy $E_{fluc}$
on this line. We conclude that enstrophy dissipation alone drives
the system towards a large scale flow without small scale fluctuations.
This large scale flow vanished when topography vanished (since $E_{mix}$
and $Z_{b}$ would also vanish). \\

Let us now consider that the enstrophy is not dissipated and look
the effect of energy dissipation. Let us first explain why one may
expect that even weak dissipation can lead to a significant decrease
of the large scale flow energy $E_{mf}$ when $E_{fluct}\ne0$ (in
the absence of small scale fluctuations, $E_{mf}$ would not decrease
significantly with weak dissipation). At equilibrium, the energy of
the small scale fluctuations $E_{fluct}$ should be equipartitioned
among all the modes of the height field and divergent field. Since
there is an infinite number of such modes, this means a loss of energy
through dissipative process, no matter how small they are. Since $E_{mf}$
decreases with $E_{fluct}$, dissipating the energy $E_{fluct}$ amounts
to diminish the energy $E_{mf}$. For a given enstrophy $Z_{2}>Z_{b}$,
as in the case of point $A$ Fig. (\ref{fig:Energy-and-enstrophy}),
this dissipative process drives the system towards a zero energy state
$B$. For a given enstrophy $Z_{2}<Z_{b},$ this process would drive
the system towards the line of minimal energy.\\

We expect to see both enstrophy and energy dissipation working together,
so the trajectory of the system in phase diagram will be somewhere
between the trajectory $A\rightarrow B$ and the trajectory $A\rightarrow C$,
e.g. the trajectory e.g. $A\rightarrow D$. We conclude that in the
presence of topography and small scale dissipation, we eventually reach a geostrophic regime 
at large time, provided that the
initial enstrophy is sufficiently low (otherwise the final state contains
no large scale flow). In addition, each time the system is perturbed
by adding a little amount of energy without changing the enstrophy,
it drives the system a bit more towards the mixed state $\left(E_{mf}=E_{mix},Z_{2}=0\right)$.

\subsection{Flow structure of the equilibrium states: application to the Zapiola
Anticyclone. \label{sub:Zapiola}}

So far we have discussed energy partition for shallow water equilibria
in the quasi-geostrophic limit. Here we focus on the structure of
the large scale flow associated with these equilibrium states. We
consider for that purpose an oceanic application to the Zapiola anticyclone.

The Zapiola anticyclone is a strong anticyclonic recirculation taking
place in the Argentine basin above a sedimentary bump known as the
Zapiola drift \cite{weatherly1993,saunders1995,mirandabarnieretal1999}.
The anticyclone is characterized by a mass transport as large as any
other major oceanic current such as the Gulf Stream. It is a quasi-barotropic
(depth independent) flow, with typical velocities of the order of
$0.1$ $m.s^{-1}$ , and a lateral extension of the order of $800$
$km$. 

It is known from the earlier statistical mechanics studies that positive
temperature states in the energy enstrophy ensemble of one layer models
lead to anticyclonic circulations above topography anomalies, see
e.g. \cite{Salmon_1998_Book} and references therein. A generalization
of this result to the continuously stratified case with application
to the Zapiola anticyclone is given in \cite{venaille2012bottom}.

\begin{figure}
\centering\includegraphics[width=0.75\textwidth]{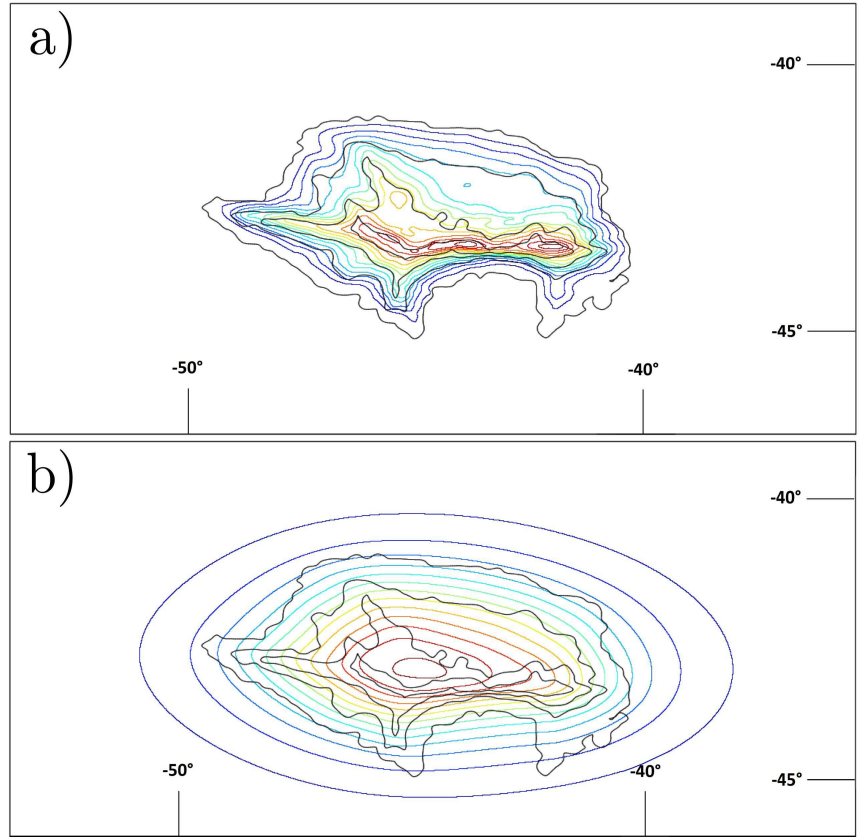}

\protect\caption{Plots of the streamfunction isolines (colored lines) from higher values
(red) to lower values (blue) over the Zapiola Drift topography iso-contours
(black lines) for a small mean-flow energy (a)) and for a high mean-flow
energy (b)). \label{fig:zapiola}}
\end{figure}

We have shown in this paper that quasi-geostrophic equilibria characterized
by positive temperature states are also shallow water equilibria.
The Zapiola anticyclone can therefore be interpreted as an equilibrium
state of the shallow water model. Let us now show the qualitative
difference between low energy states (when $Z_{2}>Z_{b}$ and $E_{mf}\rightarrow0$)
and high energy states (when $E_{mf}\rightarrow E_{mix}$), which
are shown on Fig. \ref{fig:zapiola}-a and Fig \ref{fig:zapiola}-b,
respectively. 

In both cases the bottom topography is the same, and its isolines
are visualized with thin black lines. Bottom topography has been obtained
from data available online and described in \cite{ZapiolaTopography}.
We isolated the largest close contour defining the Zapiola drift (the
sedimentary bump above which the recirculation takes place), and considered
the actual bottom topography inside this contour, and a flat bottom
outside this contour. It allows to focus on the interesting flow structure
occurring above the Zapiola drift. %
\begin{comment}
Thanks to the fast Fourier transform, we inverse the $h_{b}-\psi$
relation Eq. (\ref{eq:qg_pv_def}) and computed some isolines of the
streamfunction for a small and a large value of the mean-flow energy.
\end{comment}

Energy-enstrophy equilibrium states are then computed using Eq. (\ref{eq:Helmholtz_EnergyEnstrophy})
in Appendix \ref{sec:Energy-Enstrophy-ensemble-detail}. We have seen
that low energy limit corresponds to $\beta\rightarrow+\infty$ with
$Z_{2}>Z_{b}$. Taking this limit in Eq. (\ref{eq:Helmholtz_EnergyEnstrophy})
yields $\psi_{mf}=\left(Z_{2}-Z_{b}\right)h_{b}/\beta$ : the equilibrium
state is a Fofonoff flow \cite{Fofonoff:1954_steady_flow_frictionless},
meaning that streamlines are proportional to the isolines of topography,
just as in Fig. \ref{fig:Energy-and-enstrophy}-a.

The maximum energy state corresponds to the case $\beta=0$, which
corresponds to the mixed state defined in Eq. (\ref{eq:psi_mf_mix}).
The streamfunction is $\psi_{mix}=\left(R^{-2}-\Delta\right)^{-1}h_{b}$.
The operator $\left(R^{-2}-\Delta\right)^{-1}$ with $R\sim1$ is
expected to smooth out the small scale topography features, just as
in Fig. \ref{fig:zapiola}-a.

\section{Conclusion}

We have presented in this paper analytical computations of equilibrium
states for the shallow water system, giving thus predictions for the
energy partition into small scale fluctuations and large scale
flow. Our results rely on the definition of a discrete version
of the shallow water model. Once our semi-Lagrangian discrete model was introduced, the whole machinery of
equilibrium statistical mechanics could be applied. 

We found that equilibrium states of the shallow water system are associated
with the concomitant existence of a large scale flow which is a stationary
state of the shallow water dynamics, superimposed with small scale
fluctuations that may contain in some cases a substantial part of
the total energy. The novelty of our work was to explicitly compute
the contribution of these small scale fluctuations, and to decipher
the physical consequences of the presence of these fluctuations.\\

In particular, we found that the presence of small scale fluctuations
implies a positive temperature for the equilibrium state. This explains
a previous result by Warn \cite{Warn86}, who showed that equilibrium
states in a weak flow limit admit no large scale flow when there is
no bottom topography and no lateral boundaries. We have generalized
these results, by showing that a large scale vortical flow exists
at equilibrium when there is both rotation and bottom topography,
or when there is a non-zero circulation.

In the limit of weak height fluctuations, we found equipartition of
the small scale kinetic and potential energy. We also obtained an
interesting physical picture of the equilibrium state, which may be
interpreted in that limit as two subsystems in thermal contact. One
subsystem is the ``large scale'' potential vortical flow whose entropy is closely related to the one introduced heuristically by Chavanis and Sommeria (see \cite{Chavanis02}). Our work provides therefore a microscopic justification of their entropy with a complete statistical mechanics derivation and a generalization of this results by including the presence of small scales fluctuations of the height and divergence fields. We note however that it is wrong to interpret the "large scale" potential vortical flow entropy as the entropy of the system, it is only one part of it. The other
subsystem contains the field of height fluctuations associated with
small scale potential energy, and the field of velocity fluctuations,
associated with small scale kinetic energy. These velocity fluctuations
are due solely to the divergent part of the velocity field. Warn
obtained a similar result in the weak flow limit, by projecting the
non-linear dynamics into eigenmodes of the linearized dynamics \cite{Warn86}.
He found an energy partition into a vortical flow on the one side,
and on inertia-gravity waves on the other side, with a weak coupling
between both subsystems. Hence we may interpret local height and divergent
velocity fields fluctuations appearing in our model as inertia gravity
waves.

We studied the quasi-geostrophic limit for the large scale flow, taking
into account the presence of small scale fluctuations. We recovered
in this limit the variational problem of the Miller-Robert-Sommeria
theory, with the additional constraint that the temperature is positive.
We obtained phase diagrams in the particular case of energy-enstrophy
equilibria, with explicit prediction for the ratio of energy between
small scale fluctuations and a large scale quasi-geostrophic flow.
This allowed to discuss the qualitative effect of small scale dissipation
and shocks on the temporal evolution of the system. The main result
is that such dissipative processes drive the system towards a minimum
energy state (depending on the enstrophy), which may be non-zero.

In view of those results, the semi-Lagrangian discrete model seems to be a good discretization of the shallow water system. It has the key desired properties to take into account of the conservation of fluid
particle volume, while working in an Eulerian framework. This gives a clear framework, which allows for a rigorous derivation once the discretization is assumed. Moreover we stress that it leads to equilibrium states that are stationary states of the shallow water model, by contrast with other choices of discretization. What is not completely satisfactory however is that there is a degree of arbitrariness in the definition
of the discrete model. We have tried to have the model consistent with the geometric constraints related to the Liouville theorem, even though the link between the model and
the Liouville theorem is clearly not rigorous. There is clearly room for improvement, but we are afraid we are faced with extremely tricky mathematical problems.  Nevertheless we guess that the invariant measure of the discrete model converge to the actual invariant measure of the shallow water system in the continuous limit, but proving this is beyond the scope of this paper.\\

The statistical mechanics prediction of a vanishing large scale flow
in the absence of boundary and bottom topography seems to contradict
some numerical results performed in such configurations, in which
case long lived vortex were reported, see e.g. \cite{FargeSadourny86}.
This issue is related to the estimation of a time scale for the convergence
towards equilibrium. Indeed, if the coupling between the large scale
flow (potential vortical modes) and the small scale fluctuations (inertia-gravity
waves) is weak, and if one starts from an initially balanced and unstable
large scale flow, then this large scale flow may self-organize spontaneously
on a short time scale into an equilibrium state of the quasi-geostrophic
subsystem. This justifies the physical interest of the variational principle for the large scale flow introduced by \cite{Chavanis02} without small scale fluctuation. According to our statistical mechanics predictions, the
energy of this large scale flow should leak into small scale fluctuations,
but this process may be slow if coupling between both subsystems is
weak. This difficult issue was already raised by Warn \cite{Warn86}, and the interaction between geostrophic motion and inertia-gravity waves remains an  active field of research \cite{vanneste_review}. In particular, very interesting models of interactions between near-inertial waves and geostrophic motion have been proposed  \cite{youngjelloul,straub2009,xievanneste2014}. In the context of statistical mechanics approaches, it has been proposed to compute equilibrium states with frozen degrees of freedom \cite{salmonTherm} or restricted partition functions in order to avoid the presence of inertia-gravity waves \cite{herbertBous}.\\

One of the main interests of the present study is the prediction of
a concomitant large scale energy transfer associated with a large scale
potential vortical flow and a small scale transfer of energy that
is lost into small scale fluctuations interpreted as inertia-gravity
waves. In that respect, the shallow water model lies between three
dimensional and two-dimensional turbulence. There are other models
for which the energy may be partitioned into a large scale flow and
small scale fluctuations. There are, for instance, some strong analogies with the case of three-dimensional
axisymmetric Euler equations, that is also intermediate between 2D and 3D flows. A statistical mechanics theory has been recently derived for this system by Thalabard et al \cite{thalabard2014}. This statistical theory can also be understood as two subsystems in contact, one of them being the fluctuations, and thus leading to positive temperatures \cite{thalabard2014}. From this key observation, \cite{thalabard2014} concluded that the temperature of the system, that measures the variance of the fluctuations, is positive and uniform in space, as previously stressed in \cite{NasoAxisSym}.
More generally, and beyond those simplified flow models that allow
for analytical treatment, it is common to observe in bounded laboratory
experiment the emergence of large scale structures at the domain scale
superimposed with small scale energy fluctuations, see e.g. \cite{Thalabard_ferroturb_NJP15}.
It is not clear whether equilibrium theory may be relevant to describe
such problems, but we believe that it is at least a useful first step
to address these questions.

\appendix

\section{Invariant measure and formal Liouville theorem \label{sec:Invariant-measure-and}}

\subsection{Formal Liouville theorem for the triplet of fields $(h,hu,hv)$}

The existence of a formal Liouville theorem for the shallow water
dynamics is shown in this appendix. The shallow water system is fully
described by the triplet of fields $\left(h,hu,hv\right)$. We consider
a measure written formally as 
\[
\mathrm{d}\mu=C\mathcal{D}\left[h\right]\mathcal{D}\left[hu\right]\mathcal{D}\left[hv\right],
\]
 with uniform density in $\left(h,hu,hv\right)$-space ($C$ is a
constant). The average of any functional $A$ over this measure is 

\begin{equation}
\forall\mathcal{A}\left[h,uh,vh\right],\ \left\langle \mathcal{A}\right\rangle _{\mu}=\int\mathrm{d}\mu\ \mathcal{A}.\label{eq:avarage_def}
\end{equation}
The term $\int\mathcal{D}\left[h\right]\mathcal{D}\left[hu\right]\mathcal{D}\left[hv\right]$
means that the integral is formally performed over each possible triplet
of fields $\left(h,\ hu,\ hv\right)$. The measure is said to be invariant
if 
\begin{equation}
\forall\mathcal{A},\quad\frac{\mathrm{d}}{\mathrm{d}t}\left\langle \mathcal{A}\right\rangle _{\mu}=0\label{eq:avarage_invarience}
\end{equation}
This yields the condition
\begin{equation}
\forall\mathcal{A},\quad\int\mathcal{D}\left[h\right]\mathcal{D}\left[hu\right]\mathcal{D}\left[hv\right]\int d\mathbf{x}\ \frac{\delta\mathcal{A}}{\delta h}\partial_{t}h+\frac{\delta\mathcal{A}}{\delta\left(hu\right)}\partial_{t}\left(hu\right)+\frac{\delta\mathcal{A}}{\delta\left(hv\right)}\partial_{t}\left(hu\right)=0.\label{eq:avarage_invariance_cond}
\end{equation}
An integration by parts yields

\begin{equation}
\forall \mathcal{A}	,\quad\int\mathcal{D}\left[h\right]\mathcal{D}\left[hu\right]\mathcal{D}\left[hv\right]\mathcal{A}\int d\mathbf{x}\ \left(\frac{\delta\partial_{t}h}{\delta h}+\frac{\delta\partial_{t}\left(hu\right)}{\delta\left(hu\right)}+\frac{\delta\partial_{t}\left(hv\right)}{\delta\left(hv\right)}\right)=0.\label{eq:Avarage_inv_IPP}
\end{equation}
We say that the equation follows a formal Liouville theorem if we
formally have 
\begin{equation}
\int d\mathbf{x}\ \left(\frac{\delta\partial_{t}h}{\delta h}+\frac{\delta\partial_{t}\left(hu\right)}{\delta\left(hu\right)}+\frac{\delta\partial_{t}\left(hv\right)}{\delta\left(hv\right)}\right)=0,\label{eq:formal_lioville_def}
\end{equation}
 which ensures that the measure $\mathrm{d}\mu$ is invariant. 

The shallow water equations (\ref{eq:SW_momentum_general}) and (\ref{eq:mass_continuity})
can be written on the form 
\begin{equation}
\partial_{t}\left(hu\right)=-\partial_{x}\left(hu^{2}+\frac{1}{2}gh^{2}\right)-\partial_{y}\left(huv\right)+fhv,\label{eq:sw_equation_momentum-5}
\end{equation}
\begin{equation}
\partial_{t}\left(hv\right)=-\partial_{y}\left(hv^{2}+\frac{1}{2}gh^{2}\right)-\partial_{x}\left(huv\right)-fhu,\label{eq:sw_equation_momentum-1-3}
\end{equation}
\begin{equation}
\partial_{t}h+\partial_{x}\left(hu\right)+\partial_{y}\left(hv\right)=0.\label{eq:mass_continuity-1-1}
\end{equation}
We see that 
\begin{equation}
\frac{\delta\partial_{t}h}{\delta h}+\frac{\delta\partial_{t}\left(hu\right)}{\delta\left(hu\right)}+\frac{\delta\partial_{t}\left(hv\right)}{\delta\left(hv\right)}=-\boldsymbol{\nabla}\cdot\left(\frac{\delta\left(h\mathbf{u}\right)}{\delta h}+\frac{\delta\left( hu\mathbf{u}\right)}{\delta\left(hu\right)}+\frac{\delta\left( hv\mathbf{u}\right)}{\delta\left(hv\right)}\right).\label{eq:canonical_mesure}
\end{equation}
As the divergence operator is a linear operator, it commutes with the functional derivatives. This allows to conclude that the measure $\mu$ is invariant. This
shows formally the existence of a Liouville theorem for the fields
$\left(h,uh,vh\right)$.

\subsection{Change of variables from $(h,hu,hv)$ to $(h,q,\mu)$ \label{sec:Change-of-variables}}

The microcanonical measure can formally be written 

\begin{equation}
\mathrm{d}\mu_{h,hu,hv}=\mathcal{D}\left[h\right]\mathcal{D}\left[hu\right]\mathcal{D}\left[hv\right]\delta\left(\mathcal{E}-E\right)\prod_{k=0}^{+\infty}\delta\left(\mathcal{Z}_{k}-Z_{k}\right),\label{eq:formal_measure_hhuhv}
\end{equation}
The constraints are more easily expressed in terms of the variables
\begin{equation}
h,\quad q=\frac{-\partial_{y}u+\partial_{x}v+f}{h},\quad\mu=\Delta^{-1/2}\left(\partial_{x}u+\partial_{y}v\right).\label{eq:good_var}
\end{equation}
It will thus be more convenient to use these fields as independent
variables. We call $J\left[(h,hu,hv)/(h,q,\mu)\right]$ the Jacobian
of the transformation. We proceed step by step to compute this Jacobian.
The change of variables $(h,hu,hv)\rightarrow(h,u,v)$ involves a
upper-diagonal Jacobian matrix at each point $\mathbf{r}$:
\begin{equation}
J\left[\frac{(h,hu,hv)}{(h,u,v)}\right]=\left(\begin{array}{ccc}
1 & u & v\\
0 & h & 0\\
0 & 0 & h
\end{array}\right)\label{eq:jacobian_hhuhv_huv}
\end{equation}
which implies $\det\left(J\left[(h,hu,hv)/(h,u,v)\right]\right)=h^{2}$
and 
\begin{equation}
\mathcal{D}\left[h\right]\mathcal{D}\left[hu\right]\mathcal{D}\left[hv\right]=h^{2}\mathcal{D}\left[h\right]\mathcal{D}\left[u\right]\mathcal{D}\left[v\right].\label{eq:det_jac1}
\end{equation}
The change of variable $\left(h,u,v\right)\rightarrow(h,\omega,\mu)$
involves linear operators that do not depend on space coordinates,
thus the determinant of the Jacobian of the transformation is an unimportant
constant: 
\begin{equation}
\mathcal{D}\left[h\right]\mathcal{D}\left[u\right]\mathcal{D}\left[v\right]=C\mathcal{D}\left[h\right]\mathcal{D}\left[\omega\right]\mathcal{D}\left[\mu\right].\label{eq:change_var_lin}
\end{equation}
Using $\omega=qh-1$, the change of variable $(h,\omega,\mu)\rightarrow(h,q,\mu)$
involves an upper diagonal Jacobian matrix at each point $\mathbf{r}$:

\begin{equation}
J\left[\frac{(h,\omega,\mu)}{(h,q,\mu)}\right]=\left(\begin{array}{ccc}
1 & q & 0\\
0 & h & 0\\
0 & 0 & 1
\end{array}\right),\label{eq:jacobian_hommu_hqmu}
\end{equation}
with a determinant $\det\left(J\left[(h,\omega,\mu)/(q,h,\mu)\right]\right)=h.$
Finally, the Jacobian of the transformation is $J\left[(hu,hv,h)/(q,h,\mu)\right]=h^{3}$
and the microcanonical measure can formally be written 
\begin{equation}
\mathrm{d}\mu_{h,q,\mu}=Ch^{3}\mathcal{D}\left[h\right]\mathcal{D}\left[q\right]\mathcal{D}\left[\mu\right]\delta\left(\mathcal{E}-E\right)\prod_{k=0}^{+\infty}\delta\left(\mathcal{Z}_{k}-Z_{k}\right).\label{eq:microcanonical_measure_good}
\end{equation}
We note the presence of the pre-factor $h^{3}$ which gives the weight
of each microscopic configuration in the $\left(h,q,\mu\right)$-space.

\section{Relevance of the constraints for the discrete model\label{sec:Energy-partition}}

In this appendix, we explain how the dynamical invariants of the shallow
water model, given in Eqs. (\ref{eq:def_energy}) and (\ref{eq:casimirs_definition_moments})
respectively, are related to the constraints of the microcanonical
ensemble for the discrete model, given in Eqs. (\ref{eq:Total_moments_funct})
and (\ref{eq:total_energy_func_as_a_sum}), respectively.\\

\subsection{Areal coarse-graining for continuous fields }

Let us consider a field $g\left(\mathbf{x}\right)$ on the domain
$\mathcal{D}$ where the flow takes place, and let us consider the
uniform grid introduced in subsection \ref{sub:Definition-of-the}.
We define the local areal coarse-graining of the continuous field
$g\left(\mathbf{x}\right)$ over a site $\left(i,j\right)$ as 
\begin{equation}
\overline{g}_{ij}=N\int_{\mbox{site }\left(i,j\right)}\mathrm{d}\mathbf{x}\: g\left(\mathbf{x}\right),\label{eq:Local-Av}
\end{equation}
where $1/N$ is the area of the site $\left(i,j\right)$ and where
$\int_{\mbox{site }\left(i,j\right)}$ means that we integrate over
the site $\left(i,j\right)$ only. With an abuse of notation, we use
here the same notation $\overline{g}_{ij}$ as in Eq. (\ref{eq:CG_def}),
since the coarse-graining operator defined in Eq. (\ref{eq:Local-Av})
generalizes to the continuous case the areal coarse-graining operator
 defined in Eq. (\ref{eq:CG_def}) for the discrete microscopic
model, taking into account the fact that for a fluid particle ``n''
of area $\mathrm{d}\mathbf{x}_{n}$ and height $h_{n}$, we get $N\mathrm{d}\mathbf{x}_{n}=1/(Mh_{n})$. 

We denote $\overline{g}$ the continuous limit (large $N$) of $\overline{g}_{ij}$
. Integrating a continuous field $g$ amounts to perform the integration
over its local average field $\overline{g}$: 
\begin{equation}
\int\mathrm{d}\mathbf{x}\: g\left(\mathbf{x}\right)=\int\mathrm{d}\mathbf{x}\:\overline{g}\left(\mathbf{x}\right).\label{eq:IntoIn}
\end{equation}

\subsection{Potential vorticity moments }

Using (\ref{eq:IntoIn}), the potential vorticity moments in Eq. (\ref{eq:casimirs_definition_moments})
simply leads to 
\begin{equation}
\mathcal{Z}_{k}=\int\mathrm{d}\mathbf{x}\:\overline{hq^{k}}.\label{eq:PV_discr_final}
\end{equation}
Now that the potential vorticity moments are expressed in terms of
the areal coarse-graining of moments of $h$ and $q$, it can directly
be expressed in terms of the probability density field $\rho\left(\sigma_{h},\sigma_{q},\sigma_{\mu}\right)$
through Eq. (\ref{eq:continuous_CG}), and we recover the expression
of the constraint given in Eq. (\ref{eq:Total_moments_funct}), whose
discrete representation is given in Eq. (\ref{eq:PV_Dist_Discr_def}).%
\begin{comment}
This was one of the reason we choose to consider the independent variables
$\left(h,q,\mu\right)$. It is not straightforward to apply the same
procedure to the energy, and this will require some approximations. 
\end{comment}

\subsection{Energy }

Using (\ref{eq:IntoIn}), recalling that we restrict ourself to bottom
topographies such that $\overline{h}_{b}=h_{b}$, the total energy
of the shallow water model defined in Eq. (\ref{eq:def_energy}) can
be decomposed into a mean flow kinetic energy defined in Eq. (\ref{eq:mean_field_energy_funct}),
a potential energy term due to local height fluctuations and defined
in Eq. (\ref{eq:Potential_energy_fluct_funct}), a fluctuating kinetic
energy term 
\begin{equation}
\mathcal{E}_{c,fluct}\equiv\mathcal{E}-\mathcal{E}_{mf}-\mathcal{E}_{\delta h}
\end{equation}
 
\begin{equation}
\mathcal{E}_{c,fluct}\equiv\frac{1}{2}\int\mathrm{d}\mathbf{x}\:\left(h\mathbf{u}^{2}-\overline{h}\mathbf{u}_{mf}^{2}\right),\label{eq:Ec_Ep_dens}
\end{equation}
where the velocity fields $\mathbf{u}$ and $\mathbf{u}_{mf}$ are
computed from the triplet $\left(h,q,\mu\right)$ and from the triplet
$\left(\overline{h},\overline{hq},\overline{h\mu}\right)$, respectively
through 
\begin{equation}
\mathbf{u}=\nabla^{\bot}\psi+\nabla\phi,\quad\left(hq-f\right)=\Delta\psi,\quad\mu=\Delta^{1/2}\phi,\label{eq:u_helm}
\end{equation}
and through 

\begin{equation}
\mathbf{u}_{mf}=\nabla^{\bot}\psi_{mf}+\nabla\phi_{mf},\quad\left(\overline{hq}-f\right)=\Delta\psi_{mf},\quad\frac{\overline{h\mu}}{\overline{h}}=\Delta^{1/2}\phi_{mf}.\label{eq:umf_triplet}
\end{equation}
We want to discuss the relation between decomposition of the energy
for the discrete model, Eq. (\ref{eq:total_energy_func_as_a_sum})
and the decomposition for the actual total energy defined in Eq. (\ref{eq:def_energy}).
Our construction is relevant if these two decomposition coincide in
the continuous limit, or equivalently if $\mathcal{E}_{c,fluct}$ is
equal to $\mathcal{E}_{\delta\mu}$ (\ref{eq:kinetic_energy_fluct_funct})
in the continuous limit. In the following we show that this is the
case if some cross correlations are actually negligible. More precisely,
we assume that
\begin{enumerate}
\item For any positive integers $k,l,m$ , the coarse-grained fields $\overline{h^{k}q^{l}\mu^{m}}(\mathbf{x})$
defined through the coarse graining procedure in Eq. (\ref{eq:IntoIn})
exist. In the framework of our microscopic model introduced in subsection
\ref{sub:A-discretized-model}, this hypothesis is automatically satisfied
by assuming that the cut-off $\mu_{min},\mu_{max},q_{min},q_{max},h_{max}$
scales as $N^{\alpha}$ with $\alpha<1$. Other fields may be characterized
by local extreme values such that the limit defined in Eq. (\ref{eq:IntoIn})
does not converge. For instance, we will see that the actual divergence
$\zeta=\Delta\phi$ of the equilibrium state is not bounded, i.e. that
$\overline{\zeta}$ would have no meaning.
\item The fields $h,q,\mu$ are decorrelated (in particular, $\overline{h^{k}q^{l}\mu^{m}}=\overline{h^{k}}\overline{q^{l}}\overline{\mu^{m}}$).
This point will be shown to be self-consistent when computing the
equilibrium state.
\item The coarse-grained divergent velocity field is equal to the mean-flow
velocity field $\overline{\nabla\phi}=\nabla\phi_{mf}$. 
\item $\overline{\Delta\left[\left(\phi-\phi_{mf}\right)^{2}\right]}=0$.
While $\nabla(\phi-\phi_{mf})$ is a random vector field characterized
by wild local fluctuation, this hypothesis amounts to assume that
those fluctuations have no preferential direction.%
\begin{comment}
$\lim_{N\rightarrow+\infty}\ointop_{\partial S_{ij}}\mathrm{d}l\ \mathbf{n}\cdot\left[\nabla\left(\phi-\phi_{mf}\right)^{2}\right]=0$
, with $\partial S_{ij}$ the boundary of the surface covered by the
grid site $(i,j)$. 
\end{comment}

\end{enumerate}
We believe that these four assumptions would be satisfied by a typical
triplet of fields ($h,q,\mu)$ picked at random among all the possible
states satisfying the constraints of the dynamics. By typical, we
mean that an overwhelming number of fields would share these properties.%
\begin{comment}
A few remarks. This appendix clarifies why we have chosen the triplet
$(h,q,\mu)$ rather than the triplet $(h,q,\zeta)$ to describe the
shallow water system. While the field $\mu$ has fluctuations controlled
by the energy, which allow to apply the coarse-graining procedure
to it, the field $\zeta$
\end{comment}

We then prove that these four assumptions are sufficient to prove
that $\mathcal{E}_{c,fluct}$ is equal to $\mathcal{E}_{\delta\mu}$
(\ref{eq:kinetic_energy_fluct_funct}) in the continuous limit. According
to the assumption 1, $\overline{\omega}=\overline{hq}-1$ is well
defined. Classical arguments show that the streamfunction of the coarse-grained
vorticity field $\overline{\omega}$ is equal to the streamfunction
of the vorticity field, i.e. that%
\begin{comment}
where the operators $\nabla^{\bot}\Delta^{-1}$ and $\nabla\Delta^{-1/2}$
are expressed as a convolution with their associated kernels $\mathbf{G}^{\omega}$
and $\mathbf{G}^{\mu}$, respectively:
\begin{equation}
\nabla^{\bot}\psi\left(\mathbf{x}\right)=\int\mathrm{d}\mathbf{x}'\;\mathbf{G}^{\omega}\left(\mathbf{x}-\mathbf{x}'\right)\omega\left(\mathbf{x}'\right),\hfill\nabla\phi\left(\mathbf{x}\right)=\int\mathrm{d}\mathbf{x}'\;\mathbf{G}^{\mu}\left(\mathbf{x}-\mathbf{x}'\right)\mu\left(\mathbf{x}'\right).\label{eq:Green_Deff}
\end{equation}
\end{comment}
\begin{comment}
The same hypothesis can not be made for $\mathbf{G}^{\mu}$: in the
Fourier domain, the Green's functions $\mathbf{G}^{\omega}$ and $\mathbf{G}^{\mu}$
don't have the same behavior: $\left|\widehat{\mathbf{G}^{\omega}}\left(\mathbf{k}\right)\right|\sim1/\left|\mathbf{k}\right|$
and $\left|\widehat{\mathbf{G}^{\mu}}\left(\mathbf{k}\right)\right|\propto1$.
The kernel $\mathbf{G}^{\mu}$ is not as smooth as $\mathbf{G}^{\omega}$.
We make the following decomposition for the divergent part of the
vorticity : $\nabla\phi=\nabla\phi_{mf}+\nabla\left(\phi-\phi_{mf}\right)$.
\end{comment}
 $\nabla\psi_{mf}=\nabla\psi,$ see e.g. \cite{SommeriaRobert:1991_JFM_meca_Stat,Michel_Robert_LargeDeviations1994CMaPh.159..195M}.
Qualitatively, this is due to the fact that inverting the Laplacian
operator smooth out local fluctuations of the vorticity%
\footnote{The divergent part of the velocity field can not be treated in the
same way. Indeed, the operator $\Delta^{-1/2}$ is less smooth than
the operator $\Delta^{-1}$, and one can not derive $\phi=\phi_{mf}$
by inverting $\mu=\Delta^{1/2}\phi$. One may want to use $\zeta=\Delta\phi$,
but the result would be the same since $\zeta$ is not bounded (the
field $\mu$ is characterized by fluctuations which are controlled
by the kinetic energy, and hence by the total energy, but this is
not the case for $\zeta$). %
}. This yields%
\begin{comment}
Injecting Eq. (\ref{eq:mean_field_omega}) in Eq. (\ref{eq:u_helm})
and using the expression of the mean-field velocity given in Eq (\ref{eq:u_mf_continuous})
yields
\end{comment}
 
\begin{equation}
\mathbf{u}=\mathbf{u}_{mf}+\nabla\left(\phi-\phi_{mf}\right).\label{eq:u_mf_plus_divergences}
\end{equation}
Injecting this expression in the kinetic energy density expression
Eq. (\ref{eq:Ec_Ep_dens}), using that $h$ and $\mu$ are not correlated
(assumption 2), and using $\overline{\nabla\phi}=\nabla\phi_{mf}$
(assumption 3) yields 
\begin{equation}
\mathcal{E}_{c,fluct}=\frac{1}{2}\int\mathrm{d}\mathbf{x}\:\overline{h}\overline{\left(\nabla\left(\phi-\phi_{mf}\right)\right)^{2}}.\label{eq:Encours_app}
\end{equation}
Let us now remember the definition of the coarse-graining operator
in Eq. (\ref{eq:IntoIn}): 
\begin{equation}
\overline{\left(\nabla\left(\phi-\phi_{mf}\right)\right)^{2}}=\lim_{N\rightarrow\infty}N\intop_{S_{ij}}\mathrm{d}\mathbf{x}\ \left(\nabla\left(\phi-\phi_{mf}\right)\right)^{2},\quad\mbox{for }\mathbf{x}\in S_{ij},
\end{equation}
where $S_{ij}$ is the surface covered by a grid site $\left(i,j\right)$.
An integration by parts yields 

\begin{equation}
N\intop_{S_{ij}}\mathrm{d}\mathbf{x}\ \left(\nabla\left(\phi-\phi_{mf}\right)\right)^{2}=-N\intop_{S_{ij}}\mathrm{d}\mathbf{x}\ \left[\left(\Delta^{-1/2}\left(\mu-\overline{\mu}\right)\right)\left(\Delta^{1/2}\left(\mu-\overline{\mu}\right)\right)\right]+N\ointop_{\partial S_{ij}}\mathrm{d}l\ \mathbf{n}\cdot\left(\phi-\phi_{mf}\right)\nabla\left(\phi-\phi_{mf}\right)\label{eq:project_lapl}
\end{equation}
Projecting the first term of the rhs on Laplacian eigenmodes allows
to simplify the expression of the first term of the rhs in Eq. (\ref{eq:project_lapl}): 

\begin{equation}
-N\intop_{S_{ij}}\mathrm{d}\mathbf{x}\ \left[\left(\Delta^{-1/2}\left(\mu-\overline{\mu}\right)\right)\left(\Delta^{1/2}\left(\mu-\overline{\mu}\right)\right)\right]=N\intop_{S_{ij}}\mathrm{d}\mathbf{x}\ \left(\mu-\overline{\mu}\right)^{2}.\label{eq:projection_laplace_modes}
\end{equation}
The second term of the rhs in Eq. (\ref{eq:project_lapl}) can be
written as

\begin{equation}
N\ointop_{\partial S_{ij}}\mathrm{d}l\ \mathbf{n}\cdot\left(\phi-\phi_{mf}\right)\nabla\left(\phi-\phi_{mf}\right)=\frac{N}{2}\int_{S_{ij}}\Delta\left[\left(\phi-\phi_{mf}\right)^{2}\right]\label{eq:simplify_boundary term}
\end{equation}
which, according to assumption 4, vanishes in the large $N$ limit.
Finally, the kinetic energy density of the fluctuations is simply
expressed as 
\begin{equation}
\mathcal{E}_{c,fluct}=\frac{1}{2}\int\mathrm{d}\mathbf{x}\:\overline{h}\left(\overline{\mu^{2}}-\overline{\mu}^{2}\right).\label{eq:Ecmf_fluct}
\end{equation}
Finally, we use again the assumption 2 to get 
\begin{equation}
\mathcal{E}_{c,fluct}=\frac{1}{2}\int\mathrm{d}\mathbf{x}\:\left(\overline{h\mu^{2}}-\frac{\overline{h\mu}^{2}}{\overline{h}}\right)=\mathcal{E}_{\delta\mu},\label{eq:Ecfluc_Ecdeltamu}
\end{equation}
which is the expected result.

\section{Critical points of the mean-flow variational problem\label{sec:Critical-points-of} }

In this Appendix, we compute the critical points of the mean-flow
variational problem (\ref{eq:Max_prob_gen}) stated in section \ref{sec:Equilibrium-statistical-mechanic}.
In a first step, we solve an intermediate variational problem in order
to show the factorization of the probability density $\rho$ with
a Gaussian behavior for the divergence fluctuations. Knowing that,
we solve in a second step the original variational problem.

\subsection{Intermediate variational problem\label{intermedprob}}

As the energy and the potential vorticity moments depend only on the
coarse-grained fields $\overline{h},\,\overline{h^{2}},\,\overline{h\mu},\,\overline{h\mu^{2}}$
and the local potential vorticity moments $\left\{ \overline{hq^{k}}\right\} $,
we introduce an intermediate variational problem where these coarse-grained
fields are given as constraint:
\begin{equation}
\max_{\rho,\int\rho=1}\left\{ \mathcal{S}\left[\rho\right]\;\left|\;\overline{h},\overline{h^{2}},\overline{h\mu},\overline{h\mu^{2}},\left\{ \overline{hq^{k}}\right\} _{k\ge1}\text{ "fixed"}\right.\right\} .\label{eq:intermed_var_prob}
\end{equation}
\begin{comment}
The critical points of the general variational problem in Eq. (\ref{eq:Max_prob_gen})
must be among the critical points of the intermediate variational
problem (\ref{eq:intermed_var_prob}). 
\end{comment}
The idea of introducing the intermediate variational problem is to
find a simpler ansatz for the probability density field $\rho$ .
This ansatz will be used afterward into the general variational problem
(\ref{eq:Max_prob_gen}).

In order to compute the critical points of the variational problem
(\ref{eq:intermed_var_prob}), we introduce the Lagrange multipliers
$\alpha_{h}\left({\bf x}\right),\:\alpha_{h2}\left({\bf x}\right),\:\alpha_{h\mu}\left({\bf x}\right),\:\alpha_{h\mu2}\left({\bf x}\right),\:\left\{ \alpha_{hq,k}\left({\bf x}\right)\right\} _{k\geq0}\text{ and }\xi\left({\bf x}\right)$
associated with the constraints $\overline{h},\:\overline{h^{2}},\:\overline{h\mu},\:\overline{h\mu^{2}},\:\left\{ \overline{hq^{k}}\right\} _{k\ge1}$
and the normalization constraint, respectively. Using Eq. (\ref{eq:Mixing_Entropy}) and the first variations 
\begin{equation}
\forall\delta\rho,\ \delta S-\int\mathrm{d}{\bf x}\:\left[\alpha_{h}\delta\overline{h}+\alpha_{h2}\delta\overline{h^{2}}+\alpha_{h\mu}\delta\overline{h\mu}+\alpha_{h\mu2}\delta\overline{h\mu^{2}}+\sum_{k=1}^{+\infty}\alpha_{hq,k}\delta\overline{hq^{k}}+\xi\int\delta\rho\right]=0,
\end{equation}
leads to%
\begin{comment}
\begin{equation}
\forall\delta\rho,\ \int\mathrm{d}{\bf x}\mathrm{d}\sigma_{q}\mathrm{d}\sigma_{h}\mathrm{d}\sigma_{\mu}\;\left[\sigma_{h}\log\left(\frac{\rho}{\sigma_{h}^{\alpha-1}}\right)+\sigma_{h}+\xi+\alpha_{h}\sigma_{h}+\alpha_{h2}\sigma_{h}^{2}+\alpha_{h\mu}\sigma_{h}\sigma_{\mu}+\alpha_{h\mu2}\sigma_{h}\sigma_{\mu}^{2}+\sum_{k=1}^{+\infty}\alpha_{hq,k}\sigma_{h}\sigma_{q}^{k}\right]\delta\rho=0.
\end{equation}
This integral must vanish for any $\delta\rho$, which yields
\end{comment}
 
\begin{equation}
\sigma_{h}\log\left(\frac{\rho}{\sigma_{h}^{2}}\right)+\sigma_{h}+\xi+\alpha_{h}\sigma_{h}+\alpha_{h2}\sigma_{h}^{2}+\alpha_{h\mu}\sigma_{h}\sigma_{\mu}+\alpha_{h\mu2}\sigma_{h}\sigma_{\mu}^{2}+\sum_{k=1}^{+\infty}\alpha_{hq,k}\sigma_{h}\sigma_{q}^{k}=0\label{eq:critical_inter_appC}
\end{equation}
 We readily see from Eq. (\ref{eq:critical_inter_appC}) that the
probability density $\rho$ factorizes into three decoupled probability
densities $\rho_{q},\:\rho_{h}\text{ and }\rho_{\mu}$ corresponding
respectively to the probability densities of the potential vorticity,
the height and the divergence: 
\begin{equation}
\rho=\rho_{q}\left({\bf x},\sigma_{q}\right)\rho_{h}\left({\bf x},\sigma_{h}\right)\rho_{\mu}\left({\bf x},\sigma_{\mu}\right).\label{eq:rho_factorized-1}
\end{equation}
Using the constraints $\overline{\mu}=\int\mathrm{d}\sigma_{\mu}\ \sigma_{\mu}\rho_{\mu}$
, $\overline{\mu^{2}}=\int\mathrm{d}\sigma_{\mu}\ \sigma_{\mu}^{2}\rho_{\mu}$
, as well as the normalization constraints $\int\mathrm{d}\sigma_{\mu}\ \rho_{\mu}=\int\mathrm{d}\sigma_{h}\ \rho_{h}=\int\mathrm{d}\sigma_{q}\ \rho_{q}=1$,
we get

\begin{equation}
\begin{cases}
\begin{array}{ccc}
\rho_{q}\left({\bf x},\sigma_{q}\right) & = & \frac{{\displaystyle \exp\left(-{\displaystyle \sum_{k=1}^{+\infty}\alpha_{hq,k}\left({\bf x}\right)\sigma_{q}^{k}}\right)}}{{\displaystyle \int\mathrm{d}\sigma_{q}^{,}\:\exp\left(-{\displaystyle \sum_{k=1}^{+\infty}\alpha_{hq,k}\left({\bf x}\right)\sigma_{q}^{,k}}\right)}}\\
\\
\rho_{h}\left({\bf x},\sigma_{h}\right) & = & \frac{{\displaystyle \sigma_{h}^{2}\exp\left(-\alpha_{h2}\left({\bf x}\right)\sigma_{h}-\frac{\xi\left({\bf x}\right)}{\sigma_{h}}\right)}}{{\displaystyle \int\mathrm{d}\sigma_{h}^{,}\:\sigma_{h}^{,2}\exp\left(-\alpha_{h2}\left({\bf x}\right)\sigma_{h}^{,}-\frac{\xi\left({\bf x}\right)}{\sigma_{h}^{,}}\right)}}\\
\\
\rho_{\mu}\left({\bf x},\sigma_{\mu}\right) & = & \frac{{\displaystyle \exp\left(-\frac{1}{2}\frac{\left(\sigma_{\mu}-\overline{\mu}\left({\bf x}\right)\right)^{2}}{\overline{\mu^{2}}\left({\bf x}\right)-\overline{\mu}^{2}\left({\bf x}\right)}\right)}}{{\displaystyle \left(2\pi\right)^{1/2}\left(\overline{\mu^{2}}\left({\bf x}\right)-\overline{\mu}^{2}\left({\bf x}\right)\right)^{1/2}}}
\end{array} & .\end{cases}\label{eq:rho_q_h_mu}
\end{equation}
We could now re-inject these expressions into the main variational
problem (\ref{eq:Max_prob_gen}), but only factorization and the Gaussian
form of $\rho_{\mu}$ will be kept as an ansatz for $\rho$, which
will simplify the computations. Thanks to this intermediate variational
problem, we now know that the critical points of the original variational
problem must be of the form:
\begin{equation}
\rho\left({\bf x},\sigma_{h},\sigma_{q},\sigma_{\mu}\right)=\rho_{h}\left(\sigma_{h},{\bf x}\right)\rho_{q}\left(\sigma_{q},{\bf x}\right)\frac{\exp\left(-\frac{1}{2}\frac{\left(\sigma_{\mu}-\overline{\mu}\right)^{2}}{\overline{\mu^{2}}-\overline{\mu}^{2}}\right)}{\left(2\pi\right)^{1/2}\left(\overline{\mu^{2}}-\overline{\mu}^{2}\right)^{1/2}}\ .\label{eq:rho_ansatz}
\end{equation}
The entropy defined in Eq. (\ref{eq:Mixing_Entropy}) is therefore
(up to a constant):
\begin{equation}
\mathcal{S}\left[\rho_{h},\rho_{q},\bar{\mu},\bar{\mu^{2}}-\bar{\mu}^{2}\right]=-\int\mathrm{d}{\bf x}\mathrm{d}\sigma_{h}\;\sigma_{h}\rho_{h}\log\left(\frac{\rho_{h}}{\sigma_{h}^{2}}\right)-\int\mathrm{d}{\bf x}\:\overline{h}\int\mathrm{d}\sigma_{q}\rho_{q}\log\left(\rho_{q}\right)+\int\mathrm{d}{\bf x}\;\frac{\overline{h}}{2}\log\left(\overline{\mu^{2}}-\overline{\mu}^{2}\right).\label{eq:Entropy_factorized}
\end{equation}

As a consequence of Eq. (\ref{eq:rho_ansatz}), the height field,
the potential vorticity field and the divergence field are decorrelated.
This property allows to rewrite the energy defined in Eq. (\ref{eq:Total_energy_funct})
\begin{equation}
\mathcal{E}\left[\rho_{h},\rho_{q},\overline{\mu},\overline{\mu^{2}}-\overline{\mu}^{2}\right]=\mathcal{E}_{mf}\left[\overline{h},\overline{q},\overline{\mu}\right]+\mathcal{E}_{\delta\mu}\left[\overline{h},\overline{\mu^{2}}-\overline{\mu}^{2}\right]+\mathcal{E}_{\delta h}\left[\overline{h},\overline{h^{2}}\right],\label{eq:energy_decorelated_sum}
\end{equation}
where
\begin{equation}
\begin{cases}
\mathcal{E}_{mf}\left[\overline{h},\overline{q},\overline{\mu}\right]=\frac{1}{2}\int\mathrm{d}{\bf x}\:\left(\overline{h}{\bf u}_{mf}^{2}+g\left(\overline{h}+h_{b}-1\right)^{2}\right)\\
\mathcal{E}_{\delta\mu}\left[\overline{h},\overline{\mu^{2}}-\overline{\mu}^{2}\right]=\frac{1}{2}\int\mathrm{d}{\bf x}\:\overline{h}\left(\overline{\mu^{2}}-\overline{\mu}^{2}\right)\\
\mathcal{E}_{\delta h}\left[\overline{h},\overline{h^{2}}\right]=\frac{g}{2}\int\mathrm{d}{\bf x}\:\left(\overline{h^{2}}-\overline{h}^{2}\right)
\end{cases}\label{eq:Energy_decorelated}
\end{equation}
with ${\bf u}_{mf}=\nabla^{\bot}\Delta^{-1}\left(\overline{q}\overline{h}-f\right)+\nabla\Delta^{-1/2}\overline{\mu}$
. Similarly the potential vorticity moments (\ref{eq:Potential_energy_fluct_funct})
can be rewritten 
\begin{equation}
\forall k\quad\mathcal{Z}_{k}\left[\overline{h},\overline{q^{k}}\right]=\int\mathrm{d}{\bf x}\:\overline{h}\overline{q^{k}}\label{eq: PV_moment_decorelated}
\end{equation}
 where the coarse-grained moments are now defined as 
\begin{equation}
\overline{h^{l}}=\int\mathrm{d}\sigma_{h}\;\sigma_{h}^{l}\rho_{h},\qquad\overline{q^{m}}=\int\mathrm{d}\sigma_{q}\;\sigma_{q}^{m}\rho_{q}.\label{eq:moments_decorelated}
\end{equation}
Thus, the general variational problem of the equilibrium theory given
in Eq. (\ref{eq:Max_prob_gen}) can be recast into a new variational
problem on the independent variables $\rho_{h}\left(\mathbf{x},\sigma_{h}\right),\:\rho_{q}\left(\mathbf{x},\sigma_{q}\right),\;\overline{\mu}\left(\mathbf{x}\right)\text{ and }\left[\overline{\mu^{2}}-\overline{\mu}^{2}\right]\left(\mathbf{x}\right)$:
\begin{equation}
S\left(E,D\right)=\max_{\underset{\int\rho_{h}=1,\int\rho_{q}=1}{\rho_{h},\rho_{q},\overline{\mu},\overline{\mu^{2}}-\overline{\mu}^{2}}}\left\{ \mathcal{S}\left[\rho_{h},\rho_{q},\overline{\mu},\overline{\mu^{2}}-\overline{\mu}^{2}\right]\ |\ \mathcal{E}\left[\rho_{h},\rho_{q},\overline{\mu},\overline{\mu^{2}}-\overline{\mu}^{2}\right]=E,\ \forall k\quad\mathcal{Z}_{k}\left[\rho_{h},\rho_{q}\right]=Z_{k}\right\} .\label{eq:var_prob_main_recast}
\end{equation}

\subsection{Computation of the critical points}

In this subsection, we compute the critical points of the variational
problem defined in Eq. (\ref{eq:var_prob_main_recast}). We introduce
the Lagrange multiplier $\beta,\;\left\{ \alpha_{k}\right\} _{k\geq0},\;\xi_{q}\left({\bf r}\right)\text{ and }\xi_{h}\left({\bf r}\right)$
associated respectively with the energy, the potential vorticity moments
and the normalization constraints. Critical points of the variational
problem (\ref{eq:var_prob_main_recast}) are solutions of 
\begin{equation}
\forall\delta\rho_{q},\delta\rho_{h},\delta\overline{\mu},\delta\left(\overline{\mu^{2}}-\overline{\mu}^{2}\right),\quad\delta\mathcal{S}-\beta\delta\mathcal{E}-\sum_{k=0}^{+\infty}\:\alpha_{k}\delta\mathcal{Z}_{k}-\int\mathbf{dx}\:\left(\xi_{q}\int\delta\rho_{q}+\xi_{h}\int\delta\rho_{h}\right)=0.\label{eq:Final_Crit_points}
\end{equation}
The first variations of the macrostate entropy $\mathcal{S}$ (\ref{eq:Entropy_factorized})
are 
\begin{equation}
\begin{cases}
\frac{\delta\mathcal{S}}{\delta\rho_{h}} & =-\sigma_{h}\left(\log\left(\frac{\rho_{h}}{\sigma_{h}^{2}}\right)+1\right)-\sigma_{h}\int\mathrm{d}\sigma_{q}\rho_{q}\log\left(\rho_{q}\right)+\sigma_{h}\frac{1}{2}\log\left(\overline{\mu^{2}}-\overline{\mu}^{2}\right)\\
\frac{\delta\mathcal{S}}{\delta\rho_{q}} & =-\overline{h}\left(\log\left(\rho_{q}\right)+1\right)\\
\frac{\delta\mathcal{S}}{\delta\overline{\mu}} & =0\\
\frac{\delta\mathcal{S}}{\delta\left(\overline{\mu^{2}}-\overline{\mu}^{2}\right)} & =\frac{\overline{h}}{2\left(\overline{\mu^{2}}-\overline{\mu}^{2}\right)}
\end{cases}.\label{eq:Entropy_FV}
\end{equation}

First variations of the energy given in Eqs. (\ref{eq:energy_decorelated_sum})
and (\ref{eq:Energy_decorelated}) contain three contributions: $\delta\mathcal{E}=\delta\mathcal{E}_{mf}+\delta\mathcal{E}_{\delta\mu}+\delta\mathcal{E}_{\delta h}$.
The first contribution is 
\begin{equation}
\delta\mathcal{E}_{mf}=\int\mathrm{d}{\bf x}\;\left[B_{mf}\delta\overline{h}+\left(\overline{h}{\bf u}_{mf}\right)\cdot\delta{\bf u}_{mf}\right],
\end{equation}
where $B_{mf}={\bf u}_{mf}^{2}/2+g\left(\overline{h}+h_{b}-1\right)$
is the mean-flow Bernoulli function defined in Eq. (\ref{eq:Bernouilli_Bar}).
Then, using the Helmholtz decompositions ${\bf u}_{mf}=\nabla^{\bot}\psi_{mf}+\nabla\phi_{mf}$
and recalling that $\overline{h}{\bf u}_{mf}=\nabla^{\bot}\Psi_{mf}+\nabla\Phi_{mf}$,
two integrations by parts with the impermeability boundary condition yield 
\begin{equation}
\delta\mathcal{E}_{mf}=\int\mathrm{d}{\bf x}\;\left[B_{mf}\delta\overline{h}-\Psi_{mf}\delta\Delta\psi_{mf}-\Phi_{mf}\delta\Delta\phi_{mf}\right].
\end{equation}
Using $\Delta\psi_{mf}=\overline{h}\overline{q}-f$ and $\Delta^{1/2}\phi_{mf}=\overline{\mu}$ and the definition of the operator $\Delta^{1/2}$ leads to the final expression
\begin{equation}
\delta\mathcal{E}_{mf}=\int\mathrm{d}{\bf x}\;\left[\left(B_{mf}-\overline{q}\Psi_{mf}\right)\delta\overline{h}-\overline{h}\Psi_{mf}\delta\overline{q}-\Delta^{1/2}\Phi_{mf}\delta\overline{\mu}\right].\label{eq:Mean-field-first-var}
\end{equation}
 Finally, we get:
\begin{equation}
\begin{cases}
\frac{\delta\mathcal{E}}{\delta\rho_{h}} & =\sigma_{h}\left(B_{mf}-\overline{q}\Psi_{mf}+\frac{\overline{\mu^{2}}-\overline{\mu}^{2}}{2}+g\left(\frac{\sigma_{h}}{2}-\overline{h}\right)\right)\\
\frac{\delta\mathcal{E}}{\delta\rho_{q}} & =-\sigma_{q}\Psi_{mf}\overline{h}\\
\frac{\delta\mathcal{E}}{\delta\bar{\mu}} & =-\Delta^{1/2}\Phi_{mf}\\
\frac{\delta\mathcal{E}}{\delta\left(\overline{\mu^{2}}-\overline{\mu}^{2}\right)} & =\frac{\overline{h}}{2}
\end{cases},\label{eq:Energy_FV}
\end{equation}

First variations of the potential vorticity moments are 
\begin{equation}
\forall k\in\mathbb{N}\quad\begin{cases}
\frac{\delta\mathcal{Z}_{k}}{\delta\rho_{h}} & =\sigma_{h}\overline{q^{k}}\\
\frac{\delta\mathcal{Z}_{k}}{\delta\rho_{q}} & =\bar{h}\sigma_{q}^{k}\\
\frac{\delta\mathcal{Z}_{k}}{\delta\overline{\mu}} & =0\\
\frac{\delta\mathcal{Z}_{k}}{\delta\left(\overline{\mu^{2}}-\overline{\mu}^{2}\right)} & =0
\end{cases},\label{eq:Distrib_FV}
\end{equation}

Injecting Eqs. (\ref{eq:Entropy_FV}), (\ref{eq:Energy_FV}), and
(\ref{eq:Distrib_FV}) in Eq. (\ref{eq:Final_Crit_points}), and collecting
the term in factor of $\delta\left(\overline{\mu^{2}}-\overline{\mu}^{2}\right)$
leads to

\begin{equation}
\overline{\mu^{2}}-\overline{\mu}^{2}=\frac{1}{\beta}.\label{eq:Positiv_temperature}
\end{equation}
Injecting Eq. (\ref{eq:Positiv_temperature}) in the expression of
$\rho_{\mu}$ given in Eq. (\ref{eq:rho_q_h_mu}) yields then 
\begin{equation}
\rho_{\mu}\left({\bf x},\sigma_{\mu}\right)=\left(\frac{\beta}{2\pi}\right)^{1/2}{\displaystyle \exp\left(-\frac{1}{2}\beta\left(\sigma_{\mu}-\overline{\mu}\left({\bf x}\right)\right)^{2}\right)}.\label{eq:rho_mu_final_app}
\end{equation}

Similarly, collecting the term in factor of $\delta\rho_{q}$ in Eq.
(\ref{eq:Final_Crit_points}) leads to

\begin{equation}
0=\bar{h}\left(\log\left(p_{q}\right)+1\right)-\beta\sigma_{q}\Psi_{mf}\bar{h}+\sum_{k=0}^{+\infty}\alpha_{k}\bar{h}\sigma_{q}^{k}+\xi_{q},
\label{eq:missingnumbering}
\end{equation}
which, using the normalization constraint, leads to 

\begin{equation}
\rho_{q}\left({\bf x},\sigma_{q}\right)=\frac{1}{\mathbb{G}_{q}}{\displaystyle \exp\left(\beta\sigma_{q}\Psi_{mf}-\sum_{k=1}^{+\infty}\alpha_{k}\sigma_{q}^{k}\right)},\quad\mathbb{G}_{q}={\displaystyle \int\mathrm{d}\sigma_{q}^{,}\;\exp\left(\beta\sigma_{q}^{,}\Psi_{mf}-\sum_{k=1}^{+\infty}\alpha_{k}\sigma_{q}^{,k}\right)}.\label{eq:Final_rho_q}
\end{equation}
Note that the sum inside the exponential is performed from $k=1$
to $k=+\infty$ . The Lagrange parameter $\xi_{q}$ has been determined
using the normalization condition for the pdf. 

Collecting the term in factor of $\delta\rho_{h}$ in Eq. (\ref{eq:Final_Crit_points})
yields

\begin{equation}
-\left(\log\left(\frac{\rho_{h}}{\sigma_{h}^{2}}\right)+1\right)-\int\mathrm{d}\sigma_{q}\rho_{q}\log\left(\rho_{q}\right)-\color{black}\frac{1}{2}\log\left(\beta\right)-\beta\left(B_{mf}+\frac{\beta^{-1}}{2}+g\left(\frac{\sigma_{h}}{2}-\overline{h}\right)\right)-\frac{\xi_{h}}{\sigma_{h}}+\beta\overline{q}\Psi_{mf}-\sum_{k\ge0}\alpha_{k}\overline{q^{k}}=0,\label{eq:inter_appendic_scompute_Bmf}
\end{equation}
which, using Eq. (\ref{eq:Final_rho_q}), leads to 

\begin{equation}
-\left(\log\left(\frac{\rho_{h}}{\sigma_{h}^{2}}\right)+1\right)+\log\left(\mathbb{G}_{q}\right)-\frac{1}{2}\log\beta-\beta\left(B_{mf}+\frac{\beta^{-1}}{2}+g\left(\frac{\sigma_{h}}{2}-\overline{h}\right)\right)-\frac{\xi_{h}}{\sigma_{h}}-\alpha_{0}=0.\label{eq:Bmf_inter_rhoh_app}
\end{equation}
Using the fact that $\mathbb{G}_{q}$ and $B_{mf}$ are fields depending
only on $\mathbf{x}$, and using the normalization constraint for
the pdf $\rho_{h}(\mathbf{x},\sigma_{h})$, Eq. (\ref{eq:Bmf_inter_rhoh_app})
yields 

\begin{equation}
\rho_{h}\left({\bf x},\sigma_{h}\right)=\frac{{\displaystyle \sigma_{h}^{2}}}{{\displaystyle \mathbb{G}_{h}}}\exp\left(-\beta\frac{g}{2}\sigma_{h}-\frac{\xi_{h}}{\sigma_{h}}\right),\quad\mathbb{G}_{h}=\int\mathrm{d}\sigma_{h}^{,}\;\sigma_{h}^{2}\exp\left(-\beta\frac{g}{2}\sigma_{h}^{,}+\frac{\xi_{h}}{\sigma_{h}^{,}}\right).\label{eq:Final_rho_h}
\end{equation}
Injecting Eq. (\ref{eq:Final_rho_h}) back into Eq. (\ref{eq:Bmf_inter_rhoh_app})
yields 

\begin{equation}
B_{mf}=\beta^{-1}\log\left(\mathbb{G}_{q}\mathbb{G}_{h}\right)+g\overline{h}+\beta^{-1}\left(-\alpha_{0}-\frac{3}{2}+\frac{1}{2}\log\beta\right).\label{eq:bmf_final_app}
\end{equation}
One can notice that $\alpha_{0}$ the Lagrange parameter related to
the conservation of the total mass appears only here. Thus the last
term $\beta^{-1}\left(\alpha_{0}-3/2+\log\left(\beta\right)/2\right)$
in Eq. (\ref{eq:bmf_final_app}) can be computed from the conservation
of the total mass $\mathcal{Z}_{0}=Z_{0}$ and will be denoted $A_{0}$
in the following.

Collecting the terms in factor of $\delta\overline{\mu}$ in Eq. (\ref{eq:Final_Crit_points})
leads to

\begin{equation}
\Phi_{mf}=0.\label{eq:Undivergent_mean_flow}
\end{equation}

\section{Global maximizers of the entropy of the large scale flow\label{sec:Global-maximizers-of}}

We compute in this appendix an upper-bound for the macrostate entropy
of the large scale flow defined in Eq. (\ref{entropy_q}), for a given
set of potential vorticity moment constraints defined in Eq. (\ref{eq:Total_moments_funct})
(and arbitrary energy), and then show that when $Z_{1}=f$ and $h_{b}=0$,
this upper bound for the macroscopic entropy is reached by the rest
state. \\

This upper bound is the solution of the following variational problem: 

\begin{equation}
S_{mf,max}=\max_{\substack{\overline{h},\rho_{q}\\
\int\rho_{q}=1
}
}\left\{ \mathcal{S}_{mf}\left[\overline{h},\rho_{q}\right]\ \left|\ \forall k\quad\mathcal{Z}_{k}\left[\overline{h},\rho_{q}\right]=Z_{k}\right.\right\} .\label{eq:varprob_app_smf_max}
\end{equation}
Introducing Lagrange parameters $\left\{ \gamma_{k}\right\} _{k\ge0}$
associated with the potential vorticity moment constraints and the
Lagrange parameter $\xi(\mathbf{x})$ associated with the normalization
constraint, the cancellation of first variations yields

\begin{equation}
\forall\delta\rho_{q},\delta\overline{h},\quad\delta\mathcal{S}_{mf}-\sum_{k=0}^{+\infty}\gamma_{k}\delta\mathcal{Z}_{k}+\int\mathrm{d\mathbf{x}}\mathrm{\ }\xi\delta\overline{1}=0.\label{eq:compute_critical_point_app_Smax}
\end{equation}
The solution of this equation is

\begin{equation}
\rho_{q}=\frac{\exp^{-\sum_{k=1}^{+\infty}\gamma_{k}\sigma^{k}}}{\int\mathrm{d}\sigma\ \exp^{-\sum_{k=1}^{+\infty}\gamma_{k}\sigma^{k}}}\equiv\rho_{global}\left(\sigma\right)\label{eq:compute_rho_q_critical}
\end{equation}
 where $\rho_{global}$ depends only on the potential vorticity moments
constraints $\left\{ Z_{k}\right\} _{k\ge1}$, and is independent
from $\mathbf{x}$ and 
\begin{equation}
\overline{h}\left(\mathbf{x}\right)=1
\end{equation}

Note that the states characterized $\rho_{q}=\rho_{global}$, $\overline{h}=1$
are solutions of the variational problem in Eq. (\ref{eq:varprob_app_smf_max}),
but this is only a subclass of the solutions of the variational problem
of the equilibrium theory given in Eq. (\ref{eq:max_prob_qg_2}),
which includes an additional energy constraint \\

We have shown in subsection (\ref{sub:Critical-point-equations})
that for a given $\rho_{q}$, the large scale flow which is a solution
(\ref{eq:max_prob_qg_2}) is obtained by solving Eqs. (\ref{eq:Psi_Eq_Rel2-1})
and (\ref{eq:Psi_Eq_rel1-1}) for $\Psi_{mf}$ and $\overline{h}$.
Here we consider the particular case $\rho_{q}=\rho_{global}$ and
$\overline{h}=1$. One can compute $\overline{h}\overline{q}_{global}=\int\mathrm{d}\mathbf{x}\ \sigma\rho_{global}=Z_{1}$.
We conclude that the large scale flow of the equilibrium state is
also a global entropy maximizer, i.e. a solution of (\ref{eq:varprob_app_smf_max})
when 

\begin{equation}
Z_{1}-f=\Delta\Psi_{mf}\ ,\label{eq:global_max_large_scale_1}
\end{equation}

\begin{equation}
\frac{1}{2}\left(\nabla\Psi_{mf}\right)^{2}+gh_{b}=A_{2}.\label{eq:global_max_large_scale_2}
\end{equation}
where $A_{2}=\beta\log\mathbb{G}_{q}-A_{1}$ is a constant. We see
that in the case ($Z_{1}=f$, $h_{b}=0$) , the solution of Eqs. (\ref{eq:global_max_large_scale_1})
and (\ref{eq:global_max_large_scale_2}) is the rest state $\Psi_{mf}=cst$
(with $A_{2}=0$). We conclude that the maximum of the macroscopic
entropy of the large scale flow is reached by a flow at
rest when there is no circulation ($Z_{1}=f$) and no bottom topography
($h_{b}=0$).

{ 
\section{Comparison with a Eulerian discrete model}\label{app:eulerian}

The aim of this appendix is to discuss the construction of a possible invariant measure for the shallow water equations through an Eulerian discretization. We prove that the obtained equilibrium states differ from the one obtained through the semi-Lagrangian discretization used in the core of the paper. Moreover, we prove that the equilibrium states are not stationary states of the shallow water equations and that the statistical equilibria are not stable through coarse-graining.

We define a purely Eulerian discrete model by considering the same uniform  $N\times N$ grid as for the semi-Lagrangian model, but assuming that each node is now divided into a finer $n\times n$ uniform microscopic grid. A microscopic configuration is given by the values of the fields $(h,q,\mu)$ for all the nodes of the microscopic grid: 
\begin{equation}
y_{\mathrm{micro}}\equiv\{h_{IJ,ij},q_{IJ,ij},\mu_{IJ,ij}\}_{\substack{1\leq I,J\leq N\\1\leq i,j \leq n}},
\end{equation}
where $(I,J)$ and $(i,j)$ correspond respectively to the position on the macroscopic grid and the position on the microscopic grid within the macroscopic node. 

Contrary to the semi-Lagrangian model, the Eulerian model has the desired property to possibly be compatible with the formal Liouville theorem derived in Appendix \ref{sec:Invariant-measure-and} for the continuous dynamics (although no mathematical result exist). However, the volume of fluid varies from one microscopic grid node to another in the Eulerian model, depending on the value of the height $h_{IJ,ij}$. By comparison, our semi-Lagrangian approach respects the Lagrangian conservation laws (the height h is defined through the particle mass conservation). Because of the need to go through a discretization to build the microcanonical measure, we see that both the Eulerian and the semi-Lagrangian approaches necessarily break part of the geometric conservation laws of the continuous model. Hopefully rigorous mathematical proof of the convergence of the measures of one of the discretized model to an invariant measure of the continuous equations will settle rigorously this issue in a near future, however nobody seem to know how to attack this problem mathematically. We are thus led to the conclusion that based on current knowledge, there is no clear mathematical or theoretical \textit{a priori} argument to choose either the Eulerian or the semi-Lagrangian discretization in order to guess the microcanonical measures. For now, the use of one discrete model or another to guess the microcanonical  measure of the continuous shallow water equations can therefore only be justified \textit{a posteriori}.\\

Let us now define the empirical density field as 
\begin{equation}
d_{IJ}(\sigma_{h},\sigma_{q},\sigma_{\mu})[y_{\mathrm{micro}}]=\frac{1}{n}\sum_{i,j=1}^{n}\delta(h_{IJ,ij}-\sigma_{h})\delta(q_{IJ,ij}-\sigma_{q})\delta(\mu_{IJ,ij}-\sigma_{\mu}).\label{CG_new}
\end{equation}
One can now compute the entropy of the macrostates $\rho=\left\{ y_{micro}\left|\:\forall I,J\quad d_{IJ}\left[y_{micro}\right]=\rho_{IJ}\right.\right\}$, which, after taking first the limit $n\to\infty$  and then the limit $N\to\infty$ leads to 
\begin{equation}
\mathcal{S}_{\mathrm{Eul}}[\rho]=-\int\mathrm{d}\mathbf{x}\mathrm{d}\sigma_{h}\mathrm{d}\sigma_{q}\mathrm{d}\sigma_{\mu}\quad\rho(\mathbf{x},\sigma_{h},\sigma_{q},\sigma_{\mu})\log\left(\frac{\rho(\mathbf{x},\sigma_{h},\sigma_{q},\sigma_{\mu})}{\sigma_{h}^{3}}\right).\label{eq:entropy_new}
\end{equation}

This Eulerian macrostate entropy has to be compared with the macrostate entropy for the semi-Lagrangian discrete model given in Eq. (\ref{eq:Mixing_Entropy}). We can switch from  expression to the other by changing $\rho$ into $\sigma_{h}\rho$. We note that the two entropies become equivalent at lowest order in the limit of weak height fluctuations and weak height variations. However, in the general case, they are different, and therefore lead to different equilibrium states. In particular, is it straightforward to show that because of the absence of the factor $\sigma_h$ in the expression of this Eulerian macrostate entropy (\ref{eq:entropy_new}), the critical points $\rho(\mathbf{x},\sigma_h,\sigma_q,\sigma_{\mu})$ of the microcanonical variational problem do not factorize. Consequently,  small scale height and velocity fluctuations of the equilibrium state are correlated. One can then show that those correlations are associated with non-zero Reynolds stresses in the momentum equations.  In particular, the equilibrium state of Eulerian model satisfies 
\begin{equation}
J(\Psi,\overline{q})=-\overline{J(\Psi^{\prime},q^{\prime})}-\overline{J(\Phi^{\prime},q^{\prime})} ,\label{reynolds_stress}
\end{equation}
where the r.h.s. is non-zero. If one removes those small scale fluctuations, the large scale flow is not a stationary state of the dynamics since $J(\Psi,\overline{q})\ne 0$. In other words, the equilibrium states of the Eulerian model are not stable by coarse-graining, contrary to the equilibria of the semi-Lagrangian model. Moreover, Eq. (\ref{reynolds_stress}) and the properties  of stationary states derived in subsection \ref{sub:Stationary-states} imply that neither the potential vorticy field $\overline{q}$ nor the Bernoulli potential $B_{\mathrm{mf}}$ can simply be expressed as a function of $\Psi_{\mathrm{mf}}$.  As shown in subsection \ref{GeoBalsub}), the fact that $B_{\mathrm{mf}}$ is a function of $\Psi_{\mathrm{mf}}$ is essential to prove that the equilibrium is characterized by geostrophic balance at lowest order in the Rossby number $Ro$, when $Ro\rightarrow 0$. Consequently, the proof of geostrophic balance derived in the framework of the semi-Lagrangian model does not hold in the framework of the Eulerian model, unless the bottom topography is sufficiently small ($h_b\sim Ro$).  
\\ 

Let us finally argue that the stability by coarse-graining is a desirable physical property for the equilibrium states. 

The first argument is a body of empirical observations. In either experiments, geophysical flows or numerical simulations flows governed by the shallow water equations (or the Navier-Stokes equations or the primitive equations in a shallow water regime) in the inertial limit (when they are subjected to weak forcing and dissipation, with a clear time scale separation) do actually self-organize and form large scale coherent structures for which there is a gradual decoupling of the flow large scales and small scales. A prominent example is the velocity field of Jupiter's troposphere. 

The second argument follows. Macrostates that evolve through an autonomous equation, must increase the Boltzmann entropy.  This is a general result in statistical mechanics, which is a consequence of the definition of the macrostate entropy as a Boltzmann entropy. Indeed as the Boltzmann's entropy measure the number of microstates corresponding to a given macrostate, it must increase for most of initial conditions. When there is furthermore a concentration property (which is the case for the shallow water case, both the the Eulerian and sem--Lagrangian discretizations), the number of initial conditions for which the entropy can decrease decays exponentially with $N$ ($N$ is often the number of particles in statistical mechanics, here the number of degrees of freedom of our discretization). As a consequence, the set of equilibrium macrostates (entropy maxima) has to be stable through the dynamics for most initial conditions. In the shallow water case, in statistical equilibrium, obtained either using the semi-Lagrangian or Eulerian discretization discussed above, the stream function concentrates close to a single field (the stream function fluctuations vanish in the large $N$ limit). As a consequence the macrostate stream function, which is a single field thanks to this concentration property, has to be stationary for the dynamics. Those two properties, that follow from the definition of the Boltzmann entropy, are actually verified for the equilibrium measure constructed from a semi-Lagrangian discretization, but not for the equilibrium measure constructed from a Eulerian discretization. For this reason, we conclude that the microcanonical measure constructed from the purely Eulerian discretization is inconsistent with the shallow water dynamics.
 
We note moreover that the stability of the equilibrium macrostates through coarse graining ensures that the equilibrium states of the inviscid system are not affected by perturbations such as a weak small scale dissipation in momentum equations. This property is not a-priori required for the invariant measure of the shallow-water equations. However It is extremely interesting as it is a hint that this invariant measure may be relevant for non perfect flow in the inertial limit.
}

\section{Energy-Enstrophy ensemble\label{sec:Energy-Enstrophy-ensemble-detail}}

\subsection{Computation of the critical points}

In this Appendix, we compute the solutions of the variational problem (\ref{eq:max_prob_qg_2-1})
and describe the corresponding phase diagram. Critical points of the
variational problem (\ref{eq:max_prob_qg_2-1}) are computed through
the variational principle:
\begin{equation}
\forall\delta\rho_{g},\quad\delta\mathcal{S}_{mf,g}-\frac{1}{E_{fluct}}\delta\mathcal{E}_{mf,g}-\gamma_{2}\delta\mathcal{Z}_{g2}-\gamma_{1}\delta\mathcal{Z}_{g1}-\int\mathrm{d}\mathbf{x}\ \xi(\mathbf{x})\int\mathrm{d}\sigma_{q}\ \delta\rho_{g}=0,
\end{equation}
where $\gamma_{2},\gamma_{1}$ and $\xi(\mathbf{x}$ are Lagrange multipliers associated
with the enstrophy conservation, the circulation conservation, and the normalization respectively. Anticipating
the coupling between the large scale quasi-geostrophic flow and the
small scale fluctuations, the temperature is denoted $E_{fluct}$
(the inverse temperature is the Lagrange parameter associated with
energy conservation).This yields 
\begin{equation}
\rho_{g}\left({\bf x},\sigma_{q}\right)=\sqrt{\frac{1}{2\pi\left(Z_{2}-\overline{Z}_{2}\right)}}\exp\left[-\frac{1}{2\left(Z_{2}-\overline{Z}_{2}\right)}\left(\sigma_{q}-\left(\frac{\psi_{mf}}{E_{fluct}}-\gamma_{1}\right)\left(Z_{2}-\overline{Z}_{2}\right)\right)^{2}\right]\label{eq:rhog_energy_enstrophy}
\end{equation}
where we have introduced the enstrophy of the coarse-grained potential
vorticity 
\begin{equation}
\overline{Z}_{2}\equiv\int\mathrm{d}\mathbf{x}\overline{q}_{g}^{2}.\label{eq:enstr_Macro-1}
\end{equation}
Injecting (\ref{eq:rhog_energy_enstrophy}) in Eq. (\ref{eq:qg_moments})
, using the mass conservation constraint given in Eq. (\ref{eq:mass_conservation_psi_mf})
and the zero circulation constraint $\mathcal{Z}_{1}\left[\overline{q}_{g}\right]=0$
yields 

\begin{equation}
\overline{q}_{g}=\widetilde{\beta}\psi_{mf}\quad\mbox{with }\widetilde{\beta}\equiv\frac{\left(Z_{2}-\overline{Z_{2}}\right)}{E_{fluct}}.\label{eq:q_psi_lin}
\end{equation}
Note that $\widetilde{\beta}$ is necessarily positive given that $Z_{2}-\overline{Z_{2}}\ge0$.
Injecting Eq. (\ref{eq:q_psi_lin}) in Eq. (\ref{eq:qg_pv_def}),
the streamfunction can be computed explicitly by solving

\begin{equation}
\widetilde{\beta}\psi_{mf}=\Delta\psi_{mf}-\frac{1}{R^{2}}\psi_{mf}+h_{b}.\label{eq:Helmholtz_EnergyEnstrophy}
\end{equation}
In order to solve this equation, it is convenient to introduce the
Laplacian eigenmodes of the domain $\mathcal{D}$, with $k\in\mathbb{N}^{+}$:
\begin{equation}
\Delta e_{k}=-\lambda_{k}^{2}e_{k}\quad\mbox{with }e_{k}=0\ \mbox{on \ensuremath{\partial\mathcal{D}}},\label{eq:Laplacian_eigenmodes}
\end{equation}
where the eigenvalues $-\lambda_{k}^{2}$ are arranged in decreasing
order.  We assume those eigenvalues are pairwise distinct. We also assume that the bottom topography is sufficiently smooth such that $\sum_{k}\left| h_{bk}\right|^{2}\lambda_{k}^{2}<+\infty$. Then, given that $\widetilde{\beta}>0$ , the projection of
the mean flow streamfunction on the Laplacian eigenmode $e_{k}(\mathbf{x)}$
is obtained directly from Eq. (\ref{eq:Helmholtz_EnergyEnstrophy}):

\begin{equation}
\psi_{k}=\frac{h_{bk}}{\widetilde{\beta}+\lambda_{k}^{2}+R^{-2}}.\label{eq:Psi_EnergyEsntrophy}
\end{equation}
We see that there is a unique solution $\psi_{mf}$ for each value
of $\widetilde{\beta}$. This solution is therefore the equilibrium
state. All the large scale flows associated with statistical equilibrium
states of the shallow water system restricted to the energy-enstrophy
ensemble with zero circulation are obtained from Eq. (\ref{eq:Psi_EnergyEsntrophy})
when varying $\widetilde{\beta}$ from $0$ to $+\infty$.

\subsection{Construction of the phase diagram }

The problem is now to find which equilibrium state is associated
with the constraints $(E,Z_{2})$. In the following, we explain
how to find the equilibrium states associated with parameters $(E_{mf},Z_{2})$,
and how to compute the temperature $E_{fluc}$ for each of those states.
It is then straightforward to obtained the total energy $E=E_{mf}+E_{fluc}$.\\

\begin{figure}
\centering\includegraphics[width=\textwidth]{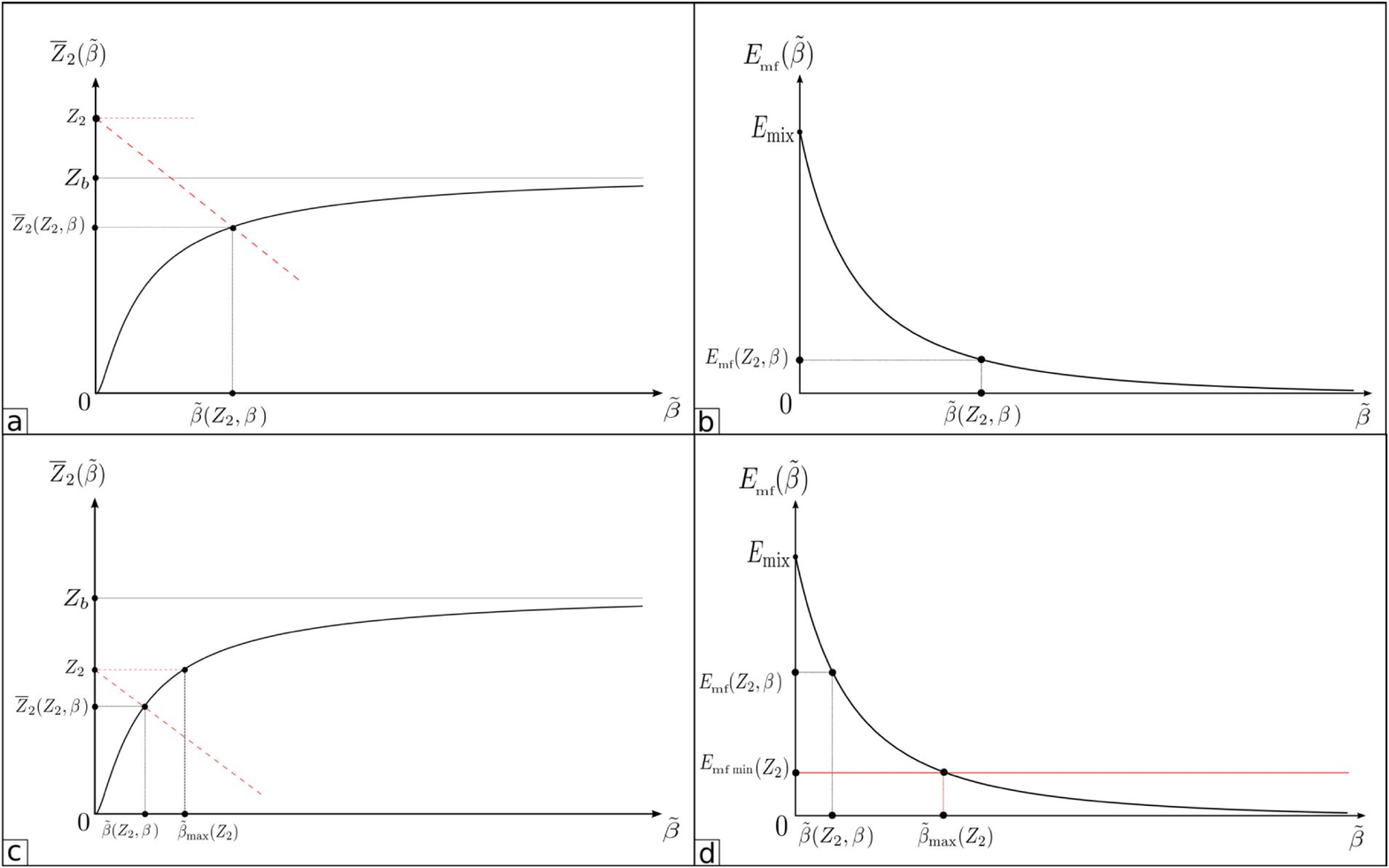}

\caption{a)  Variation of the macroscopic enstrophy $Z_{2}$ over $\widetilde{\beta}$, case $Z_2>Z_b$. b) Variation of  the mean-flow energy $E_{mf}$ with $\widetilde{\beta}$, case $Z_2<Z_b$. c) Variation of the macroscopic enstrophy $Z_{2}$ with $\widetilde{\beta}$, case $Z_2<Z_b$. d) Variation of  the mean-flow energy $E_{mf}$ with $\widetilde{\beta}$, case $Z_2<Z_b$. }\label{Emf_Z2_beta}
\end{figure}

Injecting Eq. (\ref{eq:Psi_EnergyEsntrophy}) in the expression of
the quasi-geostrophic mean-flow energy defined in Eq. (\ref{eq:Eqg_def})
yields 
\begin{equation}
E_{mf}=\frac{1}{2}\sum_{k=1}^{+\infty}\left(\lambda_{k}^{2}+R^{-2}\right)\left(\frac{\left|h_{bk}\right|}{\widetilde{\beta}+\lambda_{k}^{2}+R^{-2}}\right)^{2}.\label{eq:Eq_EnergyEnstrophy}
\end{equation}
The mixing energy $E_{mix}$ defined in Eq. (\ref{eq:Eb_def}) is
recovered for $\widetilde{\beta}=0$, given that $Z_{1}=0$. In the
range $\widetilde{\beta}>0$, the energy $E_{mf}$ is a decreasing function
of $\widetilde{\beta}$, varying from $E_{mf}=E_{mix}$ to $E_{mf}=0$,
see Fig. \ref{Emf_Z2_beta}-b and Fig. \ref{Emf_Z2_beta}-d.\\

Injecting Eq. (\ref{eq:Psi_EnergyEsntrophy}) in the expression of
the macroscopic enstrophy given in Eq (\ref{eq:enstr_Macro-1}) yields 

\begin{equation}
\overline{Z}_{2}=\sum_{k=1}^{+\infty}\left|h_{bk}\right|^{2}\left(1-\frac{\lambda_{k}^{2}+R^{-2}}{\widetilde{\beta}+\lambda_{k}^{2}+R^{-2}}\right)^{2}.\label{eq:Zmacro_EnergyEnstrophy}
\end{equation}
The potential enstrophy $Z_{b}$ defined in Eq. (\ref{eq:Zb_def})
is recovered for $\widetilde{\beta}=+\infty$. The macroscopic enstrophy
$\overline{Z}_{2}$ is an increasing function of $\widetilde{\beta}$,
varying from $\overline{Z_{2}}=0$ (for $\widetilde{\beta}=0$) to $\overline{Z_{2}}=Z_{b}$
for ($\widetilde{\beta}=+\infty$ ), see Fig. \ref{Emf_Z2_beta}-a and
Fig. \ref{Emf_Z2_beta}-c.\\

Two expressions of the macroscopic enstrophy $\overline{Z_{2}}$ in
terms of the parameters $\widetilde{\beta}$ have been obtained: one is
given by Eq. (\ref{eq:Zmacro_EnergyEnstrophy}), the other arises
from the definition of $\widetilde{\beta}$ in Eq. (\ref{eq:q_psi_lin}),
which yields 
\begin{equation}
\overline{Z_{2}}=Z_{2}-E_{fluct}\widetilde{\beta}.\label{eq:Z2_Mac_lin}
\end{equation}

For given values of $E_{fluct}$ and $Z_{2}$ , the values of $\widetilde{\beta}$
and $\overline{Z_{2}}$ are obtained by finding the intersection between
the two curves defined in Eq. (\ref{eq:Zmacro_EnergyEnstrophy}) and
(\ref{eq:Z2_Mac_lin}), respectively. Once $\widetilde{\beta}$ is obtained,
Eq. (\ref{eq:Eq_EnergyEnstrophy}) gives directly the value of the
mean-flow energy $E_{mf}$. The phase diagram presented in Fig. \ref{fig:Phase_Diag_and_E_part}
is obtained numerically by using this procedure. Graphical arguments
presented in the following allow to understand the structure of this
phase diagram. 

\subsection{Limit cases for the energy partition}

Let us first note through Figs. \ref{Emf_Z2_beta}-a and \ref{Emf_Z2_beta}-c that $\widetilde{\beta}$ is an decreasing function of $E_{fluct}$. Indeed, $\widetilde{\beta}$ is given by the intersection between the solid curve representing the expression of $\overline{Z}_{2}$ given Eq. (\ref{eq:Zmacro_EnergyEnstrophy}) and the dashed line representing the affine expression of $\overline{Z}_{2}$ given Eq. (\ref{eq:Z2_Mac_lin}) where $-E_{fluct}$ is the slope. Then we know that the total energy $E=E_{mf}(\widetilde{\beta})+E_{fluct}$ is an increasing function of $E_{fluct}$. Let us now consider different limit cases.

\textbf{The limit $\mathbf{E\to\infty}$ with $\mathbf{Z_2}$ fixed:} In this limit, we have also $E_{fluct}\to\infty$. and $\widetilde{\beta}\to0$ Hence, one gets from Eq. (\ref{eq:Eq_EnergyEnstrophy}) (see also Figs. \ref{Emf_Z2_beta}-b and \ref{Emf_Z2_beta}-d):
\begin{equation}
\lim_{E\to+\infty}E_{mf}=E_{mix},\quad\lim_{E\to+\infty}\frac{E_{mf}}{E}=0.\label{eq:Conclude_large_E_appendix}
\end{equation}

\textbf{The lowest $\mathbf{E}$ limit with $\mathbf{Z_2<Z_{b}}$ fixed:} In this limit, we have also $E_{fluct}\to0$. One gets from Fig. \ref{Emf_Z2_beta}-c that $\widetilde{\beta}\to\widetilde{\beta}_{max}\left(Z_2\right)$. Hence, $E_{mf}$ reaches a minimum admissible energy $E_{min}\left(Z_2\right)=E_{mf}\left(\widetilde{\beta}_{max}\left(Z_2\right)\right)$. Then:
\begin{equation}
\lim_{E\to E_{min}\left(Z_2\right)}\frac{E_{mf}}{E}=1.\label{eq:Conclude_small_z2lzb}
\end{equation}

\textbf{The limit $\mathbf{E\to0}$ with $\mathbf{Z_2>Z_{b}}$ fixed:} In this limit, we have also $E_{fluct}\to0$. One gets from Fig. \ref{Emf_Z2_beta}-a that $\widetilde{\beta}\to\infty$ and that $\overline{Z}_2 \to Z_b$. Hence, from Eqs. (\ref{eq:Eq_EnergyEnstrophy}) and (\ref{eq:Z2_Mac_lin}), we obtain:
\begin{equation}
\begin{cases}
E_{mf}&=\;C_b\widetilde{\beta}^{-2}+o\left(\widetilde{\beta}^{-2}\right)\\
\widetilde{\beta}&\underset{E\to 0}{\sim}\;\left(Z_2-Z_b\right)E_{fluct}^{-1}
\end{cases},\quad\text{with}\quad C_b=\frac{1}{2}\sum_{k=1}^{+\infty}\left|h_{bk}\right|^{2}\left(\lambda_{k}^{2}+R^{-2}\right).\label{eq:Cb}
\end{equation}
Here, $C_b$ is a constant depending on the topography only. Thus we have $E_{mf}\sim E_{fluct}^{2}C_{b}/\left(Z_2-Z_b\right)^{2}$, which leads to:
\begin{equation}
\lim_{E\to 0}\frac{E_{mf}}{E_{fluct}}=0,\quad\lim_{E\to 0}\frac{E_{mf}}{E}=0.\label{eq:Conclude_small_z2gzb}
\end{equation}

\textbf{The limit $\mathbf{E\to0}$ with $\mathbf{Z_2-Z_{b}\sim C_{\alpha}E^{\alpha}}$ with $\mathbf{\alpha\geq0}$:} In this limit, we have $E_{fluct}\to0$. One gets from Fig. \ref{Emf_Z2_beta}-a that $\widetilde{\beta}\to\infty$. Hence, from Eqs. (\ref{eq:Eq_EnergyEnstrophy}) and (\ref{eq:Zmacro_EnergyEnstrophy}), we obtain:
\begin{equation}
\begin{cases}
E_{mf}&=\;C_b\widetilde{\beta}^{-2}+o\left(\widetilde{\beta}^{-2}\right)\\
Z_b-\overline{Z}_2&=\;4C_b\widetilde{\beta}^{-1}+o\left(\widetilde{\beta}^{-1}\right)
\end{cases},\label{eq:emf_cb}
\end{equation}
where $C_b$ is defined in Eq. (\ref{eq:Cb}). From those two equations along with Eq. (\ref{eq:Z2_Mac_lin}) and using $Z_2-Z_b\sim C_{\alpha}E^{\alpha}$,  we can extract:
\begin{equation}
\widetilde{\beta}\underset{E\to 0}{\sim}\frac{C_{\alpha}E^{\alpha}+\sqrt{C_{\alpha}^{2}E^{2\alpha}+20C_{b}E}}{2E}.\label{eq:betaequiv}
\end{equation}
Now, we have to consider different cases for the value of $\alpha$.

For $\alpha>1/2$, we have from Eq. (\ref{eq:betaequiv}) that $\widetilde{\beta}\sim\sqrt{5C_b}E^{-1/2}$. Injecting this in Eq. (\ref{eq:emf_cb}), we gets:
\begin{equation}
\lim_{E\to 0}\frac{E_{mf}}{E}=\frac{1}{5}\label{eq:Conclude_small_z2gzb_ag12}
\end{equation}

For $\alpha<1/2$, we have from Eq. (\ref{eq:betaequiv}) that $\widetilde{\beta}\sim C_{\alpha}E^{\alpha-1}$. Injecting this in Eq. (\ref{eq:emf_cb}), we gets:
\begin{equation}
\lim_{E\to 0}\frac{E_{mf}}{E}=0\label{eq:Conclude_small_z2gzb_al12}
\end{equation}

For $\alpha=1/2$, we have from Eq. (\ref{eq:betaequiv}) that $\widetilde{\beta}\sim C_{1/2}E^{-1/2}\left(1+\sqrt{1+20C_{b}/C_{\alpha}^{2}}\right)/2$. Injecting this in Eq. (\ref{eq:emf_cb}), we gets:
\begin{equation}
\lim_{E\to 0}\frac{E_{mf}}{E}=\frac{2C_{b}/C_{\alpha}^{2}}{\left(1+\sqrt{1+20C_{b}/C_{\alpha}^{2}}\right)^{2}}\label{eq:Conclude_small_z2gzb_ae12}.
\end{equation}
Contrary to the previous cases, here, the partition of the energy depends on the bottom topography.

\begin{acknowledgements}
The authors warmly thank M. Potters and A. Licari for their preliminary work on this subject during a traineeship under the supervision of FB. The research leading to these results has received funding from the European Research Council under the European Union's Seventh Framework Programme (FP7/2007-2013 Grant Agreement no. 616811) (FB and AV). We warmly thank the three referees for their very positive evaluation of our work and for the numerous remark that helped us to improve our work. The level of commitment of the three referee in reading the detail of our computations has been extremely high, and we thank them for this important work.
\end{acknowledgements}

\bibliographystyle{spmpsci}      % mathematics and physical sciences
%\bibliography    % name your BibTeX data base
%\bibliographystyle{amsplain}
\bibliography{shallow_water,Ocean,Meca_Stat_Euler}

\end{document}